\documentclass[a4paper,11pt]{article}
\pdfoutput=1 % if your are submitting a pdflatex (i.e. if you have
\usepackage{jcappub} % for details on the use of the package, please
\usepackage[T1]{fontenc} % if needed
\usepackage{graphicx,multirow}
\usepackage{subcaption}
\usepackage{longtable}
\usepackage{dcolumn}% Align table columns on decimal point
\usepackage{bm}% bold math
\usepackage[all]{xy}
\usepackage{color}
\usepackage[usenames,dvipsnames]{xcolor}
\usepackage{lscape}
\usepackage{tikz}
\usepackage{pgf}
\usepackage{booktabs}
\usepackage{subcaption}
\usepackage{float}

\usepackage[normalem]{ulem} 

\mathchardef\minus="002D

\def\LCDM{$\Lambda \text{CDM }$}

\newcommand{\CLASS}{\texttt{CLASS}}
\newcommand{\BDMCLASS}{\texttt{BdmCLASS}}
\newcommand{\HyRec}{\texttt{HyRec}}

\newcommand{\FirstClass}{\texttt{21cmFirstCLASS}}
\newcommand{\Sense}{\texttt{21cmSense}}

\title{\Large
21 cm Cosmology of Boosted Dark Matter}

\author[a]{Min Sub Lee,}
\author[a]{Jeong Han Kim,}
\author[b]{Kyoungchul Kong,}
\author[c]{Jong-Chul Park}

\affiliation[a]{Department of Physics, Chungbuk National University, Cheongju, Chungbuk 28644, Republic of Korea}
\affiliation[b]{Department of Physics and Astronomy, University of Kansas, Lawrence, KS 66045, USA}
\affiliation[c]{Department of Physics and Institute for Sciences of the Universe, Chungnam National University, Daejeon 34134, Republic of Korea}

\emailAdd{eminsub@chungbuk.ac.kr}
\emailAdd{jeonghan.kim@cbu.ac.kr}
\emailAdd{kckong@ku.edu}
\emailAdd{jcpark@cnu.ac.kr}

\abstract{
The formation history of low-mass halos is sensitive to the properties of dark matter. At low redshift, nonlinear evolution and baryonic processes can obscure this information. 
The 21\,cm signal provides access to a high-redshift stage of structure formation, when the abundance and clustering of low-mass halos retain direct sensitivity to the mass and interactions of dark matter.
Boosted dark matter (BDM) realizes this possibility through a two-component scenario in which the annihilation of the heavier component converts part of the mass difference into kinetic energy of the lighter component. 
Self-interactions redistribute this energy and generate pressure support, suppressing small-scale density perturbations even for dark matter masses in the MeV range. 
We calculate the global 21\,cm brightness temperature and its power spectrum in BDM cosmology. 
The reduced abundance of low-mass halos delays the global absorption feature and shifts the characteristic redshift evolution of the power spectrum.
We assess the projected sensitivity of SKA1-Low to these signatures.
The sensitivity is strongest at intermediate light-component fractions, where both components remain sufficiently abundant for self-heating to affect structure formation. 
For $\sigma_{\rm self1}/m_1\simeq0.1\,{\rm cm^2\,g^{-1}}$, models with $0.2\lesssim r_1\lesssim0.7$ can be distinguished from the fiducial \LCDM\ prediction at the $95\%$ confidence level under the optimistic forecast.
We compare our results with existing direct-detection bounds from SENSEI and prospects for BDM searches at large-volume neutrino experiments, including Super-/Hyper-Kamiokande, JUNO, and DUNE. This comparison highlights the unique and complementary role of high-redshift surveys in testing BDM models.
}

\begin{document}
\maketitle
\flushbottom

%%%%%%%%%%%%%%%%%%%%%%%%%%%%%%%%%%%%%%%%%%%%%%%%%%%
\section{Introduction}
\label{sec:intro}
%%%%%%%%%%%%%%%%%%%%%%%%%%%%%%%%%%%%%%%%%%%%%%%%%%%

The cosmic dawn marks the epoch when the first stars and galaxies emerged from the gravitational scaffolding of dark matter (DM)~\cite{Barkana:2000fd, Bromm:2011cw,Klessen:2023qmc}. In this era, primordial density fluctuations seeded during inflation grew nonlinear, collapsed into the first DM halos, and provided the potential wells in which baryons could cool and form the earliest luminous objects~\cite{Abel:2001pr,Bromm_2013,Bromm:2009}. Since the abundance and internal structure of these halos can retain imprints of the underlying DM properties, observations of high-redshift galaxies offer a powerful way to test a broad class of dark-sector scenarios~\cite{Frenk:2012ph,Wechsler:2018pic}.

This theoretical motivation is now being matched by rapid observational progress, as the James Webb Space Telescope (JWST)~\cite{Castellano_2022,Naidu_2022, Harikane_2023,Carniani_2024} has opened a new window onto early galaxy populations over roughly $7\lesssim z\lesssim14$.
At the same time, planned wide-field surveys with the Nancy Grace Roman Space Telescope~\cite{Spergel:2015sza,Yung_2022} will complement these deep JWST observations by mapping much larger sky areas and identifying rare galaxy candidates over roughly $6\lesssim z\lesssim10$.

A complementary probe is provided by the redshifted 21 cm signal from neutral hydrogen~\cite{Madau:1996cs,Barkana:2000fd,Furlanetto:2006jb,Morales:2009gs,Pritchard:2012,Furlanetto:2015apc,Mesinger:2010ne}. This signal arises from the hyperfine transition of the ground state of neutral hydrogen and is measured relative to the cosmic microwave background (CMB)~\cite{Planck:2018vyg}. A notable aspect of this probe is that a microscopic quantum transition can be observed as a cosmological field over macroscopic volumes. More importantly, the same signal provides an indirect probe of DM, since any change in the growth of small-scale structure modifies the abundance and clustering of the DM halos that host the first luminous sources shaping the 21 cm signal during the cosmic dawn ($15\lesssim z\lesssim 30$) and the epoch of reionization ($6\lesssim z\lesssim 12$).
Current and upcoming radio telescopes offer new opportunities for 21\,cm cosmology. Ground-based searches for the 21 cm signal include the LOw-Frequency ARray (LOFAR)~\cite{LOFAR:van2013,Patil:2017zqk,Gehlot_2019} and the Hydrogen Epoch of Reionization Array (HERA)~\cite{DeBoer:2016tnn,Berkhout:2024,HERA:2021bsv,HERA:2022wmy}. Their main redshift ranges are  $6\lesssim z\lesssim 11$ and $6\lesssim z\lesssim 13$, respectively. LOFAR also targets the 21 cm signal at $z\simeq20$--$25$, and the upgraded HERA Phase II configuration covers the observing frequencies corresponding to redshifts up to $z\simeq27$. The low-frequency component of the Square Kilometre Array (SKA-Low)~\cite{Mellema:2013,Cosmology-SWG:2015tjb,Koopmans:2015sua,Greig:2019tcg} is expected to probe $6\lesssim z\lesssim 27$ in the future.
Observations at still lower frequencies are limited by the Earth's ionosphere~\cite{Jester:2009dw}. Proposed telescopes on the lunar far side could overcome this limitation. These include the Lunar Crater Radio Telescope (LCRT)~\cite{Goel:2022jgw}, targeting $46\lesssim z\lesssim472$, and FARSIDE~\cite{Burns:2021pkx}, targeting $35\lesssim z\lesssim200$.

This expanding experimental program makes the 21 cm signal a promising probe of DM models at high-$z$. In the standard $\Lambda$CDM framework, cold DM (CDM) has negligible primordial velocity dispersion, allowing primordial density perturbations to grow efficiently into low-mass halos. This efficient assembly of low-mass halos brings forward the formation of the first luminous sources and the radiation backgrounds that shape the 21 cm signal.\footnote{Although CDM itself has negligible velocity
dispersion, baryons and CDM retain a coherent relative streaming velocity
after recombination, which can suppress gas accretion into minihalos and
modify early structure formation and reionization, with possible imprints
on the 21 cm signal~\cite{Ahn:2018rkq,Park:2020oqg,Park:2020ydt}.}
Although CDM remains the standard DM scenario, its predictions have not yet been directly tested by current 21 cm observations. Radio interferometric searches have placed upper bounds on the 21 cm power spectrum~\cite{Munshi:2025hgk,Sims:2025hfm,Mertens:2025pvk,HERA:2025ajm}, but these limits remain above the amplitudes expected in standard CDM. 
The EDGES report of a global 21 cm absorption feature at $z\simeq17$~\cite{Bowman_2018}\footnote{In this work, we conservatively restrict our use of 21 cm constraints to those derived from measurements of the 21 cm power spectrum. For recent reviews and broader discussions of DM scenarios proposed to explain the EDGES result, see Refs.~\cite{Asadi:2022njl,Bechtol:2022koa,Liu:2022iyy}.} corresponds to a brightness-temperature contrast of approximately $\simeq -500\;{\rm mK}$, substantially more negative than standard astrophysical expectations within $\Lambda$CDM. However, this cosmological interpretation remains unsettled, particularly after the SARAS 3 non-detection of an EDGES-like profile~\cite{Singh:2021mxo}.

The status of CDM as the baseline DM scenario is well motivated by the success of the $\Lambda$CDM framework in describing the CMB and large-scale structure. From a particle physics perspective, a canonical realization of CDM is provided by weakly interacting massive particles (WIMPs)~\cite{Jungman:1995df,Bertone:2004pz}, with masses typically in the GeV to TeV range and historically motivated in part by supersymmetric extensions of the Standard Model (SM). This paradigm has driven extensive searches through direct detection~\cite{Schumann:2019eaa}, indirect detection~\cite{Gaskins:2016cha}, and collider experiments~\cite{Boveia:2018yeb}, but its mass and non-gravitational interactions remain largely unknown. At the same time, CDM predictions on subgalactic scales remain subject to possible tensions~\cite{Bullock:2017xww,Perivolaropoulos:2021jda}. These considerations, therefore, motivate searches for nonstandard DM scenarios,with high-redshift 21 cm observations now offering a new avenue.

A well-studied example is warm dark matter (WDM), which is typically associated with keV scale masses.
In thermal relic WDM models, the keV scale mass of the DM particle leads to a non-negligible residual velocity dispersion, which induces free streaming and suppresses the formation of low-mass halos relative to CDM. This suppression delays the emergence of the first stars and galaxies, thereby shifting the thermal and ionization history of the intergalactic medium (IGM) and modifying the 21 cm signal.\footnote{Beyond WDM, 21 cm signatures arising from similar suppressed small-scale structure have been studied in fuzzy DM models~\cite{Nebrin:2018vqt,Jones:2021mrs,Giri:2022nxq} and interacting DM models within the effective theory of
structure formation (ETHOS) framework~\cite{Das:2017nub,Munoz:2019hjh,Munoz:2020mue, Verwohlt:2024efh, Plombat:2024kla}.
Similar suppression can also arise from nonstandard early-Universe dynamics, such as thermal inflation~\cite{HongZoeAhn2017}, delaying structure formation and shifting the associated 21 cm signatures toward lower redshifts. Conversely, scenarios with enhanced small-scale density fluctuations can promote earlier structure formation and lead to corresponding 21 cm signatures at higher redshifts~\cite{Weon:2026tqy,Kadota:2020ybe}. For 21 cm studies encompassing a broader range of DM scenarios, see, e.g., Refs.~\cite{Munoz:2015bca,Lopez-Honorez:2018ipk,Tashiro:2021kcg,Barkana:2022hko}.} 
Existing small-scale structure probes already place strong lower limits on the WDM mass.
Ly$\alpha$ forest measurements~\cite{Viel:2005qj,Viel:2013fqw,Irsic:2023equ} constrain $m_{\rm WDM}\gtrsim3.3$--$5.7\,{\rm keV}$, while strong-lensing
analyses of the JWST quasar survey~\cite{Keeley:2024brx,Gilman:2025fhy} give $m_{\rm WDM}\gtrsim6.1$--$7.4\,{\rm keV}$. Milky Way satellite galaxy counts~\cite{DES:2020fxi,Tan:2024cek} provide comparable bounds, $m_{\rm WDM}\gtrsim6.5\,{\rm keV}$. The timing of the EDGES global absorption~\cite{Schneider:2018xba,Safarzadeh:2018hhg,Boyarsky:2019fgp} claim has also been used to infer lower bounds on WDM, reaching $m_{\rm WDM}\gtrsim6.1\,{\rm keV}$.
Future 21 cm observations are expected to become competitive with these
limits. Forecasts for 1000 h HERA power-spectrum measurements~\cite{Decant:2024bpg,Dandoy:2025twl} suggest a projected reach of $m_{\rm WDM}\gtrsim9\,{\rm keV}$.
Forecasts for SKA-Low~\cite{Giri:2022nxq} can reach $m_{\rm WDM}\gtrsim15\,{\rm keV}$.
A recent study~\cite{Park:2025phj} argued that non-linear gas and DM clumping during the dark ages can imprint a measurable correction on the sky-averaged 21 cm signal, providing a potential probe of WDM masses of order $m_{\rm WDM}\gtrsim7\,{\rm keV}$.

For mixed CDM--WDM scenarios, the timing of the EDGES global absorption~\cite{Schneider:2018xba} has been used to constrain a sufficiently hot subcomponent, $m_{\rm WDM}\lesssim1\,{\rm keV}$, to contribute no more than $f_{\rm WDM}\lesssim0.17$ of the total DM abundance. Future SKA-Low power-spectrum forecasts~\cite{Giri:2022nxq} suggest that 1000 h observations could constrain the mixed fraction to $f_{\rm WDM}\lesssim0.02$ for $m_{\rm WDM}\lesssim1\,{\rm keV}$.
The WDM example illustrates how a nonzero velocity dispersion can leave observable imprints on the 21 cm signal, but this effect is tied to the keV-scale masses. 
It is then natural to ask whether similar small-scale suppression can arise in heavier DM, including the sub-GeV range relevant for direct-detection searches~\cite{Essig:2011nj, Essig:2015cda, Battaglieri:2017aum, Knapen:2017xzo, Kim:2020coy}. 
Since the thermal velocity of such heavier particles is negligible, a different mechanism is needed to generate velocity dispersion or pressure support.
Boosted DM (BDM)~\cite{Belanger:2011ww,Agashe:2014yua,Kong:2014mia,Kim:2016zjx,Kamada:2021muh} provides such a possibility. 
In BDM scenarios, two DM species with a mass splitting can convert the rest
mass energy into the kinetic energy of the lighter species through annihilation processes, producing a boosted component that suppresses small-scale perturbations even for particle masses at the $\mathcal{O}({\rm MeV})$ scale.
Earlier studies implemented this idea in $N$-body simulations using Monte Carlo prescriptions and showed that nearly degenerate spectra can reduce substructure and flatten central halo density profiles~\cite{Medvedev:2013vsa,Todoroki:2017pdh,Todoroki:2017kge,Todoroki:2020num}. 
More recently, Refs.~\cite{Kim:2023onk,Kim:2024ltz} developed a cosmological treatment of BDM beyond the degenerate limit by solving the linear perturbation evolution, computing the matter power spectrum, and using the resulting spectra to initialize $N$-body simulations.

\begin{figure}[t!]
    \centering
    \includegraphics[width=1.0\linewidth,clip]
    {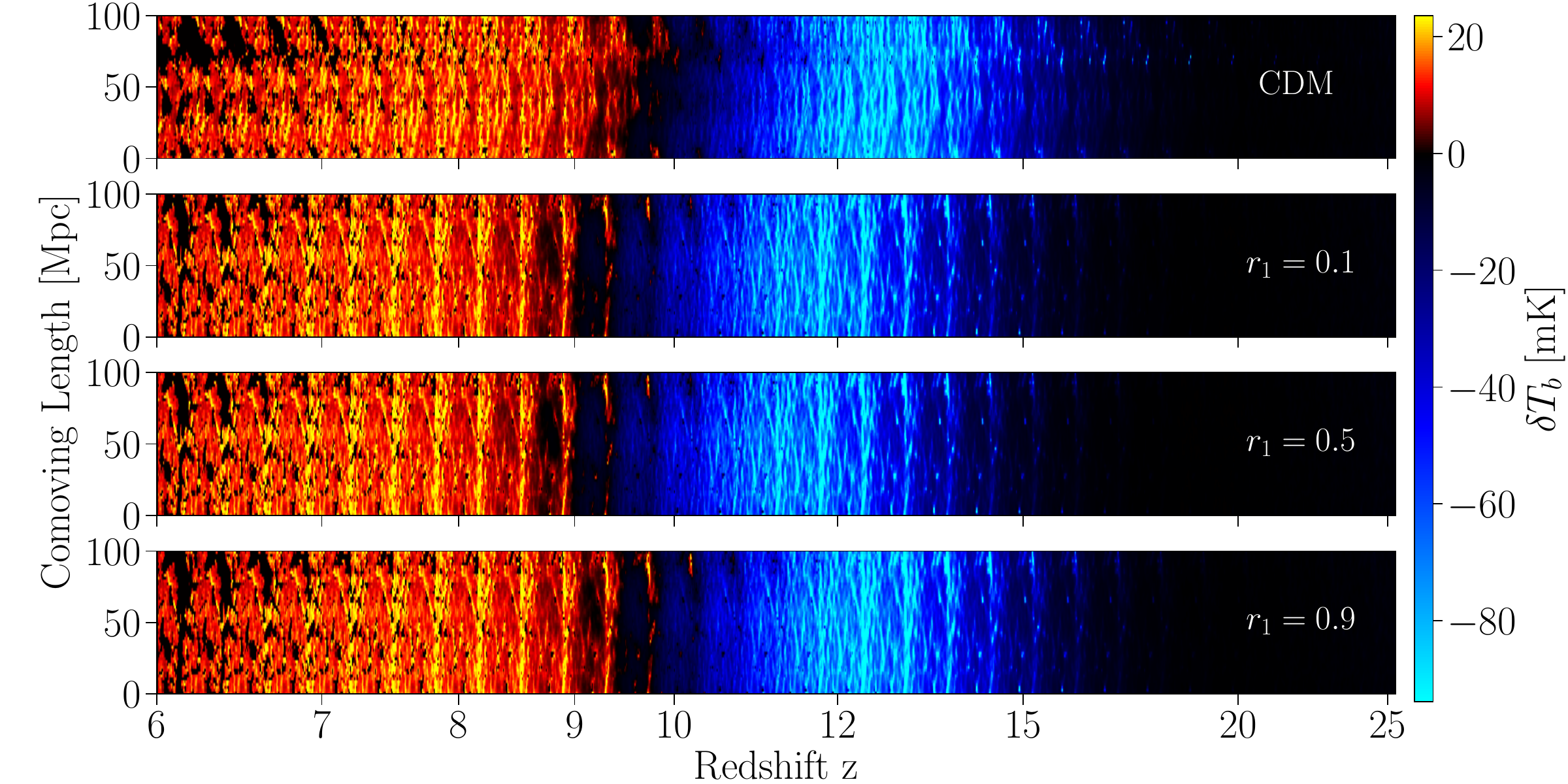}
    \caption{
    Light-cone slices of the differential 21\,cm brightness
    temperature defined in Eq.~\eqref{eq:deltaTbObs}, shown over $6\leq z\leq25$ for \LCDM\ (top) and BDM models
    with $r_1=0.1$, $0.5$, and $0.9$ (lower panels). Each panel spans
    $100\,{\rm Mpc}$ along the displayed spatial direction, and the
    color scale denotes the differential brightness temperature
    relative to the CMB in units of mK. The BDM parameters are fixed
    to $m_1=5\,{\rm MeV}$, $m_2=30\,{\rm MeV}$, and
    $\sigma_{\rm self1}/m_1=1\,{\rm cm^2\,g^{-1}}$, except for the
    indicated values of $r_1$. Relative to \LCDM, BDM delays the
    redshift evolution of the spatial 21\,cm signal, with the largest
    shift occurring for the intermediate fraction $r_1\sim0.5$.
    }
    \label{fig:21cm_lightcone}
\end{figure}

These studies focused primarily on late-time structure formation and observables at $z\simeq0$. 
At low redshift, nonlinear evolution can partially wash out features already present in the linear matter power spectrum, and baryonic processes can further obscure the DM origin of the signal. 
This motivates a high-redshift test of BDM using the 21 cm signal, where the suppression of low-mass halos affects the IGM before late time nonlinear evolution becomes dominant. 
Figure~\ref{fig:21cm_lightcone} provides a visual overview of the resulting 21\,cm light-cone evolution. 
Relative to \LCDM, the suppression of low-mass halo formation in BDM delays the evolution of the spatial brightness-temperature field. 
Among the BDM models shown, this shift is most pronounced for the intermediate light-component fraction $r_1\equiv\Omega_1/\Omega_{\rm DM}\sim0.5$, where $\Omega_1$ and $\Omega_{\rm DM}$ denote the present-day abundances of the lighter component and total DM, respectively.
In this work, we compute the 21 cm power spectrum in BDM cosmology and compare it with the \LCDM prediction. 
Using projected SKA1-Low sensitivities, we examine whether BDM imprints remain distinguishable after accounting for observational limitations and astrophysical parameter uncertainties.
We also compare the resulting reach with sub-GeV direct-detection constraints from SENSEI~\cite{SENSEI:2020dpa,SENSEI:2023zdf} and the search prospects for relativistic DM produced by the annihilation of the heavier relic component at large-volume neutrino experiments such as Super-Kamiokande (SK)~\cite{Super-Kamiokande:2017dch}, Hyper-Kamiokande (HK)~\cite{Hyper-Kamiokande:2018ofw, Hyper-Kamiokande:2016srs}, JUNO~\cite{JUNO:2021vlw}, and DUNE~\cite{DUNE:2020ypp, DUNE:2020fgq}. 
These comparisons demonstrate that high-redshift 21 cm observations can access BDM parameter space that lies beyond the reach of current and upcoming terrestrial searches.

The remainder of this paper is organized as follows. 
In Section~\ref{sec:twoDM}, we review the background and linear perturbation evolution of BDM cosmology. 
In Section~\ref{sec:21cm}, we describe how these cosmological inputs are incorporated into the semi-numerical calculation of the 21\,cm signal. 
In Section~\ref{sec:result}, we present the resulting global 21\,cm signal and 21\,cm power spectrum and assess the projected distinguishability of BDM from \LCDM\ after marginalizing over the selected astrophysical parameters. 
We compare our projected reach with sub-GeV DM direct-detection bounds and the search prospects for BDM at large-volume neutrino experiments.
In Section~\ref{sec:summary}, we
summarize our main results and discuss future directions. 
Additional technical details of the semi-numerical calculation and
forecast methodology are provided in the Appendix.

%%%%%%%%%%%%%%%%%%%%%%%%%%%%%%%%%%%%%%%%%%%%%%%%%%%
\section{Theoretical Framework}
\label{sec:twoDM}
%%%%%%%%%%%%%%%%%%%%%%%%%%%%%%%%%%%%%%%%%%%%%%%%%%%

In this section, we summarize the BDM cosmology that provides the cosmological input for the 21 cm calculation. 
We first review the homogeneous evolution of the two-component DM sector, with particular emphasis on the self-heating of the light component $\chi_1$. 
We then describe how this temperature enhancement generates pressure support in the $\chi_1$ fluid and
modifies the growth of linear perturbations. 
These effects determine the small-scale matter power spectrum, which in turn controls the abundance and spatial distribution of early radiative sources relevant for the 21 cm signal.

%%%%%%%%%%%%%%%%%%%%%%%%%%%%%%%%%%%%%%%%%%%%%%%%%%%
\subsection{Background evolution}
\label{subsec:background_evolution}
%%%%%%%%%%%%%%%%%%%%%%%%%%%%%%%%%%%%%%%%%%%%%%%%%%%

Dark sectors containing two DM species, with the heavier component able to annihilate into the lighter one, arise in a variety of models~\cite{Belanger:2011ww, Agashe:2014yua, Kong:2014mia, Kim:2016zjx, Kamada:2021muh, Kim:2023onk, Kim:2024ltz}. 
In this work, following Refs.~\cite{Belanger:2011ww, Kamada:2021muh, Kim:2023onk, Kim:2024ltz}, we consider a phenomenological model consisting of two Dirac fermions, $\chi_1$ and $\chi_2$, with a mass hierarchy $m_2>m_1$. 
We take the dark gauge group to be ${\rm U}(1)^\prime\times{\rm U}(1)^{\prime\prime}$.
Both $\chi_1$ and $\chi_2$ are charged under
${\rm U}(1)^{\prime\prime}$, whereas only $\chi_1$ is charged under ${\rm U}(1)^\prime$. 
We denote the gauge couplings associated with
${\rm U}(1)^\prime$ and ${\rm U}(1)^{\prime\prime}$ by $g'$ and $g''$, respectively. 
These dark gauge symmetries are spontaneously
broken when the corresponding dark-Higgs fields acquire vacuum expectation values. 
The resulting massive dark gauge bosons are denoted by $\gamma'$ and $\gamma''$, with masses $m_{\gamma'}$ and $m_{\gamma''}$, respectively. 
For simplicity, we neglect mixing between the dark Higgs fields and the SM Higgs and assume that the dark-Higgs fields are sufficiently heavy compared with the two DM species so that they do not enter the phenomenology considered in this work. 
The ${\rm U}(1)^\prime$ gauge field communicates with the visible sector through kinetic mixing with the hypercharge ${\rm U}(1)_Y$ gauge field, parameterized by $\epsilon$.\footnote{The induced kinetic mixing between ${\rm U}(1)^{\prime\prime}$ and ${\rm U}(1)_Y$ is suppressed by both a loop factor and the small kinetic mixing between
${\rm U}(1)^\prime$ and ${\rm U}(1)_Y$~\cite{Agashe:2014yua}. 
We therefore neglect this mixing in the rest of this paper.}
The model is therefore specified by
\begin{equation}
\left\{
m_1,\,
m_2,\,
m_{\gamma'},\,
m_{\gamma''},\,
g'',\,
g',\,
\epsilon
\right\}.
\label{eq:fundamental_BDM_parameters}
\end{equation}

With this setup, the heavier component $\chi_2$ is secluded from the SM sector and predominantly annihilates into the lighter component through
\begin{equation}
\chi_2\bar{\chi}_2
\xrightarrow{\gamma''^{\ast}}
\chi_1\bar{\chi}_1\,.
\end{equation}
The lighter component can annihilate into SM particles through the kinetically mixed $\gamma'$ mediator,
\begin{equation}
\chi_1\bar{\chi}_1
\xrightarrow{\gamma'^{\ast}}
X\bar X\,,
\end{equation}
where $X$ denotes an SM particle. 
For concreteness, we assume that this annihilation is dominated by the electron channel, i.e.,\
$X=e$.
We denote the thermally averaged annihilation cross sections for these two processes by $\langle\sigma v\rangle_{22\rightarrow11}$ and $\langle\sigma v\rangle_{11\rightarrow e^+e^-}$, respectively. 
In the heavy-mediator limit, their dependence on the gauge couplings and mediator masses is
\begin{equation}
\left\langle\sigma v\right\rangle_{22\rightarrow11}
\propto
\frac{g''^{\,4}}{m_{\gamma''}^{4}}\,,
\qquad
\left\langle\sigma v\right\rangle_{11\rightarrow e^+e^-}
\propto
\frac{g'^{\,2}\epsilon^2}{m_{\gamma'}^4}\,.
\label{eq:annihilation_cross_section_scalings}
\end{equation}
Here, $v$ is the relative velocity of the annihilating particles, and the brackets denote a thermal average. 
The proportionalities in Eq.~\eqref{eq:annihilation_cross_section_scalings} display only the dependence on the gauge couplings and mediator masses.

These annihilation processes exchange energy between the two DM
components and the SM sector. 
Including this energy exchange, the homogeneous evolution of the energy densities $\rho_1$, $\rho_2$, and $\rho_X$ is governed by
\begin{align}
    \dot{\rho}_2 + 3H\rho_2
    &=
    -\frac{\langle \sigma v\rangle_{22\to 11}}{m_2}
    \left[
        \rho_2^2
        -
        \frac{\rho_{2,{\rm eq}}^2}{\rho_{1,{\rm eq}}^2}
        \rho_1^2
    \right] ,
    \label{eq:rho2_boltzmann}
    \\
    \dot{\rho}_1 + 3H\rho_1
    &=
    -\frac{\langle \sigma v\rangle_{11\to XX}}{m_1}
    \left[
        \rho_1^2-\rho_{1,{\rm eq}}^2
    \right]
    +
    \frac{m_1}{m_2}
    \frac{\langle \sigma v\rangle_{22\to 11}}{m_2}
    \left[
        \rho_2^2
        -
        \frac{\rho_{2,{\rm eq}}^2}{\rho_{1,{\rm eq}}^2}
        \rho_1^2
    \right] ,
    \label{eq:rho1_boltzmann}
    \\
    \dot{\rho}_X + 4H\rho_X
    &=
    \frac{\langle \sigma v\rangle_{11\to XX}}{m_1}
    \left[
        \rho_1^2-\rho_{1,{\rm eq}}^2
    \right] .
    \label{eq:rhoX_boltzmann}
\end{align}
Here $H$ is the Hubble expansion rate, and the equilibrium energy density of species $M$ is given by
\begin{eqnarray}  
\rho_{M, {\text{eq}}} &=& \int \frac{d^3p}{(2\pi)^3} E_{M} ~e^{-E_{M}/T} \\
&=& \frac{m^4_{M}}{2 \pi^2} \left( \frac{T}{m_{M}} \right) \left[ K_1 (m_{M}/T) + 3\left( \frac{T}{m_{M}}\right) K_2 (m_{M}/T) \right] \;, \nonumber
\label{eq:rho_eq}
\end{eqnarray} 
where $M=1,2,X$. 
Here $K_n$ is the modified Bessel function of the second kind, and $T$ denotes the temperature of the SM radiation bath.

The present-day relic abundances of the two DM components are denoted by $\Omega_1$ and $\Omega_2$, with
\begin{equation}
\Omega_{\rm DM}
=
\Omega_1+\Omega_2\,.
\label{eq:total_DM_abundance}
\end{equation}
We define the fractional abundance of the light component as
\begin{equation}
r_1
\equiv
\frac{\Omega_1}{\Omega_{\rm DM}}\,.
\label{eq:r1_definition}
\end{equation}
The individual abundances are therefore $\Omega_1=r_1\Omega_{\rm DM}$ and $\Omega_2=(1-r_1)\Omega_{\rm DM}$. 
The relative sizes of $\langle\sigma v\rangle_{22\to11}$ and $\langle\sigma v\rangle_{11\to e^+e^-}$ in Eq.~\eqref{eq:annihilation_cross_section_scalings} determine how the total relic abundance is distributed between $\chi_1$ and $\chi_2$.

For cosmological analysis, it is convenient to reexpress the fundamental BDM parameters in Eq.~\eqref{eq:fundamental_BDM_parameters} in terms of parameters that directly specify the relic abundances and the self-interaction strength.
We first specify the three mass parameters. 
We fix the ${\rm U}(1)^\prime$ gauge-boson mass to $m_{\gamma'}=30\,{\rm MeV}$ and consider MeV-scale values of $m_1$ satisfying $m_1<m_{\gamma'}$. 
This condition kinematically forbids the annihilation channel $\chi_1\bar{\chi}_1\to\gamma'\gamma'$. 
The heavy-component mass is taken in the range $m_1<m_2\leq100\,{\rm MeV}$.

Among the remaining four parameters $\{m_{\gamma''},g'',g',\epsilon\}$, the parameters $g''$ and $m_{\gamma''}$ appear in the relevant cross sections only through the combination $g''/m_{\gamma''}$.
The four parameters therefore reduce to the three independent combinations $\{g''/m_{\gamma''},g',\epsilon\}$.
The abundance parameters $\Omega_{\rm DM}$ and $r_1$ jointly determine the two annihilation cross sections in Eq.~\eqref{eq:annihilation_cross_section_scalings}. 
The first relation fixes $g''/m_{\gamma''}$, while the second fixes $g'^{\,2}\epsilon^2$ for the adopted value of $m_{\gamma'}$.
Next, the $\chi_1$ self-scattering cross section is approximated by~\footnote{In our phenomenological study, to focus on the effects of $\chi_1$ self-interactions, we neglect $\chi_2$ self-interactions and the $\gamma''$-mediated contribution to $\chi_1$ self-scattering.
These contributions scale as $\sigma_{\rm self2}\propto g''^{\,4}/m_{\gamma''}^{4}$ and $\sigma_{\rm self1}^{(\gamma'')}
\propto g''^{\,4}/m_{\gamma''}^{4}$, respectively.}
\begin{equation}
\sigma_{\rm self1}
\simeq
\sigma_{\rm self1}^{(\gamma')}
\propto
\frac{g'^{\,4}}{m_{\gamma'}^4}.
\label{eq:self1_cross_section_scaling}
\end{equation}
Since $m_{\gamma'}$ is fixed, the specified value of $\sigma_{\rm self1}$ determines $g'$. 
For a given $r_1$, the relic-abundance calculation fixes the combination $g'^{\,2}\epsilon^2$.\footnote{For fixed $\Omega_{\rm DM}$ and particle masses, $r_1$ specifies the target abundances $\Omega_1=r_1\Omega_{\rm DM}$ and $\Omega_2=(1-r_1)\Omega_{\rm DM}$. 
Solving the coupled freeze-out equations determines the two annihilation cross sections in Eq.~\eqref{eq:annihilation_cross_section_scalings}, thereby fixing $g''/m_{\gamma''}$ and $g'^{\,2}\epsilon^2$. 
In particular, $g'^{\,2}\epsilon^2$ is fixed by the required light-component abundance.}
Combining the resulting value of $g'^{\,2}\epsilon^2$ with $g'$ determines the magnitude of $\epsilon$.

We therefore trade the three combinations $\{g''/m_{\gamma''},g',\epsilon\}$ for the effective parameters $\{\Omega_{\rm DM},r_1,\sigma_{\rm self1}\}$.
Fixing the total DM abundance to $\Omega_{\rm DM}\simeq0.27$, the two BDM parameters of primary interest in this study are
\begin{equation}
\left\{
r_1,\,
\frac{\sigma_{\rm self1}}{m_1}
\right\}.
\label{eq:BDM_primary_parameters}
\end{equation}
In our numerical study, we vary the fraction of light-component in the range $0\leq r_1\leq1$ and the strength of self-interaction in $\sigma_{\rm self1}/m_1=0.1\text{--}10\,{\rm cm^2\,g^{-1}}$. 
We also examine the dependence on $m_1$ and $m_2$ using representative parameter choices. 
Our reference model is defined by $r_1=0.5$, $\sigma_{\rm self1}/m_1=1\,{\rm cm^2\,g^{-1}}$, $m_1=5\,{\rm MeV}$, and $m_2=30\,{\rm MeV}$.

After kinetic decoupling from the SM bath, the annihilation of the
heavy species injects kinetic energy into the light species through the mass
splitting $\delta m \equiv m_2-m_1$.
If the $\chi_1$ self-interaction rate is larger than the Hubble expansion rate,
this injected energy is redistributed among the light particles and raises the
effective temperature of the $\chi_1$ fluid. We denote this temperature by
$T_1$, whose evolution is approximately described by
\begin{equation}
    \dot{T}_1
    \simeq
    -2H T_1
    + \gamma_{\rm heat} T
    -2\gamma_{\chi_1 X}\left(T_1-T\right) .
    \label{eq:T1_evolution}
\end{equation}
The first term represents adiabatic cooling due to the expansion of the
Universe. The second term describes the heating caused by
$\chi_2\bar{\chi}_2\to\chi_1\bar{\chi}_1$, with
\begin{equation}
    \gamma_{\rm heat}
    \simeq
    \frac{
        2 n_2^2
        \langle\sigma v\rangle_{22\to 11}
        \delta m
    }{
        3 n_1 T
    } .
    \label{eq:gamma_heat}
\end{equation}
Here $n_M$ denotes the number density of species $M$, with $M=1,2, X$. 
The last term accounts for energy exchange between $\chi_1$ and the SM bath,
\begin{equation}
    \gamma_{\chi_1 X}
    \simeq
    \left(\frac{\delta E}{T}\right)
    n_X
    \langle\sigma v\rangle_{\chi_1 X} ,
    \label{eq:gamma_chiX}
\end{equation}
where $\delta E$ is the change of the $\chi_1$ kinetic energy per elastic
scattering event, and $\langle \sigma v \rangle_{\chi_1 X}$ is the thermally averaged scattering cross section of $\chi_1$ off SM particles $X$.

Eq.~\eqref{eq:gamma_heat} shows that the heating of the light component
is controlled not only by the available energy $\delta m$, but also by the
combination $n_2^2/n_1$. This point is important for the interpretation of
the model. A larger mass splitting tends to increase the injected energy per
annihilation, but increasing $m_2$ at fixed energy density lowers the number
density of the heavy component. Therefore, the heating efficiency does not
grow monotonically with $m_2$. Similarly, increasing the light component
fraction $r_1$ increases the number of particles that must share the injected
energy, and can lower $T_1$.

\begin{figure}[t]
    \centering
    \includegraphics[width=1.0\linewidth,clip]{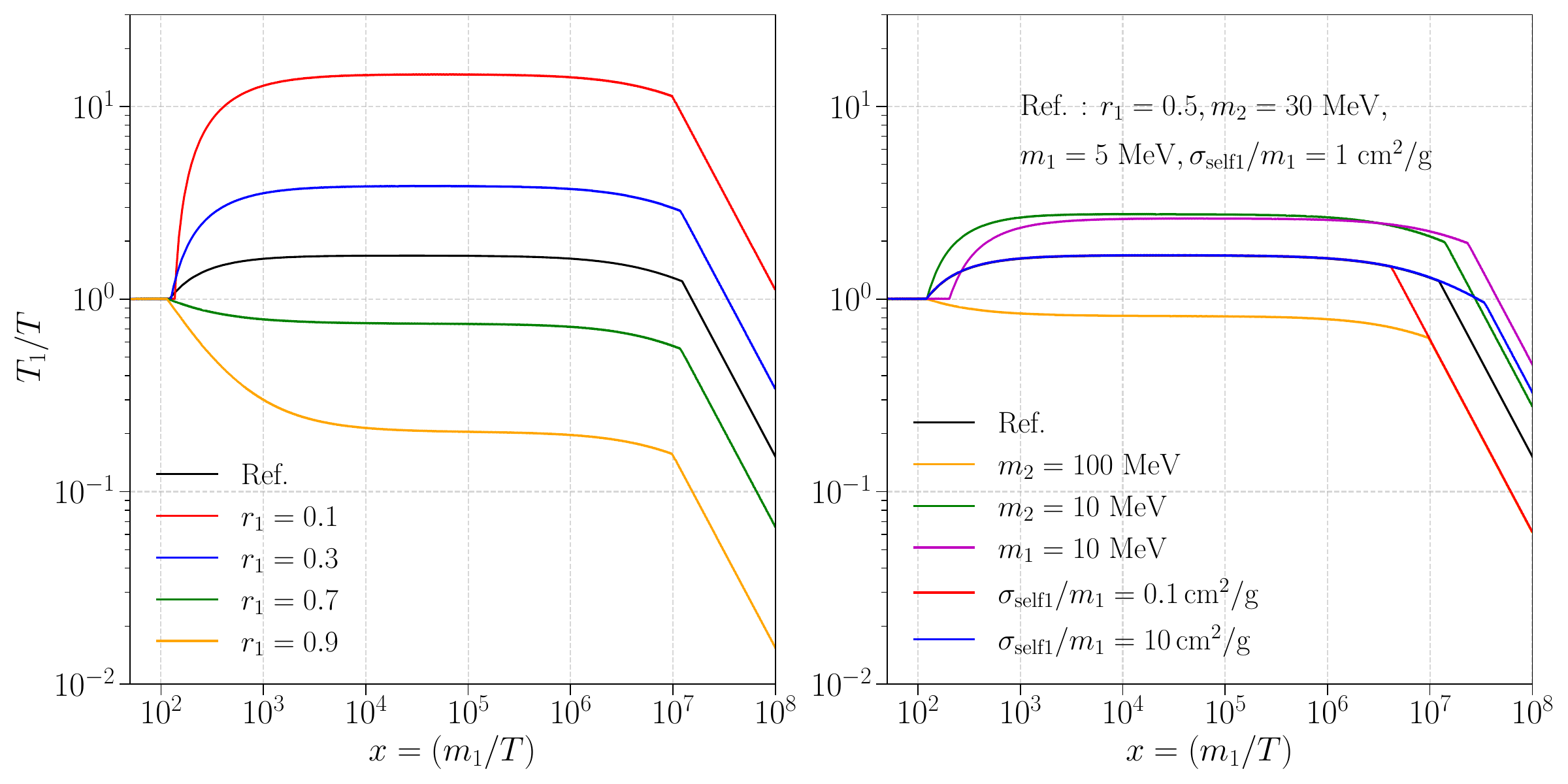}
    
    \caption{
    Evolution of the light dark matter temperature $T_1$, obtained by solving Eq.~\eqref{eq:T1_evolution} and normalized by the
    SM bath temperature $T$, shown as a function of
    $x=m_1/T$. The reference model is
    $r_1=0.5$, $\sigma_{\rm self1}/m_1=1~{\rm cm^2/g}$,
    $m_2=30~{\rm MeV}$, and $m_1=5~{\rm MeV}$.
    The left panel varies $r_1$, while the right panel varies
    $\sigma_{\rm self1}/m_1$ and the DM masses.
    }
    \label{fig:Temperature}
\end{figure}

Figure~\ref{fig:Temperature} illustrates the parameter dependence of
$T_1/T$. A smaller $r_1$ both increases the abundance of the heavy component $\chi_2$ and reduces the number of light particles that share the injected energy. As a result, the annihilation process $\chi_2\bar{\chi}_2\to\chi_1\bar{\chi}_1$ can raise the temperature of the
$\chi_1$ fluid more efficiently.
In the opposite limit, $r_1\to 1$, the abundance of $\chi_2$ becomes small while
the number of $\chi_1$ particles becomes large. The annihilation
$\chi_2\bar{\chi}_2\to\chi_1\bar{\chi}_1$ therefore provides less kinetic energy, and that energy is shared by a larger light-particle population. This reduces the temperature enhancement of the $\chi_1$ fluid.

The self-interaction cross section mainly controls how long the heating stage lasts. A larger $\sigma_{\rm self1}/m_1$ keeps the light component thermally
redistributed for a longer time and delays the subsequent cooling of $T_1$.
The mass dependence follows the scaling in Eq.~\eqref{eq:gamma_heat}. At
fixed $m_1$, increasing $m_2$ increases $\delta m$ but decreases $n_2$, so the
net temperature enhancement can become smaller. At fixed $m_2$, increasing
$m_1$ reduces the number density of the light component and can raise $T_1$.
The elevated value of $T_1$ gives the $\chi_1$ fluid a non-negligible pressure
support. As discussed in the next subsection, this pressure term suppresses
the growth of small-scale perturbations.

%%%%%%%%%%%%%%%%%%%%%%%%%%%%%%%%%%%%%%%%%%%%%%%%%%%
\subsection{Perturbation evolution}
\label{subsec:perturbation_evolution}
%%%%%%%%%%%%%%%%%%%%%%%%%%%%%%%%%%%%%%%%%%%%%%%%%%%

We next summarize the evolution of linear perturbations in the two-component
DM sector. The full set of perturbed Boltzmann equations was derived in Refs.~\cite{Kim:2023onk, Kim:2024ltz}. Here we introduce the equations and
variables needed for the 21\,cm physics.

We use the Newtonian gauge, in which the scalar metric perturbations $\Psi$ and $\Phi$ are introduced through
\begin{equation}
    ds^2
    =
    -(1+2\Psi)dt^2
    +
    a^2(t)(1-2\Phi)\delta_{ij}dx^idx^j \, .
    \label{eq:newtonian_gauge}
\end{equation}
For each DM component, we define
\begin{equation}
    \rho_M=\bar\rho_M(1+\delta_M),
    \qquad
    \theta_M\equiv \nabla\cdot{\bf v}_M ,
    \qquad
    M=1,2 ,
    \label{eq:dm_pert_def}
\end{equation}
where barred quantities denote homogeneous background values,\footnote{
In Eqs.~\eqref{eq:rho2_boltzmann}--\eqref{eq:rhoX_boltzmann}, the bars on
homogeneous background densities were omitted for notational simplicity.
} $\delta_M$ is
the density contrast, and $\theta_M$ is the velocity divergence. The
equilibrium-density perturbations are related to the photon perturbation by
\begin{equation}
    \delta_{M,{\rm eq}}
    =
    \frac{1}{4}
    \left(
        \frac{3}{2}+\frac{m_M}{T}
    \right)
    \delta_\gamma ,
    \qquad
    M=1,2 ,
    \label{eq:delta_eq}
\end{equation}
in the non-relativistic limit.

The linear perturbations of the two DM fluids obey
\begin{align}
    \dot{\delta}_2
    +\frac{\theta_2}{a}
    -3\dot{\Phi}
    &=
    \frac{\langle\sigma v\rangle_{22\to 11}}{m_2\bar\rho_2}
    \Bigg[
        -\Psi
        \left(
            \bar\rho_2^2
            -
            \frac{\bar\rho_{2,{\rm eq}}^2}
                 {\bar\rho_{1,{\rm eq}}^2}
            \bar\rho_1^2
        \right)
        -\bar\rho_2^2\delta_2
        \label{eq:delta2_pert}\\
    &\hspace{4.0cm}
        +
        \frac{\bar\rho_{2,{\rm eq}}^2}
             {\bar\rho_{1,{\rm eq}}^2}
        \bar\rho_1^2
        \left(
            2\delta_{2,{\rm eq}}
            -\delta_2
            -2\delta_{1,{\rm eq}}
            +2\delta_1
        \right)
    \Bigg],
    \nonumber
    \\
    \dot{\theta}_2
    +H\theta_2
    +\frac{\nabla^2\Psi}{a}
    &=
    \frac{\langle\sigma v\rangle_{22\to 11}}{m_2\bar\rho_2}
    \frac{\bar\rho_{2,{\rm eq}}^2}
         {\bar\rho_{1,{\rm eq}}^2}
    \bar\rho_1^2
    \left(
        \theta_1-\theta_2
    \right),
    \label{eq:theta2_pert}
    \\
    \dot{\delta}_1
    +\frac{\theta_1}{a}
    -3\dot{\Phi}
    &=
    -\frac{\langle\sigma v\rangle_{22\to 11}}{m_2\bar\rho_1}
    \Bigg[
        -\Psi
        \left(
            \bar\rho_2^2
            -
            \frac{\bar\rho_{2,{\rm eq}}^2}
                 {\bar\rho_{1,{\rm eq}}^2}
            \bar\rho_1^2
        \right)
        -\bar\rho_2^2
        \left(
            2\delta_2-\delta_1
        \right)
        \label{eq:delta1_pert}\\
    &\hspace{4.0cm}
        +
        \frac{\bar\rho_{2,{\rm eq}}^2}
             {\bar\rho_{1,{\rm eq}}^2}
        \bar\rho_1^2
        \left(
            2\delta_{2,{\rm eq}}
            +\delta_1
            -2\delta_{1,{\rm eq}}
        \right)
    \Bigg]
    \nonumber\\
    &\quad
    +
    \frac{\langle\sigma v\rangle_{11\to XX}}{m_1\bar\rho_1}
    \Bigg[
        -\Psi
        \left(
            \bar\rho_1^2-\bar\rho_{1,{\rm eq}}^2
        \right)
        -\bar\rho_1^2\delta_1
        +
        \bar\rho_{1,{\rm eq}}^2
        \left(
            2\delta_{1,{\rm eq}}-\delta_1
        \right)
    \Bigg],
    \nonumber
    \\
    \dot{\theta}_1
    +H\theta_1
    +\frac{\nabla^2\Psi}{a}
    +c_{s,1}^2\frac{\nabla^2\delta_1}{a}
    &=
    \frac{\langle\sigma v\rangle_{22\to 11}}{m_2\bar\rho_1}
    \bar\rho_2^2
    \left(
        \theta_2-\theta_1
    \right),
    \label{eq:theta1_pert}
\end{align}
where an overdot denotes a derivative with respect to time $t$.
In the limit where the annihilation and pressure terms are absent, these
equations reduce to the usual CDM perturbation equations. The terms
proportional to $\langle\sigma v\rangle_{22\to 11}$ describe the energy and
momentum transfer associated with
$\chi_2\bar\chi_2\to\chi_1\bar\chi_1$, while the terms proportional to
$\langle\sigma v\rangle_{11\to XX}$ arise from the annihilation of $\chi_1$
into SM particles.

The distinctive feature of the model is the pressure term in
Eq.~\eqref{eq:theta1_pert}. The self-heating discussed in
Section~\ref{subsec:background_evolution} gives the light component $\chi_1$ a
non-negligible pressure support, which can be characterized by an effective
sound speed determined by the temperature evolution of $\chi_1$ as
\begin{equation}
    c_{s,1}^2
    =
    \frac{T_1}{m_1}
    \left[
        1
        -
        \frac{1}{3}
        \frac{\partial \ln T_1}{\partial \ln a}
    \right] .
    \label{eq:sound_speed_chi1}
\end{equation}
Since $T_1$ can be enhanced by the self-heating mechanism, the resulting
pressure support can affect the growth of $\chi_1$ perturbations. In Fourier
space, after combining the continuity and Euler equations, this pressure
contribution enters the density evolution through a term proportional to
$c_{s,1}^2 k^2\delta_1/a^2$. It therefore becomes more
important at larger wavenumber $k$. For modes with sufficiently large $k$, the
pressure support competes with gravitational collapse and drives oscillatory
behavior in the light-component perturbation. We refer to this feature as dark
acoustic oscillations (DAO). By contrast, the heavy component $\chi_2$ is
treated as pressureless, so no analogous pressure term is included for
$\chi_2$.

The metric perturbations are evolved consistently with the perturbed Einstein
equations. In particular, we use
\begin{equation}
    \nabla^2\left(\dot{\Phi}+H\Psi\right)
    =
    -4\pi G a
    \sum_N
    \left(
        \bar\rho_N+\bar P_N
    \right)
    \theta_N ,
    \label{eq:einstein_0i}
\end{equation}
and
\begin{equation}
    \left(
        \partial_i\partial_j
        -
        \frac{1}{3}\delta_{ij}\nabla^2
    \right)
    \left(
        \Phi-\Psi
    \right)
    =
    8\pi G a^2\Pi_{ij} \;.
    \label{eq:einstein_anisotropic}
\end{equation}
Here the summation runs over all species $N$ contributing to the total
energy-momentum tensor, including photons, baryons, neutrinos, and the two
DM components $\chi_1$ and $\chi_2$. The quantity $\bar P_N$ denotes the
background pressure of species $N$. For the light DM component, we use
$\bar P_1\simeq c_{s,1}^2\bar\rho_1$, whereas the heavy component is treated
as pressureless, $\bar P_2=0$. The tensor $\Pi_{ij}$ denotes the anisotropic
stress. In our setup, the anisotropic stress is supplied by the standard
relativistic species, while the two DM components are modeled as perfect fluids
with no intrinsic anisotropic stress.

In this work, we use the implementation of the BDM cosmology developed in Ref.~\cite{Kim:2023onk,Kim:2024ltz} based on Cosmic Linear Anisotropy Solving System, \CLASS~(v3.2)~\cite{Blas:2011rf}. We refer to this implementation as \BDMCLASS.
The code solves the background evolution of
the two DM components according to
Eqs.~\eqref{eq:rho2_boltzmann}--\eqref{eq:rhoX_boltzmann}, together with the
Friedmann equation, and evolves their linear perturbations using
Eqs.~\eqref{eq:delta2_pert}--\eqref{eq:theta1_pert}. The metric perturbations
are obtained from the perturbed Einstein equations, including the contributions
of photons, baryons, neutrinos, and the two DM components, as in
Eqs.~\eqref{eq:einstein_0i} and \eqref{eq:einstein_anisotropic}. Photons,
baryons, and neutrinos are otherwise evolved using the standard \CLASS\
equations. For initial inputs, we adopt the Planck
2018~\cite{Planck:2018vyg} $\Lambda$CDM parameters
\begin{equation}
    \{\Omega_{\rm m},\Omega_{\rm b},h,10^9A_s,n_s,\tau_{\rm reio}\}=\{0.3111,0.04897,0.6766,2.105,0.9665,0.0561\}\;.
\label{eq:CosmoParameters}
\end{equation}

Focusing on the matter sector, we initialize the linear perturbations at an early redshift $z_{\rm in}=10^{13}$ using adiabatic initial conditions
\begin{equation}
    \delta_{b}(\mathbf k,z_{\rm in})
    =
    \delta_{1}(\mathbf k,z_{\rm in})
    =
    \delta_{2}(\mathbf k,z_{\rm in})
    =
    -{\cal R}(\mathbf k),
    \label{eq:matter_initial_seed}
\end{equation}
where ${\cal R}(\mathbf k)$ denotes the gauge-invariant curvature perturbation. The primordial curvature perturbation is specified by
\begin{equation}
    \left\langle
    {\cal R}(\mathbf{k})
    {\cal R}^{\ast}(\mathbf{k}')
    \right\rangle
    =
    (2\pi)^3
    \delta_D^{(3)}(\mathbf{k}-\mathbf{k}')
    P_{\cal R}(k),
    \label{eq:PR_def}
\end{equation}
with
\begin{equation}
    \Delta_{\cal R}^2(k)
    \equiv
    \frac{k^3}{2\pi^2}P_{\cal R}(k)
    =
    A_s
    \left(
        \frac{k}{k_\ast}
    \right)^{n_s-1}
    \qquad
    \big({\rm with}\;\; k_\ast=0.05\,{\rm Mpc}^{-1}\big).
    \label{eq:primordial_power}
\end{equation}
Here $\delta_D^{(3)}$ denotes the three-dimensional Dirac delta function, and
$A_s$ and $n_s$ are the amplitude and scalar spectral index of the primordial
curvature power spectrum, respectively, as given in Eq.~\eqref{eq:CosmoParameters}.

Starting from these adiabatic initial conditions, \BDMCLASS\ evolves the linear perturbations of baryons and the two DM components down to the redshifts relevant for the 21\,cm calculation. Since the semi-numerical 21\,cm calculation uses a single effective DM density field, we combine the two DM components into
\begin{equation}
    \delta_{\rm dm}(\mathbf k,z)
    =
    \frac{
        \bar\rho_1(z)\delta_1(\mathbf k,z)
        +
        \bar\rho_2(z)\delta_2(\mathbf k,z)
    }{
        \bar\rho_1(z)+\bar\rho_2(z)
    } .
    \label{eq:delta_dm_eff}
\end{equation}
We also define the total matter density contrast,
\begin{equation}
    \delta_m(\mathbf k,z)
    =
    \frac{
        \bar\rho_b(z)\delta_b(\mathbf k,z)
        +
        \bar\rho_1(z)\delta_1(\mathbf k,z)
        +
        \bar\rho_2(z)\delta_2(\mathbf k,z)
    }{
        \bar\rho_b(z)+\bar\rho_1(z)+\bar\rho_2(z)
    } .
    \label{eq:delta_m_total_components}
\end{equation}

The linear evolution of the density contrasts from the initial conditions to redshift $z$ can be written in terms of transfer functions,
\begin{equation}
    \delta_i(\mathbf k,z)
    =
    {\cal T}_i(k,z)\delta_{i}(\mathbf k,z_{\rm in})
    \qquad
    \big({\rm with}\;\; i\in\{b,{\rm dm},m\}\big).
    \label{eq:linear_transfer_from_initial}
\end{equation}
The corresponding growth functions describe the same evolution normalized to the present-day density contrasts at $z=0$,
\begin{equation}
    D_i(k,z)
    \equiv
    \frac{\delta_i(\mathbf k,z)}
    {\delta_i(\mathbf k,0)}
    =
    \frac{{\cal T}_i(k,z)}
    {{\cal T}_i(k,0)}
    \qquad
    \big({\rm with}\;\; D_i(k,0)=1\big).
    \label{eq:growth_from_transfer}
\end{equation}
The linear matter power spectrum is defined through the two-point correlation function of the total matter density contrast,
\begin{equation}
    \left\langle
    \delta_m(\mathbf{k},z)
    \delta_m^{\ast}(\mathbf{k}',z)
    \right\rangle
    =
    (2\pi)^3
    \delta_D^{(3)}(\mathbf{k}-\mathbf{k}')
    P_m(k,z).
    \label{eq:Pm_def}
\end{equation}
Using Eqs.~\eqref{eq:matter_initial_seed}, \eqref{eq:PR_def}, and \eqref{eq:linear_transfer_from_initial}, we then obtain
\begin{equation}
    P_m(k,z)
    =
    {\cal T}^2_{m}(k,z) P_{\cal R}(k) = {\cal T}^2_{m}(k,z)\frac{2\pi^2}{k^3} A_s
    \left(
        \frac{k}{k_\ast}
    \right)^{n_s-1}.
    \label{eq:Pm_from_seed}
\end{equation}
The corresponding dimensionless matter power spectrum can therefore be written as
\begin{equation}
    \Delta_m^2(k,z)
    \equiv
    \frac{k^3}{2\pi^2}P_m(k,z)
    =
    {\cal T}_m^2(k,z)
    A_s
    \left(
        \frac{k}{k_\ast}
    \right)^{n_s-1}.
    \label{eq:Delta_m_from_PR}
\end{equation}
Equivalently, using Eqs.~\eqref{eq:matter_initial_seed}, \eqref{eq:PR_def}, and \eqref{eq:growth_from_transfer}, the matter power spectrum can also be written as
\begin{equation}
    P_m(k,z)
    =
    D_m^2(k,z)P_m(k,0),
    \label{eq:Pm_from_Pm0}
\end{equation}
and the corresponding dimensionless matter power spectrum as
\begin{equation}
    \Delta_m^2(k,z)
    =
    D_m^2(k,z)\Delta_m^2(k,0)
    \qquad
    \Big({\rm with}\;\; \Delta_m^2(k,0)
    \equiv
    \frac{k^3}{2\pi^2}P_m(k,0) \Big).
    \label{eq:Delta_m_from_Pm0}
\end{equation}

\begin{figure}[t]
    \centering
    \includegraphics[width=1\linewidth]{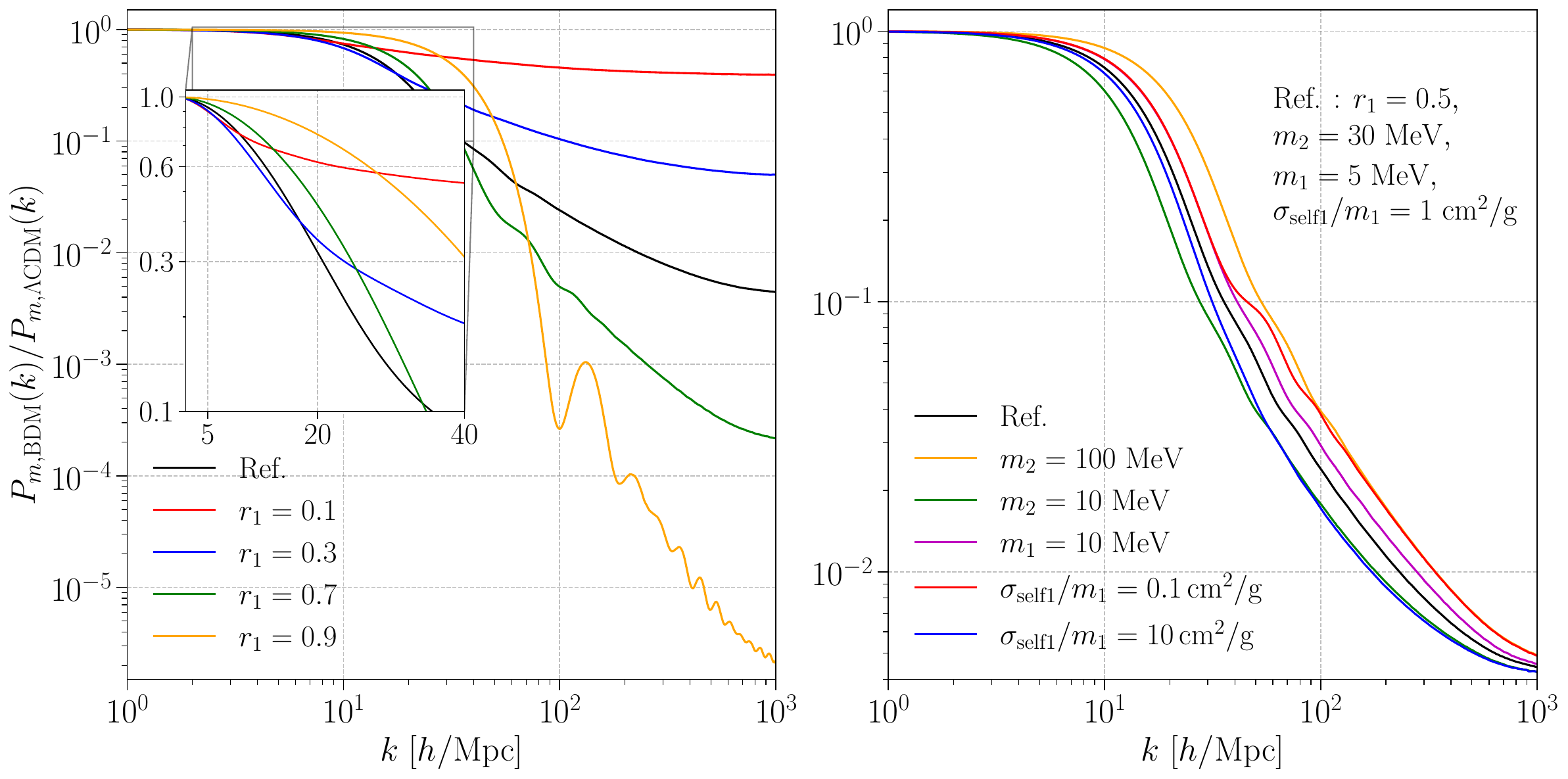}
    \caption{
    Linear matter power spectra, calculated from Eq.~\eqref{eq:Pm_from_seed} at $z=10$, for representative BDM parameter choices, normalized to the \LCDM model. The left panel shows the dependence on the light-component fraction $r_1$, while the right panel shows the effects of varying $m_2$, $m_1$, and $\sigma_{\rm self1}/m_1$. 
    The suppression at large wavenumber is caused by the pressure support of the self-heated $\chi_1$ fluid, and reduces the abundance of low-mass halos that can host the first radiative sources.
    }
    \label{fig:bdm_pk}
\end{figure}

Figure~\ref{fig:bdm_pk} shows the ratio of the BDM linear matter power
spectrum to the $\Lambda$CDM one at $z=10$,
$P_m(k)/P_{m,\Lambda{\rm CDM}}(k)$. In the left panel, we vary the
light-component fraction $r_1$ while keeping the other BDM parameters fixed to
the reference values. When $r_1$ is small, the light component is strongly
heated, but its contribution to the total DM density is limited.
Consequently, the suppression of the total matter power spectrum remains
moderate. As $r_1$ increases, the pressure support of the self-heated $\chi_1$ component
becomes more relevant for the total density perturbation. 
Around $k=\mathcal{O}(10)\,h{\rm Mpc}^{-1}$, the strongest suppression occurs for an
intermediate admixture, where $\chi_1$ is abundant enough to affect the total
matter field while the annihilation of the remaining heavy component still
provides efficient heating. In the opposite limit, $r_1\to 1$, the abundance of
$\chi_2$ becomes small, and the energy injection from
$\chi_2\bar{\chi}_2\to\chi_1\bar{\chi}_1$ is reduced. The temperature
enhancement of $\chi_1$ is therefore weakened, and the suppression at these
intermediate wavenumbers becomes less pronounced.

At larger wavenumbers, $k\gtrsim 100\,h{\rm Mpc}^{-1}$, the pressure term is
enhanced by the factor $c_{s,1}^2 k^2/a^2$. Therefore, even a relatively small sound
speed can affect the growth of perturbations. This explains why models with a
large light-component fraction can still show strong suppression and residual
oscillatory features at high $k$. These features are the DAO generated by the competition between gravitational collapse and the pressure support of the $\chi_1$ fluid.

The right panel shows the dependence on the other BDM parameters. Increasing
$\sigma_{\rm self1}/m_1$ extends the period over which the injected kinetic
energy is redistributed within the light component, leading to stronger
small-scale suppression. The dependence on $m_2$ follows from the heating rate
$\gamma_{\rm heat}\propto n_2^2\delta m/n_1$. Although a larger $m_2$ increases
the mass splitting $\delta m$, it also reduces the number density
$n_2\simeq \rho_2/m_2$ of the heavy component. In the benchmark shown here, the
reduction in $n_2$ dominates, so the $m_2=100\,{\rm MeV}$ case exhibits weaker
suppression than the reference model. Conversely, lowering $m_2$ increases the
heavy-particle number density and can enhance the suppression despite the
smaller mass splitting.

The effect of varying $m_1$ should be interpreted with some care. At fixed
$m_2$ and $r_1$, increasing $m_1$ lowers the number density
$n_1\simeq \rho_1/m_1$ of the light component. The injected kinetic energy is
therefore shared by fewer $\chi_1$ particles, which can lead to a larger
temperature $T_1$. However, the suppression of density perturbations is governed
by the sound speed $c_{s,1}^2\sim T_1/m_1$, rather than by $T_1$ alone. Thus the increase in $T_1$ can be compensated by the larger particle mass. In addition, increasing
$m_1$ at fixed $m_2$ reduces the mass splitting $\delta m=m_2-m_1$, and hence lowers the kinetic energy injected per annihilation event. As a result, the $m_1=10\,{\rm MeV}$ case can exhibit weaker small-scale suppression than the reference model, despite having a
larger light-sector temperature.

%%%%%%%%%%%%%%%%%%%%%%%%%%%%%%%%%%%%%%%%%%%%%%%%%%%
\section{21 cm Physics in BDM Cosmology}
\label{sec:21cm}
%%%%%%%%%%%%%%%%%%%%%%%%%%%%%%%%%%%%%%%%%%%%%%%%%%%

In this section, we describe how the BDM cosmology is incorporated into the
21\,cm calculation. Its main effect enters through the growth of density perturbations. The self-heated light component suppresses small-scale matter power, which changes the abundance and spatial distribution of early halos. This in turn modifies the Ly$\alpha$, X-ray, and ionizing radiation backgrounds that determine the observable 21\,cm signal.
We detail below how these changes are implemented in the semi-numerical 21\,cm calculation.

%%%%%%%%%%%%%%%%%%%%%%%%%%%%%%%%%%%%%%%%%%%%%%%%%%%
\subsection{Semi-numerical Realization and Simulation Setup}
\label{ssec:simsetup}
%%%%%%%%%%%%%%%%%%%%%%%%%%%%%%%%%%%%%%%%%%%%%%%%%%%

In this subsection, we describe how the background and perturbation quantities computed by the \BDMCLASS\ code are passed to \FirstClass~\cite{Flitter:2023mjj,Flitter:2023rzv,Flitter:2024eay} as inputs for the semi-numerical 21\,cm calculation.

The background quantities used in \FirstClass\ are
\begin{equation}
\left\{
H(z),\,
\bar\rho_b(z),\,
\bar\rho_{\rm dm}(z),\,
Y_p,\,
z_{\rm rec},\,
T_k(z_{\rm rec}),\,
\bar x_e(z_{\rm rec})
\right\},
\label{eq_bdm_background_outputs}
\end{equation}
where $H(z)$ denotes the Hubble parameter, and $\bar\rho_b(z)$ and $\bar\rho_{\rm dm}(z)$ are the mean energy densities of baryons and total DM, respectively. The primordial helium mass fraction is defined by $Y_p=4\bar{n}_{\rm He}/(\bar{n}_{\rm H}+4\bar{n}_{\rm He})$, where $\bar{n}_{\rm H}$ and $\bar{n}_{\rm He}$ denote the mean number densities of hydrogen and helium nuclei, respectively. In the nonrelativistic limit, the energy densities of hydrogen and helium nuclei are given by $\bar\rho_{\rm H}\simeq m_p\bar n_{\rm H}$ and $\bar\rho_{\rm He}\simeq 4m_p\bar n_{\rm He}$. Their sum gives the mean baryon energy density
\begin{equation}
    \bar\rho_b(z)
    =
    \bar\rho_{\rm H}(z)+\bar\rho_{\rm He}(z)
    \simeq
    m_p
    \left[
    \bar n_{\rm H}(z)+4\bar n_{\rm He}(z)
    \right].
    \label{rhob_def_with_n}
\end{equation}
Combining the definition of $Y_p$ with Eq.~\eqref{rhob_def_with_n}, we can express the mean number densities of hydrogen and helium nuclei as
\begin{equation}
    \bar n_{\rm H}(z)
    =
    \frac{(1-Y_p)\bar\rho_b(z)}{m_p},
    \qquad
    \bar n_{\rm He}(z)
    =
    \frac{Y_p\bar\rho_b(z)}{4m_p}.
    \label{eq_mean_nH_nHe}
\end{equation}

The homogeneous gas kinetic temperature is initialized as $T_k(z_{\rm rec})$ at the recombination redshift $z_{\rm rec}$, and subsequently evolved by \FirstClass\ according to Eq.~\eqref{eq_Tk_evolution}. The free electron fraction is defined as the number of free electrons per hydrogen nucleus,  
\begin{equation}
    \bar x_e(z_{\rm rec})
    \equiv
    \frac{\bar n_e(z_{\rm rec})}{\bar n_{\rm H}(z_{\rm rec})}\;.
    \label{eq_xe_initial}
\end{equation}
Its value is also initialized at $z_{\rm rec}$ and subsequently evolved according to Eq.~\eqref{eq_xeX_evolution_z}. The initial redshift $z_{\rm rec}=1069$ is chosen as the redshift at which the residual mean free-electron fraction satisfies $\bar x_e(z_{\rm rec})=0.1$.

The background inputs listed in
Eq.~\eqref{eq_bdm_background_outputs} are kept identical to those of the
reference \LCDM\ cosmology. 
The BDM dependence enters through the perturbation quantities. The modified
linear outputs include
\begin{equation}
\left\{
D_b(k,z),\,
D_{\rm dm}(k,z),\,
\theta_b(\mathbf{k},z),\,
\theta_{\rm dm}(\mathbf{k},z),\,
P_m(k,z),\,
\sigma(M_R,z)
\right\}.
\label{eq:bdm_perturb_outputs}
\end{equation}
The growth functions $D_b(k,z)$ and $D_{\rm dm}(k,z)$ are defined in
Eq.~\eqref{eq:growth_from_transfer}. The effective DM velocity divergence is given by
\begin{equation}
    \theta_{\rm dm}(\mathbf{k},z)
    =
    \frac{
        \bar\rho_1(z)\theta_1(\mathbf{k},z)
        +
        \bar\rho_2(z)\theta_2(\mathbf{k},z)
    }{
        \bar\rho_1(z)+\bar\rho_2(z)
    } ,
    \label{eq:theta_dm_eff_21cm}
\end{equation}
where $\theta_1(\mathbf{k},z)$ and $\theta_2(\mathbf{k},z)$ are the velocity divergences of $\chi_1$ and $\chi_2$, as defined in Eq.~\eqref{eq:dm_pert_def}.
The baryon velocity divergence $\theta_b(\mathbf{k},z)$ is also used when including the feedback associated with the relative motion between baryons and DM. We characterize this effect through the relative velocity-divergence field as
\begin{equation}
    \theta_{{\rm dm}b}(\mathbf{k},z)
    \equiv
    \theta_{\rm dm}(\mathbf{k},z)-\theta_b(\mathbf{k},z) .
\end{equation}

The matter power spectrum $P_m(k,z)$, defined in Eq.~\eqref{eq:Pm_from_seed}, is used to compute the variance of the smoothed linear density field over a scale $R$,
\begin{equation}
    \sigma^2(M_R,z)
    =
    \int d\ln k\,
    \frac{k^3}{2\pi^2}
    P_m(k,z)
    \left|W(kR)\right|^2 .
    \label{eq_sigma_R}
\end{equation}
The smoothing radius $R$ corresponds to a mass $M_R=4\pi R^3 \bar\rho_m(0)/3$, where $\bar\rho_m(0)$ is the mean matter density at $z=0$. The real-space top-hat window function and its Fourier transform are defined as
\begin{equation}
W_{R}(r)
=
\frac{3}{4\pi {R}^3}
\Theta({R}-r),
\qquad
W(kR)
=
\frac{3}{(kR)^3}
\left[
    \sin(kR)-kR\cos(kR)
\right].
\label{eq_tophat_window}
\end{equation}
The top-hat window function filters out fluctuations on scales smaller than $R$, suppressing modes with $k\gtrsim 1/R$. Increasing $R$ therefore reduces the contribution of small-scale fluctuations. As a result, $\sigma(M_R,z)$ decreases. Conversely, a smaller $R$ leaves more small-scale fluctuations, increasing $\sigma(M_R,z)$.
In BDM cosmology, the linear matter power spectrum is suppressed relative to \LCDM\ for modes with large $k$, as shown in Figure~\ref{fig:bdm_pk}. Therefore, $\sigma(M_R,z)$ is smaller than in \LCDM. 

The spatial distribution of baryon and effective DM is realized on grids within a periodic cubic volume of comoving side length $L$. The fiducial box size and grid resolutions are set to
\begin{equation}
    L=100\,{\rm Mpc},\qquad
    N_{\rm hi}=150^3,\qquad
    N_{\rm lo}=50^3 .
    \label{eq:box_setup}
\end{equation}
A high-resolution grid with $N_{\rm hi}$ cells is used to calculate perturbed density and velocity fields, whereas a low-resolution grid with $N_{\rm lo}$ cells is used for 21\,cm calculations. The fundamental mode of the simulation is determined by the size of the box,
\begin{equation}
    k_{\rm f}=\frac{2\pi}{L}
    \simeq 0.063\,{\rm Mpc}^{-1}.
\end{equation}
The cell sizes for the high and low resolution grids are $\Delta x_{\rm hi}=L/150\simeq 0.67\,{\rm Mpc}$ and $\Delta x_{\rm lo}=L/50=2\,{\rm Mpc}$, respectively. The Nyquist wavenumber is given by
\begin{equation}
    k_{\rm Ny,lo}
    =
    \frac{\pi}{\Delta x_{\rm lo}}
    \simeq
    1.57\,{\rm Mpc}^{-1}.
\end{equation}

Primordial density fluctuations of baryon and effective DM density fields at each redshift are constructed using the second-order Lagrangian perturbation theory (2LPT). In this approach, particles initially located at uniform Lagrangian coordinates $\mathbf q$ are mapped to Eulerian coordinates $\mathbf x$ by displacement field $\mathbf s_i$,
\begin{equation}
    \mathbf x_i(\mathbf q,z)
    =
    \mathbf q
    +
    \mathbf s_i(\mathbf q,z),
    \qquad
    i\in\{b,{\rm dm}\},
    \label{eq:lpt_mapping}
\end{equation}
where its leading-order contribution in Fourier space is given by
\begin{equation}
    \mathbf s_i^{(1)}(\mathbf k,z)
    =
    \left[
        D_i(k,z)-D_i(k,z_0)
    \right]
    \frac{i\mathbf k}{k^2}
    \delta_{m}(\mathbf k,0),
    \qquad
    i\in\{b,{\rm dm}\},
    \label{eq:lpt_first_order}
\end{equation}
where $z_0=z_{\rm rec}\simeq 1069$. The next-leading order displacement fields are computed using the 2LPT implementation~\cite{Bouchet:1994xp} inherited from \FirstClass. The resulting 3D non-linear density fields for baryons and effective DM are denoted by
\begin{equation}
\left\{
\delta_b^{\rm NL}(\mathbf r,z),
\delta_{\rm dm}^{\rm NL}(\mathbf r,z)
\right\}.
\label{eq_nonlinear_density_fields}
\end{equation}

Unless otherwise stated, the astrophysical parameters are fixed to the fiducial EOS2021~\cite{Munoz:2021psm} values,
\begin{eqnarray}
    \Theta_{\rm astro}
    &=&
    \left\{
    \log_{10}f_{\ast}^{\rm II},\,
    \alpha_{\ast}^{\rm II},\,
    \log_{10}f_{\rm esc}^{\rm II},\,
    \alpha_{\rm esc}^{\rm II},\,
    \log_{10}\left(\frac{L_X^{\rm II}}{{\rm SFR}}\right),\,
    \log_{10}\left(\frac{M_{\rm turn}}{M_\odot}\right),\,
    N_{\gamma},\,
    t_\ast
    \right\}
    \nonumber\\
    &=&
    \left\{
    -1.25,\,
    0.5,\,
    -1.35,\,
    -0.3,\,
    40.5,\,
    8.7,\,
    5000,\,
    0.5
    \right\}.
    \label{eq:astro_params_schematic}
\end{eqnarray}
The parameters $f_{\ast}^{\rm II}$ and $\alpha_{\ast}^{\rm II}$ set the normalization and halo-mass dependence of Pop II star formation efficiency, respectively. Similarly, $f_{\rm esc}^{\rm II}$ sets the normalization of the fraction of ionizing photons that escape from galaxies into the IGM, while $\alpha_{\rm esc}^{\rm II}$ determines its halo-mass dependence. The soft-band X-ray luminosity per unit star formation rate (SFR) is characterized by $L_X^{\rm II}/{\rm SFR}$, while $M_{\rm turn}$ sets the halo mass scale below which star formation is exponentially suppressed. The quantity $N_\gamma$ denotes the number of ionizing photons produced per stellar baryon, and $t_\ast$ specifies the star-formation timescale as a fraction of the Hubble time. All parameters except $L_X^{\rm II}/{\rm SFR}$ and $M_{\rm turn}$ are dimensionless. In Eq.~\eqref{eq:astro_params_schematic}, $L_X^{\rm II}/{\rm SFR}$ is expressed in units of ${\rm erg}\,{\rm s}^{-1}(M_\odot\,{\rm yr}^{-1})^{-1}$, and $M_{\rm turn}$ is given in units of $M_\odot$.

For computational tractability, we vary only the following three astrophysical parameters in the forecasts presented in Section~\ref{sec:bayesian},
\begin{eqnarray}
    \Theta_{\rm astro}^{\rm scan}
    &=&
    \left\{
    \log_{10}\left(\frac{L_X^{\rm II}}{{\rm SFR}}\right),\,
    \log_{10}\left(\frac{M_{\rm turn}}{M_\odot}\right),\,
    \log_{10}N_\gamma
    \right\}.
    \label{eq:astro_params_varied}
\end{eqnarray}
These parameters have the largest impact on the timing and amplitude of the 21\,cm signal. The remaining astrophysical parameters are fixed to their fiducial values.

%%%%%%%%%%%%%%%%%%%%%%%%%%%%%%%%%%%%%%%%%%%%%%%%%%%
\subsection{21 cm brightness temperature}
\label{sec:21cm_brightness}
%%%%%%%%%%%%%%%%%%%%%%%%%%%%%%%%%%%%%%%%%%%%%%%%%%%

In this subsection, we describe how the BDM-modified perturbation quantities listed in Eq.~\eqref{eq:bdm_perturb_outputs} enter the semi-numerical calculation of the 21\,cm signal. 
The theoretical framework adopted in this subsection follows Refs.~\cite{Madau:1996cs,Barkana:2000fd,Furlanetto:2006jb,Morales:2009gs,Pritchard:2012,Furlanetto:2015apc,Mesinger:2010ne} and the numerical implementation is based on the publicly available code \FirstClass~\cite{Flitter:2023mjj,Flitter:2023rzv,Flitter:2024eay}.

We study the redshifted 21\,cm signal arising from the hyperfine transition of neutral hydrogen in the IGM.\footnote{Throughout this subsection, we use cgs based astrophysical units for spatial quantities, densities, and rates. Number densities are measured in ${\rm cm}^{-3}$, rate coefficients in ${\rm cm}^{3}\,{\rm s}^{-1}$, and transition rates in ${\rm s}^{-1}$. Temperatures are measured in Kelvin. Energies are quoted in eV, with $k_B$ understood in ${\rm eV}\,{\rm K}^{-1}$.} 
A transition from the triplet to the singlet state emits a photon with a rest-frame wavelength $\lambda_{21}\simeq 21.1~\mathrm{cm}$, corresponding to a frequency $\nu_{21}\simeq 1420.4~\mathrm{MHz}$. For a signal originating at redshift $z$, the observed frequency $\nu_{\rm obs}$ at Earth is related to the rest-frame frequency $\nu_{21}$ by
\begin{equation}
    \nu_{\mathrm{obs}} = \frac{\nu_{21}}{1+z} ,
\end{equation}

At a comoving position $\mathbf r=\chi(z)\hat{\mathbf n}$, where $\hat{\mathbf n}$ specifies the line-of-sight direction and $\chi(z)$ is the radial comoving distance, the relative populations of the two hyperfine levels are characterized by the spin temperature $T_s(\mathbf r,z)$,
\begin{equation}
    \frac{n_1(\mathbf r,z)}{n_0(\mathbf r,z)}
    \equiv
    \frac{g_1}{g_0}
    \exp\left[
        -\frac{T_\star}{T_s(\mathbf r,z)}
    \right],
    \label{eq:Ts_def}
\end{equation}
where $n_1(\mathbf r,z)$ and $n_0(\mathbf r,z)$ are the local number densities of hydrogen atoms in the triplet and singlet states, respectively. Their statistical degeneracies are $g_1=3$ and $g_0=1$, and
$T_\star\equiv h\nu_{21}/k_B\simeq 68.2~{\rm mK}$ is the 21\,cm transition temperature.

The spin temperature $T_s(\mathbf r,z)$ enters the calculation of the 21\,cm brightness temperature obtained from radiative transfer through the IGM. With CMB photons servaing as the background radiation and $\tau_{21}(\mathbf r,z)$ denoting the 21\,cm optical depth of the IGM, the brightness temperature is given by
\begin{equation}
    T_{b,{\rm loc}}(\mathbf r,z)
    =
    \bar T_\gamma(z)e^{-\tau_{21}(\mathbf r,z)}
    +
    T_s(\mathbf r,z)
    \left[
        1-e^{-\tau_{21}(\mathbf r,z)}
    \right],
    \label{eq:Tb_loc}
\end{equation}
where $\bar T_\gamma(z)\simeq 2.726\,(1+z)\,{\rm K}$ is the background CMB temperature. The brightness temperature observed on Earth is redshifted according to $T_{b}=T_{b,{\rm loc}}/(1+z)$.
Subtracting the CMB temperature from $T_{b}$, we obtain the observed differential brightness temperature,
\begin{align}
    \delta T_{b}(\mathbf r,z)
    &\equiv
    \frac{
        T_{b,{\rm loc}}(\mathbf r,z)-\bar T_\gamma(z)
    }{
        1+z
    }
    \nonumber\\
    &=
    \frac{
        \left(T_s(\mathbf r,z)-\bar T_\gamma(z)\right)
    }{
        1+z
    }
    \left[
        1-e^{-\tau_{21}(\mathbf r,z)}
    \right].
    \label{eq:deltaTbObs}
\end{align}
The optical depth represents a dimensionless absorption probability of 21\,cm photons along the line of sight
\begin{equation}
    \tau_{21}(\mathbf r,z)
    =
    \frac{3}{16}
    \frac{\hbar c^3 A_{10}}
    {k_B T_s(\mathbf r,z)\nu_{21}^2}
    \frac{
        n_{\rm HI}(\mathbf r,z)
    }{
        (1+z)\,dv_\parallel/dr_\parallel(\mathbf r,z)
    },
    \label{eq_tau21_general}
\end{equation}
where $A_{10}\simeq 2.85\times 10^{-15}~{\rm s}^{-1}$ is the Einstein
coefficient for spontaneous decay of the hyperfine triplet state. 
Along the line-of-sight coordinate $r_\parallel$\footnote{We use the subscript $\parallel$ for the component along $\hat{\mathbf n}$ and the subscript $\perp$ for the transverse component.}, the velocity gradient is given by the sum of the Hubble-flow and baryon peculiar-velocity contributions,
\begin{equation}
\frac{dv_\parallel}{dr_\parallel}(\mathbf r,z)
=
\frac{H(z)}{1+z}
+
\frac{\partial v_{b,\parallel}}
{\partial r_\parallel}(\mathbf r,z),
\label{eq:los_velocity_gradient}
\end{equation}
where $v_{b,\parallel}\equiv \mathbf v_b\cdot\hat{\mathbf n}$ is the
line-of-sight component of the baryon peculiar velocity. In BDM cosmology, this velocity field is indirectly modified through gravitational interactions with DM.

The local neutral hydrogen number density in Eq.~\eqref{eq_tau21_general} can be written in terms of the local neutral fraction $x_{\rm HI}(\mathbf r,z)$ and the local number density of hydrogen nuclei $n_{\rm H}(\mathbf r,z)$ as
\begin{equation}
n_{\rm HI}(\mathbf r,z)
=
x_{\rm HI}(\mathbf r,z)n_{\rm H}(\mathbf r,z).
\label{eq_nHI_local}
\end{equation}
The local density $n_{\rm H}(\mathbf r,z)$ is related to its mean value $\bar n_{\rm H}(z)$ by
\begin{equation}
n_{\rm H}(\mathbf r,z)
=
\bar n_{\rm H}(z)
\left(
1+\delta_b^{\rm NL}(\mathbf r,z)
\right),
\label{eq_local_nH}
\end{equation}
where $\bar n_{\rm H}(z)$ and $\delta_b^{\rm NL}(\mathbf r,z)$ are defined in Eqs.~\eqref{eq_mean_nH_nHe} and \eqref{eq_nonlinear_density_fields}, respectively. 
In BDM cosmology, $\bar n_{\rm H}(z)$ remains the same as in the reference \LCDM cosmology, while the local density field is modified through $\delta_b^{\rm NL}(\mathbf r,z)$. The suppression of DM perturbations on small scales reduces the corresponding gravitational potentials and thereby weakens small-scale baryon clustering.

The neutral fraction $x_{\rm HI}(\mathbf r,z)$ is related to the local ionized hydrogen fraction $x_{\rm HII}(\mathbf r,z)$ by
\begin{equation}
x_{\rm HI}(\mathbf r,z)
=
1-x_{\rm HII}(\mathbf r,z),
\label{eq_xHI_from_xHII}
\end{equation}
where $x_{\rm HII}(\mathbf r,z)$ accounts for ionization by both UV and X-ray photons,
\begin{equation}
x_{\rm HII}(\mathbf r,z)
=
x_{\rm HII}^{\rm UV}(\mathbf r,z)
+
x_{\rm HII}^{\rm X}(\mathbf r,z).
\label{eq_xHII_total}
\end{equation}
The UV contribution $x_{\rm HII}^{\rm UV}(\mathbf r,z)$ and the X-ray contribution $x_{\rm HII}^{\rm X}(\mathbf r,z)\simeq x_e^{\rm X}(\mathbf r,z)$ are described in Appendices~\ref{app:UV_fraction} and \ref{subsec:xe_evolution}, respectively.

In the hydrogen-dominated approximation, the local free-electron fraction per hydrogen nucleus is
\begin{equation}
x_e(\mathbf r,z)
\simeq
x_{\rm HII}(\mathbf r,z).
\label{eq_xe_total}
\end{equation}
Therefore, the associated local number densities are
\begin{eqnarray}
n_e(\mathbf r,z)
&=&
x_e(\mathbf r,z)n_{\rm H}(\mathbf r,z),
\label{eq_ne}
\\
n_{\rm HII}(\mathbf r,z)
&=&
x_{\rm HII}(\mathbf r,z)n_{\rm H}(\mathbf r,z),\label{eq_nHII} \\
n_{\rm HI}(\mathbf r,z)
&=&
x_{\rm HI}(\mathbf r,z)n_{\rm H}(\mathbf r,z).
\label{eq_nHI}
\end{eqnarray}

\begin{figure}[t!]
    \centering
    \includegraphics[width=0.98\linewidth]{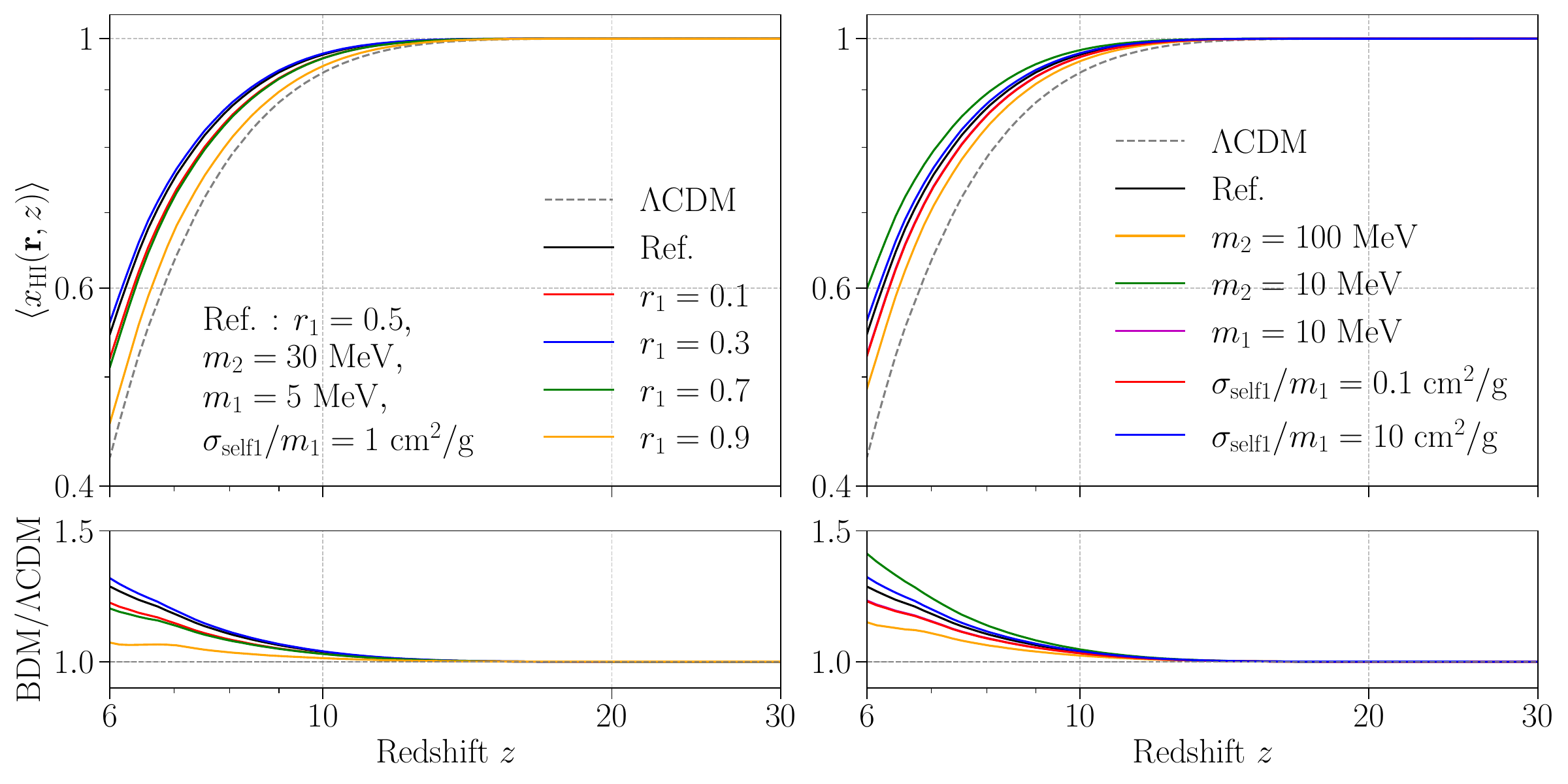}
    \includegraphics[width=0.98\linewidth]{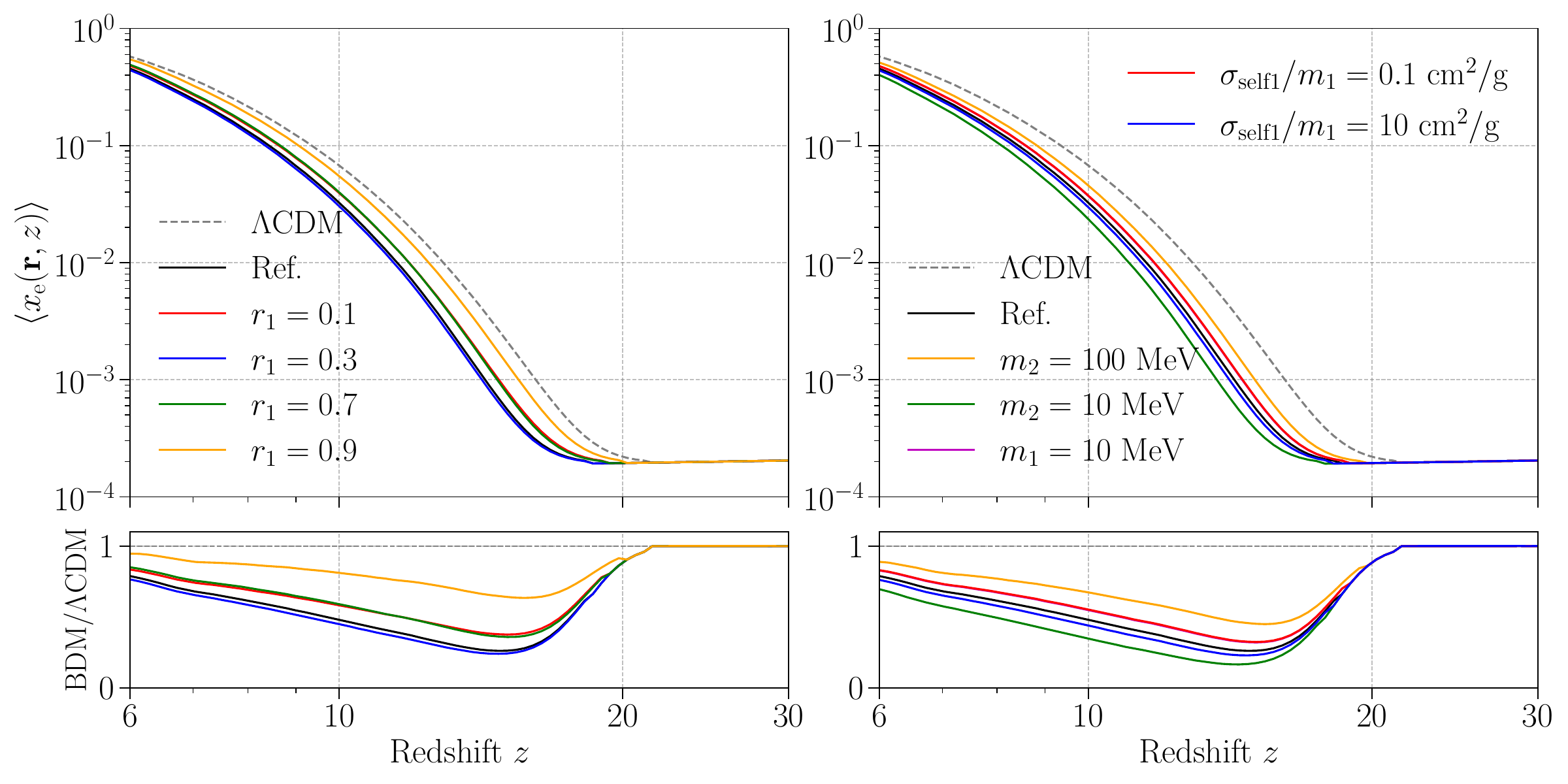}
    \caption{
    Redshift evolution of the volume averaged neutral hydrogen fraction
    $\langle x_{\rm HI}(\mathbf r,z)\rangle$ (upper panels; see Eq.~\eqref{eq_xHI_from_xHII}) and the
    free electron fraction per hydrogen nucleus
    $\langle x_e(\mathbf r,z)\rangle$ (lower panels; see Eq.~\eqref{eq_xe_total}) for \LCDM\ (dashed gray) and representative BDM
    parameter choices. For each quantity, the lower subpanels show the ratio of the BDM prediction to the corresponding
\LCDM\ result. The left panels show the dependence
    on the light component fraction $r_1$, while the right panels illustrate
    the effects of varying $m_2$, $m_1$, and $\sigma_{\rm self1}/m_1$ around
    the reference model. In BDM cosmology, the suppression of small-scale
    matter fluctuations delays the buildup of ionizing sources. As a result, the neutral fraction remains larger,
    and the free electron fraction remains smaller than in \LCDM\ at the
    same redshift.
    }
    \label{fig:xHI_bdm}
\end{figure}

We show in Figure~\ref{fig:xHI_bdm} the volume averaged neutral hydrogen fraction $\langle x_{\rm HI}(\mathbf r,z)\rangle$ and the  free electron fraction per hydrogen nucleus $\langle x_e(\mathbf r,z)\rangle$ for representative BDM parameter choices. At a given redshift, BDM retains a larger neutral fraction and a smaller free-electron fraction than \LCDM, reflecting a delayed ionization history. This delay originates from the suppression of small-scale matter power shown in Figure~\ref{fig:bdm_pk}, which inhibits the formation of source-hosting halos and reduces the effective collapsed fraction, as shown in Figure~\ref{fig:eff_f_coll_ratio}. The resulting reduction in ionizing sources lowers the cumulative number of ionizing photons computed in Eq.~\eqref{eq_ionizing_photon_per_baryon}.

The ratio panels make this delayed ionization history explicit. 
For the neutral fraction, $\langle x_{\rm HI}\rangle_{\rm BDM}/\langle x_{\rm HI}\rangle_{\Lambda{\rm CDM}}$ stays close to unity before reionization becomes efficient, and then rises above unity at lower redshift. Around $z\simeq6$, the BDM models shown here retain about $10\%$ to $40\%$ more neutral hydrogen than \LCDM. 
On the other hand, the averaged free-electron fraction is
suppressed in BDM over $8\lesssim z\lesssim12$, with
$\langle x_e\rangle_{\rm BDM}/
\langle x_e\rangle_{\Lambda{\rm CDM}}$
ranging approximately from $0.5$ to $0.8$.

The parameter dependence follows the same physical trend as the linear power
spectrum. In the left panels, the delay is most pronounced for intermediate
values of $r_1$, where the light component is abundant enough to affect the
total matter perturbation while the remaining heavy component still provides
efficient self heating. The effect is weaker toward the CDM like limits,
where either the heated light component has a small abundance or the heavy
component becomes too depleted to inject sufficient kinetic energy. In the
right panels, models with stronger small-scale suppression in
Figure~\ref{fig:bdm_pk}, such as smaller $m_2$ or larger
$\sigma_{\rm self1}/m_1$, retain a larger neutral fraction and a smaller free
electron fraction at fixed redshift. By contrast, parameter choices that
produce milder suppression of the linear matter power spectrum lead to a
reionization history closer to \LCDM.

The discussion above specifies the BDM dependence of the neutral hydrogen
density $n_{\rm HI}(\mathbf r,z)$ and the line-of-sight velocity gradient
$dv_\parallel/dr_\parallel(\mathbf r,z)$, which enter the 21\,cm optical
depth in Eq.~\eqref{eq_tau21_general}.
The remaining ingredient is the
spin temperature $T_s(\mathbf r,z)$. Since $T_s$ appears both in the optical
depth and directly in the differential brightness temperature in
Eq.~\eqref{eq:deltaTbObs}, we discuss its evolution separately in
Section~\ref{sec_21cm_spin}. 

For comparison with the usual analytic limit, we also note that the optical
depth is typically small in the diffuse IGM over the redshift range mainly
considered in this work, roughly $6\lesssim z\lesssim30$. In the optically
thin limit, $\tau_{21}(\mathbf r,z)\ll1$, Eq.~\eqref{eq:deltaTbObs} reduces to
\begin{align}
\delta T_{b}(\mathbf r,z)
&\simeq
\frac{
\left(T_s(\mathbf r,z)-\bar T_\gamma(z) \right)
}{
1+z
}
\tau_{21}(\mathbf r,z).
\label{eq_deltaTbObs_thin}
\end{align}
This approximation can become less accurate at higher redshifts, or in very
cold or highly overdense regions, where the neutral hydrogen density is large
or the spin temperature is low. We therefore use the full expression in
Eq.~\eqref{eq:deltaTbObs} in our numerical calculation, while the optically
thin form is useful for interpreting the lower-redshift behavior of the
21 cm signal.

%%%%%%%%%%%%%%%%%%%%%%%%%%%%%%%%%%%%%%%%%%%%%%%%%%%
\subsection{Spin temperature}
\label{sec_21cm_spin}
%%%%%%%%%%%%%%%%%%%%%%%%%%%%%%%%%%%%%%%%%%%%%%%%%%%

In this section, we summarize how the spin temperature $T_s(\mathbf r,z)$, defined in Eq.~\eqref{eq:Ts_def}, is related to the gas temperature, ionization state, and Ly$\alpha$ radiation field. The detailed calculations are provided in Appendices~\ref{sec_21cm_Tk}--\ref{subsec_xalpha}. Here we present the resulting redshift evolution of the relevant temperatures and coupling coefficients.

Three competing processes govern the spin temperatures. Interactions with CMB photons tend to couple $T_s(\mathbf r,z)$ to the CMB temperature $\bar T_\gamma(z)$, while collisions of neutral hydrogen with other hydrogen atoms, electrons, and protons couple it to the gas kinetic temperature $T_k(\mathbf r,z)$. Ly$\alpha$ scattering induces transitions between the hyperfine states through the Wouthuysen-Field effect, providing a coupling to the color temperature $T_\alpha(\mathbf r,z)$ of the Ly$\alpha$ radiation field.

For $T_s\gg T_\star$, the Boltzmann equations governing the hyperfine level populations yield the quasi-equilibrium relation,
\begin{equation}
T_s^{-1}(\mathbf r,z)
=
\frac{
x_{\rm CMB}(\mathbf r,z)\bar T_\gamma^{-1}(z)
+
x_{\rm coll}(\mathbf r,z)T_k^{-1}(\mathbf r,z)
+
x_\alpha(\mathbf r,z)T_\alpha^{-1}(\mathbf r,z)
}{
x_{\rm CMB}(\mathbf r,z)
+
x_{\rm coll}(\mathbf r,z)
+
x_\alpha(\mathbf r,z)
} .
\label{eq_Ts_spin_bdm}
\end{equation}
The dimensionless coefficient $x_{\rm{CMB}}(\mathbf{r},z) \equiv (1-e^{-\tau_{21}(\mathbf{r},z)})/\tau_{21}(\mathbf{r},z)$ denotes how much of the background CMB at the 21\,cm frequency actually couples to the hyperfine transition after transfer through the IGM of the optical depth $\tau_{21}(\mathbf{r},z)$. For the redshift range mainly considered in this work, $6\lesssim z\lesssim30$, the optical depth is typically small $\tau_{21}(\mathbf r,z)\ll1$. 
In this optically thin limit, one has $x_{\rm CMB}(\mathbf r,z)\simeq1$.
In BDM cosmology, $x_{\rm CMB}(\mathbf r,z)$ inherits BDM dependence through the optical depth in Eq.~\eqref{eq_tau21_general}.

The collisional coupling coefficient accounts for the collisions of neutral hydrogen with electrons, protons, and other hydrogen atoms
\begin{equation}
x_{\rm coll}(\mathbf r,z)
=
\frac{T_\star}{A_{10}\bar T_\gamma(z)}
\left[
\kappa_{10}^{\rm HH}\!\left(T_k(\mathbf r,z)\right)
n_{\rm HI}(\mathbf r,z)
+
\kappa_{10}^{\rm eH}\!\left(T_k(\mathbf r,z)\right)
n_e(\mathbf r,z)
+
\kappa_{10}^{\rm pH}\!\left(T_k(\mathbf r,z)\right)
n_p(\mathbf r,z)
\right] .
\label{eq_xcoll_bdm}
\end{equation}
The coefficients $\kappa_{10}^{i{\rm H}}$ are the standard collisional
de-excitation rates for scattering between hydrogen atoms and the species
$i\in\{{\rm H},e,p\}$, where $p$ denotes free protons, equivalently ionized
hydrogen $\rm HII$. These rates have units of ${\rm cm}^{3}\,{\rm s}^{-1}$ and they depend on the local gas temperature $T_k(\mathbf r,z)$.
In this work, we use the
tabulated rates adopted in \FirstClass.
In BDM cosmology, $x_{\rm coll}(\mathbf r,z)$ inherits its dependence on the modified structure formation through $T_k(\mathbf r,z)$ in Eq.~\eqref{eq_Tk_evolution} and the local quantities $n_{\rm HI}(\mathbf r,z)$, $n_e(\mathbf r,z)$, and $n_p(\mathbf r,z)\simeq n_{\rm HII}(\mathbf r,z)$, defined in Eqs.~\eqref{eq_nHI_local}, \eqref{eq_ne}, and \eqref{eq_nHII}, respectively.

The Wouthuysen-Field coupling coefficient $x_\alpha$ is
\begin{equation}
x_\alpha(\mathbf r,z)
=\frac{T_\star}
{A_{10}\bar T_\gamma(z)}P_{10}(\mathbf r,z)
=
\frac{4T_\star}
{27A_{10}\bar T_\gamma(z)}
P_\alpha(\mathbf r,z),
\label{eq_xalpha_bdm}
\end{equation}
where $P_{10}(\mathbf r,z)$ is the triplet-to-singlet transition rate induced by Ly$\alpha$ scattering, and $P_\alpha(\mathbf r,z)$, defined in Eq.~\eqref{eq:Palpha_narrowline}, is the total Ly$\alpha$ scattering rate per neutral hydrogen atom, both with units of $\mathrm{s}^{-1}$. The selection rules and the relative transition probabilities give $P_{10}=4P_\alpha/27$ in the second equality. Since $P_\alpha$ is proportional to the Ly$\alpha$ background, $x_\alpha$ increases as the Ly$\alpha$ background generated by the radiative source grows. Further details of the Ly$\alpha$ background and scattering rate, including their BDM dependency, are provided in the Appendix~\ref{subsec_xalpha}.

\begin{figure}
    \centering
    \includegraphics[width=1.0\linewidth]{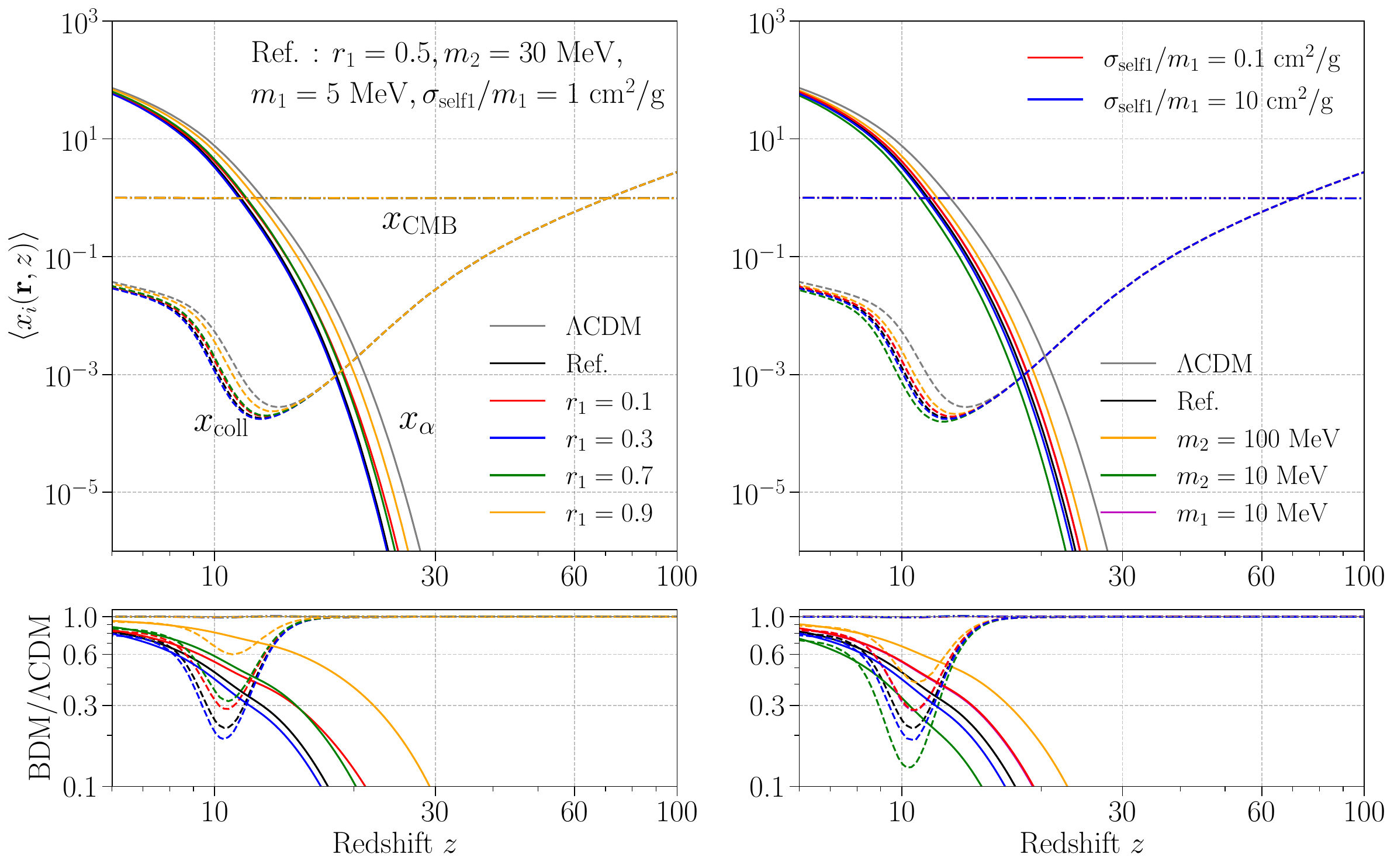}
    \caption{Redshift evolution of the volume-averaged coupling coefficients entering the spin-temperature relation in Eq.~\eqref{eq_Ts_spin_bdm}. The upper panels show the Wouthuysen-Field coupling $x_{\alpha}$ (solid; see Eq.~\eqref{eq_xalpha_bdm}), collisional coupling $x_{\rm coll}$ (dashed; see Eq.~\eqref{eq_xcoll_bdm}), and CMB coupling $x_{\rm CMB}$ (dot-dashed; defined below Eq.~\eqref{eq_Ts_spin_bdm}), while the lower panels show the ratios of the corresponding BDM predictions to their \LCDM values. The gray and black curves denote \LCDM and the reference BDM model, respectively. The left panels show the dependence on the light component fraction $r_1$, while the right panels show one-parameter variations of $m_2, m_1$, and $\sigma_{\rm self1}/m_1$ around the reference model. BDM alters the redshift evolution of both the source-driven Wouthuysen-Field coupling and the collisional coupling through its effects on early structure formation.}
    \label{fig:coefficients}
\end{figure}

Figure~\ref{fig:coefficients} shows the redshift evolution of the CMB, collisional, and Wouthuysen-Field coupling coefficients. The coefficient $x_{\rm coll}$ becomes important at high redshift, roughly $z\gtrsim70$, where the gas density is high enough. As the Universe expands, the number densities entering Eq.~\eqref{eq_xcoll_bdm} decrease rapidly, and
the products $n_i\kappa_{10}^{i{\rm H}}$ become inefficient.
In the intermediate range, roughly $30\lesssim z\lesssim70$, both
$x_{\rm coll}$ and $x_\alpha$ are small, and the CMB coupling
$x_{\rm CMB}$ dominates the spin-temperature evolution. In this regime,
$T_s(\mathbf r,z)$ remains close to $\bar T_\gamma(z)$.
At lower redshift, the first radiative sources build up the Ly$\alpha$
background and $x_\alpha$ increases. At redshifts $10\lesssim z\lesssim 12$, the Wouthuysen--Field coupling becomes the
dominant mechanism once $x_\alpha$ exceeds unity. In the strong-coupling
limit, $x_\alpha\gg1$, repeated Ly$\alpha$ scattering drives the spin
temperature toward the color temperature of the Ly$\alpha$ radiation field.
In BDM cosmology, suppression of small-scale structure formation reduces the abundance of low-mass halos that host the first radiative sources, delaying the development of Ly$\alpha$ backgrounds. The $x_\alpha$ coefficient becomes smaller than its \LCDM at $z\lesssim30$, where the source of radiation begins to form. The $x_{\rm coll}$ coefficient begins to deviate from \LCDM at $z\lesssim 15$, when the heating of the IGM becomes important.

The Ly$\alpha$ color temperature $T_{\alpha}$ entering Eq.~\eqref{eq_Ts_spin_bdm} is defined as
\begin{equation}
    \frac{P_{01}(\mathbf r,z)}{P_{10}(\mathbf r,z)}\equiv3 \exp\left[-\frac{T_\star}{T_\alpha(\mathbf r,z)}\right].
    \label{eq:Talpha}
\end{equation}
Here, $P_{01}(\mathbf r,z)$ represents the corresponding singlet-to-triplet transition rate induced by Ly$\alpha$ scattering with units of $\mathrm{s}^{-1}$. A larger $T_\alpha$ favors a higher relative triplet population.
Repeated Ly$\alpha$ scatterings modify the frequencies of Ly$\alpha$ photons. The recoil effect gradually shifts photon frequencies to lower values, while the thermal motion of hydrogen atoms leads to Doppler diffusion. Together, these effects drive the color temperature toward the gas kinetic temperature, $T_\alpha \rightarrow T_k$. The Wouthuysen–Field effect also influences the photon frequency distribution, slightly adjusting $T_\alpha$ towards $T_s$. In the default implementation of \FirstClass, $T_\alpha$ is evaluated using the fitting formula calibrated to the numerical Fokker--Planck solutions of Ref.~\cite{Hirata:2006}.
\begin{equation}
    T_\alpha^{-1}(\mathbf r,z)\simeq T_k^{-1}(\mathbf r,z)\left[1 + 0.405535\,{\rm K}\,(T_s^{-1}(\mathbf r,z)-T_k^{-1}(\mathbf r,z))\right].
    \label{eq:Talpha_approx}
\end{equation}
\begin{figure}
    \centering
    \includegraphics[width=1\linewidth]{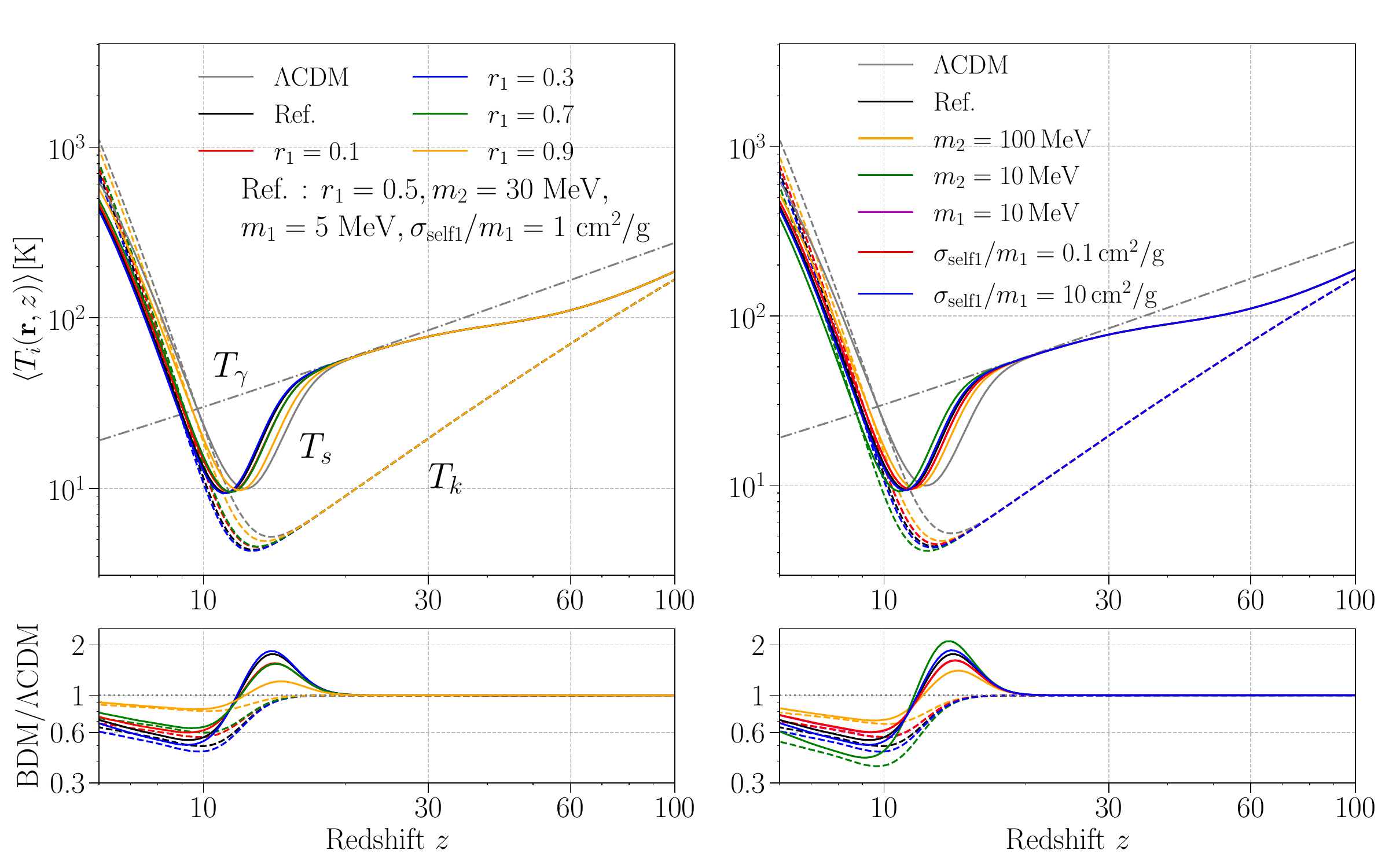}
    \caption{Redshift evolution of the volume-averaged spin temperature $T_s$ (solid; see Eq.~\eqref{eq_Ts_spin_bdm}), gas kinetic temperature $T_k$ (dashed; see Eq.~\eqref{eq_Tk_evolution}), and CMB temperature $T_\gamma$ (dot-dashed). The lower panels show the ratios of the corresponding BDM predictions to their \LCDM values. The left panels show the dependence on the light-component fraction $r_1$, while the right panels show variations of $m_2, m_1$ and $\sigma_{\rm self1}/m_1$ around the reference model. The suppression of low-mass halos in BDM delays both the Wouthuysen-Field coupling of $T_s$ to $T_k$ and the rise of $T_k$.}
    \label{fig:TsTk}
\end{figure}

To sum up, Figure~\ref{fig:TsTk} illustrates how the volume-averaged spin temperature $T_s$ (solid) shifts between the gas kinetic temperature $T_k$ (dashed) and the CMB temperature $\bar T_\gamma$ (dot-dashed) as the dominant coupling changes with redshift. At $z\gtrsim70$, the high gas density makes the coefficient $x_{\rm coll}$ large, so $T_s$ closely tracks $T_k$. As the Universe expands, $x_{\rm coll}$ declines, allowing $x_{\rm CMB}$ to dominate over both $x_{\rm coll}$ and $x_\alpha$ before the first radiative sources appear. As a result, $T_s$ is driven toward $\bar T_\gamma$, even though the gas itself cools below it. At redshifts $10\lesssim z\lesssim 15$, the accumulation of the Ly$\alpha$ background enhances $x_\alpha$, and the Wouthuysen–Field effect again couples $T_s$ to $T_\alpha\simeq T_k$. Subsequently at lower redshifts $z\lesssim 10$, the X-ray heating increases $T_k$, and the strong Ly-$\alpha$ coupling causes $T_s$ to increase in tandem with $T_k$. In BDM cosmology, deviations from \LCDM emerge once the first radiative sources begin to form. The suppression of small-scale structure reduces the abundance of low-mass halos, delaying the development of the Ly$\alpha$ background. Consequently, the onset of Wouthuysen-Field coupling is shifted to lower redshift around $z\sim20$, and $T_s$ remains close to $\bar T_\gamma$ for longer before approaching $T_k$. The reduced abundance of X-ray sources also postpones the heating of the gas, causing the rise of $T_k$ to occur at lower redshift.

%%%%%%%%%%%%%%%%%%%%%%%%%%%%%%%%%%%%%%%%%%%%%%%%%%%
\section{Results}
\label{sec:result}
%%%%%%%%%%%%%%%%%%%%%%%%%%%%%%%%%%%%%%%%%%%%%%%%%%%

The modifications to the ionization and spin-temperature histories discussed above are imprinted on the 21\,cm signal.
In this section, we study how BDM cosmology modifies the global 21\,cm
brightness temperature and the 21\,cm power spectrum during cosmic dawn and
reionization, and compare the resulting predictions with those of \LCDM.
Using projected SKA1-Low sensitivities, we examine whether BDM imprints
remain distinguishable after accounting for observational limitations and
astrophysical parameter uncertainties. We also compare the resulting reach
with sub-GeV direct detection constraints from
SENSEI.

%%%%%%%%%%%%%%%%%%%%%%%%%%%%%%%%%%%%%%%%%%%%%%%%%%%
\subsection{Global signal and 21 cm power spectrum}
\label{sec:global_signal}
%%%%%%%%%%%%%%%%%%%%%%%%%%%%%%%%%%%%%%%%%%%%%%%%%%%

\begin{figure}[t!]
    \centering
    \includegraphics[width=1.\linewidth]{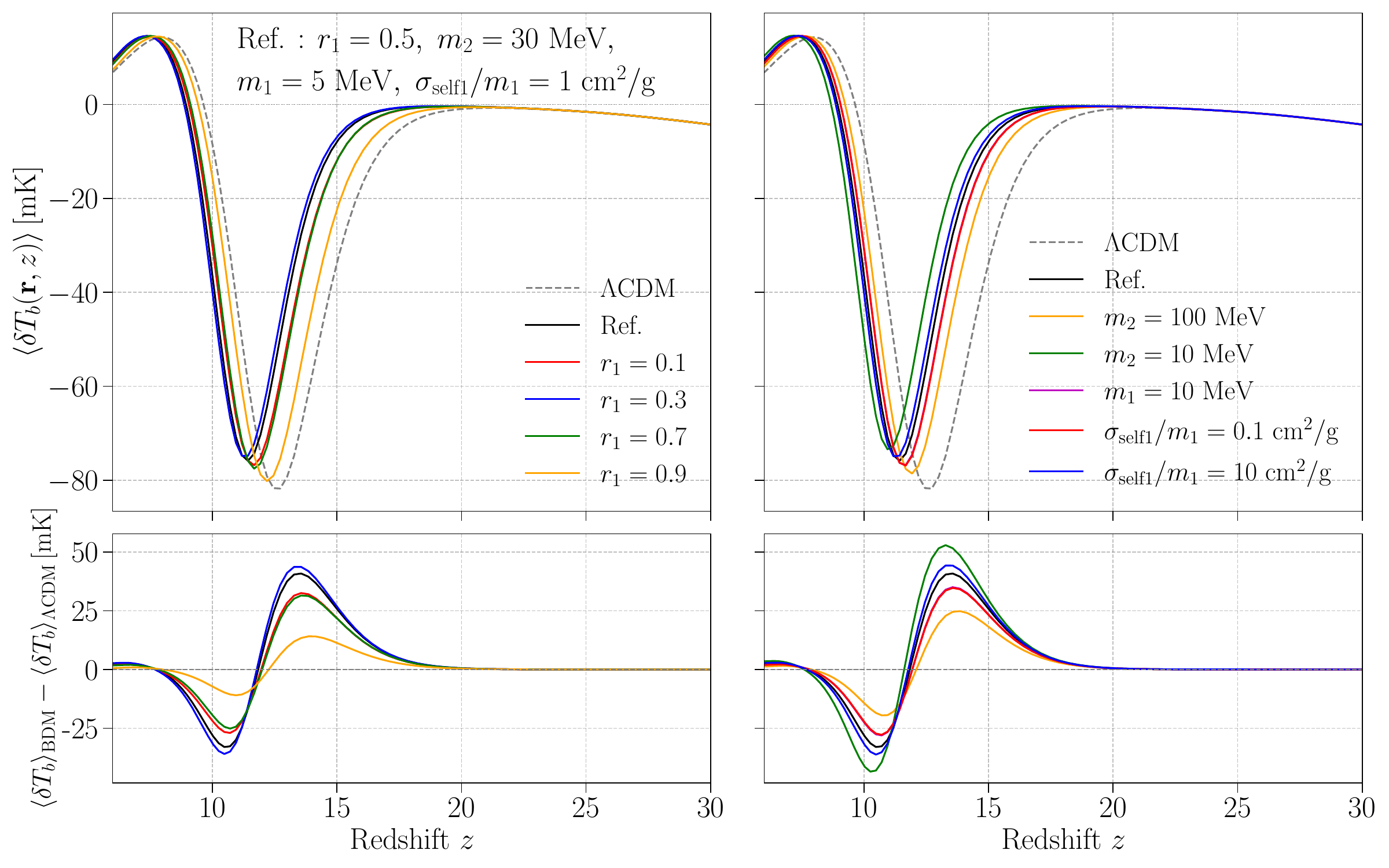}
    \caption{
    Redshift evolution of the volume averaged 21\,cm brightness temperature $\langle \delta T_{b}(\mathbf r,z)\rangle$ (see Eq.~\eqref{eq:deltaTbObs}) for \LCDM\ and representative BDM parameter choices (upper panels). The lower panels show the corresponding differences relative to \LCDM. The left panel shows the dependence on the light component fraction $r_1$, while the right panel illustrates the effects of varying $m_2$, $m_1$, and $\sigma_{\rm self1}/m_1$ around the reference model. BDM models shift the absorption feature to lower redshift because the buildup of Ly$\alpha$ coupling, X-ray heating, and ionization occurs later than in \LCDM. The absorption trough is also slightly shallower in BDM. These results are discussed in more detail in the main text.
    }
    \label{fig:T21}
\end{figure}

We first consider the global 21\,cm signal $\left\langle\delta T_{b}(\mathbf r,z)\right\rangle$ in Eq.~\eqref{eq:deltaTbObs}, where the average is taken over the simulation volume. 
As shown in Figure~\ref{fig:T21}, the global signal is weak  at $z\gtrsim25$ because the spin temperature
remains close to the CMB temperature.
Collisional coupling is already
inefficient in this redshift range, while the Ly$\alpha$
background has not yet become strong enough to couple $T_s$ to
$T_k$. 
As the first sources form at $z\sim20$, Ly$\alpha$ scattering
activates the Wouthuysen--Field coupling and drives the spin temperature
away from the CMB temperature toward the colder gas temperature. This
produces the absorption feature, whose minimum occurs around
$z\simeq11$--$13$. At later
times, roughly $z\lesssim12$, X-ray heating becomes efficient, raises the gas
kinetic temperature, and drives the signal back toward zero.

The difference between BDM and \LCDM\ begins to appear only
when the first radiative sources build up an appreciable Ly$\alpha$
background, around $z\sim20$.
Since BDM suppresses the formation of low-mass source hosting halos, the
growth of $\widehat J_\alpha$ and hence of $x_\alpha$ is delayed. The
absorption trough therefore appears later than in \LCDM. For the parameter choices shown in
Figure~\ref{fig:T21}, the position of the trough is shifted by
$\Delta z_{\rm trough}\equiv
z_{\rm trough}^{\Lambda{\rm CDM}}-z_{\rm trough}^{\rm BDM}
\simeq 1$--$3$, with the largest shifts occurring for the models with the
strongest small-scale suppression. In the left panel, the strongest shifts occur for
intermediate values of $r_1$. In the right panel, smaller
$m_2$ or larger $\sigma_{\rm self1}/m_1$ generally leads to a stronger delay,
while parameter choices that more closely approach the \LCDM\ matter power
spectrum yield a global signal closer to the \LCDM\ prediction.

The shift is not a simple horizontal translation of the \LCDM\ curve. The
absorption trough is also mildly shallower in BDM, with the minimum brightness
temperature higher than the \LCDM\ value by up to $\sim 10\,{\rm mK}$ for the
models shown in Figure~\ref{fig:T21}.
A lower gas kinetic
temperature $T_k$ would deepen the absorption if the spin temperature $T_s$ were already
strongly coupled to the gas temperature. In BDM, however, the Ly$\alpha$
background is weaker during the coldest stage. The Wouthuysen--Field coupling $x_\alpha$
is therefore not yet strong enough to drive
$T_s$ fully toward $T_k$, and the spin temperature
remains closer to the CMB temperature. This reduces the absorption contrast.

The delayed emergence of radiation backgrounds from the first luminous sources affects both the global absorption feature and the spatial fluctuations in the 21\,cm brightness temperature field. We define the fluctuation field by
\begin{equation}
\Delta T_{21}(\mathbf r,z)
\equiv
\delta T_{b}(\mathbf r,z)
-
\left\langle
\delta T_{b}(\mathbf r,z)
\right\rangle .
\label{eq_DeltaT21_field}
\end{equation}
The 21\,cm power spectrum $P_{21}(k,z)$, with units of
${\rm mK}^2\,{\rm Mpc}^3$, is defined through
\begin{equation}
\left\langle
\Delta T_{21}(\mathbf k,z)
\Delta T_{21}^{*}(\mathbf k',z)
\right\rangle
=
(2\pi)^3
\delta_D^{(3)}(\mathbf k-\mathbf k')
P_{21}(k,z).
\label{eq_P21_def}
\end{equation}
We also use the dimensionless power spectrum
\begin{equation}
\Delta_{21}^2(k,z)
\equiv
\frac{k^3}{2\pi^2}
P_{21}(k,z),
\label{eq_Delta21_def}
\end{equation}
which has units of ${\rm mK}^2$.

\begin{figure}[t!]
    \centering
    \includegraphics[width=0.95\linewidth]{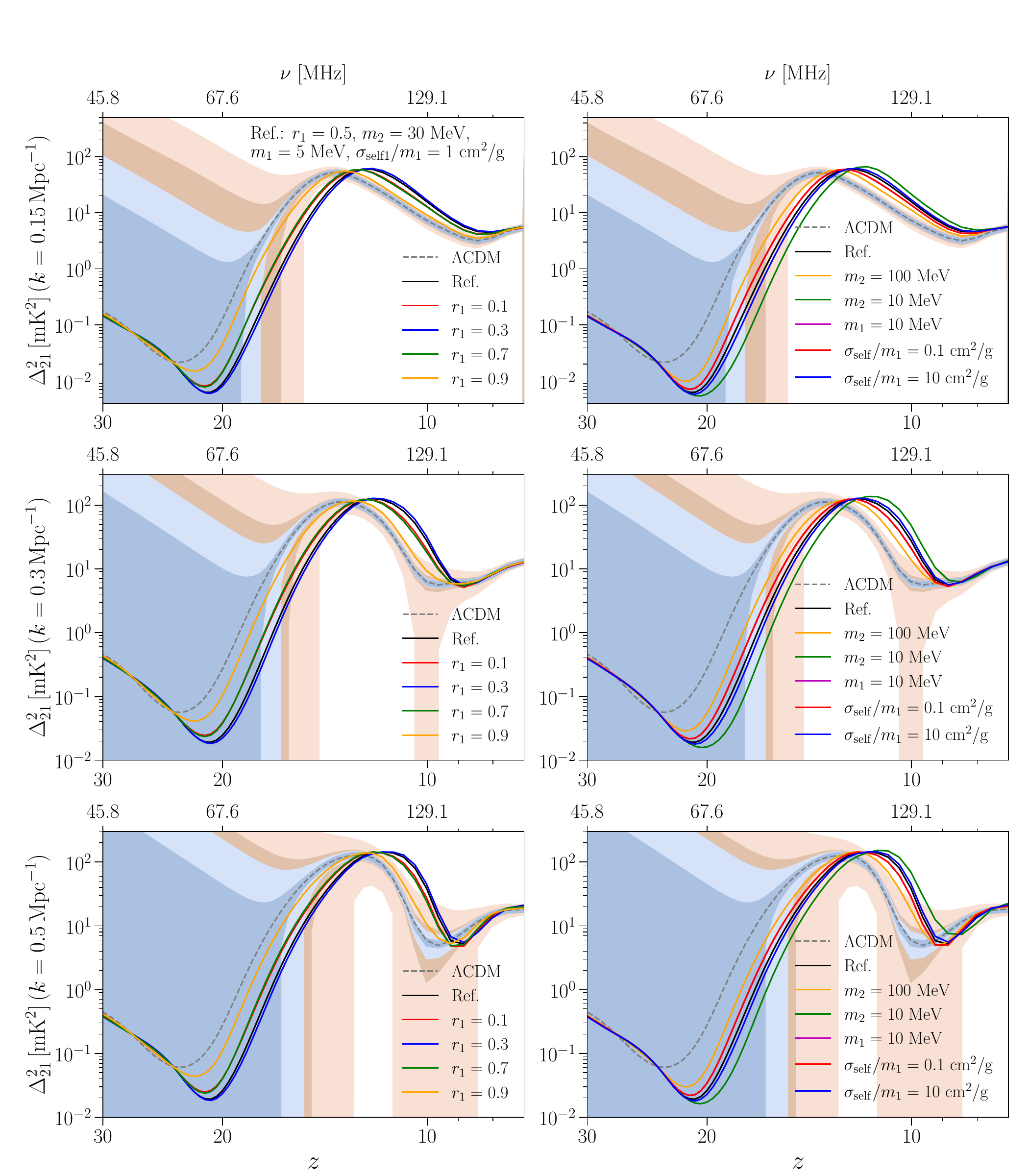}
    \vspace*{-0.1cm}
    \caption{
    Redshift evolution of the 21\,cm dimensionless power spectrum
    $\Delta_{21}^2(k,z)$ (see Eq.~\eqref{eq_Delta21_def}) at fixed wavenumbers
    $k=0.15,\,0.3,\,0.5\,{\rm Mpc}^{-1}$ from top to bottom. The left panels
    show the dependence on the light component fraction $r_1$, while the
    right panels show variations of $m_2$, $m_1$, and
    $\sigma_{\rm self1}/m_1$ around the reference model. 
    The shaded bands show the projected total uncertainties, combining the SKA thermal noise with a theoretical modeling uncertainty of $f_{\rm theo}=0.1$. Orange and blue shading correspond to the moderate and optimistic forecast scenarios summarized in Table~\ref{tab:observational_setting}, respectively. The lighter and darker shades within each color correspond to observing times of $1080\,{\rm hr}$ and $5000\,{\rm hr}$, respectively.
    BDM shifts the main features
    of the power spectrum to lower redshift, or equivalently to higher
    observed frequency, reflecting the delayed onset of Ly$\alpha$ coupling,
    X-ray heating, and reionization.}
    \label{fig:21power_fix_k}
\end{figure}

Figure~\ref{fig:21power_fix_k} shows $\Delta_{21}^2(k,z)$ for fixed
$k=0.15,\,0.3,\,0.5\,{\rm Mpc}^{-1}$. The horizontal axis is shown in terms
of the redshift, with the corresponding frequency displayed on the
upper axis. At high redshift, roughly $z\gtrsim25$, radiative sources are still rare and the 21\,cm fluctuations are small. 
As the first sources build up the background radiations, spatial
variations in Wouthuysen--Field coupling and early heating generate a rapid
rise in the power spectrum, especially over $13\lesssim z\lesssim20$. The
power spectrum then reaches a broad peak around $z\simeq13$--$15$.
In this redshift range, the Ly$\alpha$ and the X-ray backgrounds are still highly inhomogeneous. Regions near the first radiative sources experience stronger Wouthuysen--Field
coupling and earlier X-ray heating, while more distant regions remain weakly coupled and colder. This contrast enhances the spatial fluctuations in the 21\,cm brightness temperature.
At lower redshift, roughly $ z\lesssim12$, X-ray heating
becomes more widespread and partially erases temperature fluctuations, so the power decreases. Toward the end of the redshift range shown, $z\lesssim8$, the growth of ionized regions introduces an additional contribution to the 21\,cm power spectrum through spatial fluctuations in the neutral hydrogen fraction during patchy reionization.

BDM cosmology changes this pattern mainly by shifting the timing of these features.
At high redshift, around $z\simeq20$--$27$, the BDM power spectrum
can be smaller than the \LCDM\ prediction. In this regime, \LCDM\ has already
started to develop spatial fluctuations in Ly$\alpha$ coupling, while the BDM
source fields are still less evolved. The Wouthuysen--Field coupling and
early X-ray heating fluctuations are therefore weaker, leading to a deeper
high-redshift dip in $\Delta_{21}^2(k,z)$.

At lower redshift, the delayed BDM source population enters the coupling and
heating stages. Consequently, the subsequent rise, broad peak, and dip of
$\Delta_{21}^2(k,z)$ occur at lower redshift than in \LCDM. For the
parameter choices shown in Figure~\ref{fig:21power_fix_k}, the displacement of
these features is typically of order $\Delta z\sim1$--$3$, with the largest
shifts occurring for the models with the strongest small-scale suppression.

The broad peak around $z\simeq12$--$15$ can also be slightly enhanced in BDM. This can arise because star formation is concentrated in rare, more massive halos, producing larger spatial variations in the radiation backgrounds. In addition, the later onset of X-ray heating increases the temperature contrast between heated regions near the sources and the colder surrounding IGM.
At later redshifts, $z\lesssim8$--$9$, the BDM curves approach the
\LCDM\ prediction. This is mainly because reionization has already ionized a
large fraction of the volume, reducing the amount of neutral hydrogen that
can contribute to the 21\,cm signal.

To assess whether the BDM deviations from the \LCDM\ prediction are
observable, we assign an effective uncertainty to each $(k,z)$ bin of the
21\,cm dimensionless power spectrum,
\begin{equation}
\sigma_{\rm obs}^2(k,z)
=
\sigma_{\rm SKA}^2(k,z)
+
\sigma_{\rm theo}^2(k,z),
\qquad
\sigma_{\rm theo}(k,z)
=
f_{\rm theo}\,
\Delta^2_{21,\Lambda{\rm CDM}}(k,z),
\label{eq_noise_obs}
\end{equation}
where $\sigma_{\rm SKA}(k,z)$ is the SKA1-Low uncertainty on $\Delta_{21}^2(k,z)$ in each $(k,z)$ bin. 
It is computed with \Sense~\cite{Murray:2024} using the AA4 layout~\cite{SKAO:2023}, and the details of the baseline mapping, thermal-noise model, observing setup, and system-temperature choices are summarized in Appendix~\ref{app_ska_sensitivity}.
The theoretical uncertainty $\sigma_{\rm theo}(k,z)$ is included to account
for residual modeling uncertainties in semi-numerical predictions of the
21\,cm power spectrum~\cite{Zahn:2011,Schneider:2023ciq}. 

In Figure~\ref{fig:21power_fix_k}, the shaded regions show the
projected total uncertainty bands for the forecast configurations
summarized in Table~\ref{tab:observational_setting} of
Appendix~\ref{app_ska_sensitivity}.
Orange and blue shading correspond to the moderate and optimistic
forecast scenarios, respectively. Within each color, the light
and dark shaded regions indicate observing times of $1080\,{\rm hr}$ and
$5000\,{\rm hr}$, respectively. 
The size of the uncertainty band depends on the wavenumber. The
large-scale mode $k=0.15\,{\rm Mpc}^{-1}$ has a smaller uncertainty band than
the smaller-scale mode $k=0.5\,{\rm Mpc}^{-1}$. 
The theoretical modeling uncertainty is fixed
to $f_{\rm theo}=0.1$. 
Roughly at $z\gtrsim18$, these uncertainty bands become
larger than the difference between the BDM and \LCDM\ power
spectra. This behavior is mainly driven by the SKA thermal noise, which
increases rapidly at the low observing frequencies corresponding to high
redshifts due to diffuse Galactic synchrotron emission.

%%%%%%%%%%%%%%%%%%%%%%%%%%%%%%%%%%%%%%%%%%%%%%%%%%%
\subsection{Bayesian analysis}
\label{sec:bayesian}
%%%%%%%%%%%%%%%%%%%%%%%%%%%%%%%%%%%%%%%%%%%%%%%%%%%

We perform a Bayesian analysis to determine which regions of the BDM parameter space can be distinguished from \LCDM\ in the presence of astrophysical uncertainties. We use the \LCDM\ 21\,cm power spectrum, calculated with the astrophysical parameters fixed to the fiducial values in Eq.~\eqref{eq:astro_params_schematic}, as mock data and fit it with BDM predictions by jointly varying the BDM and selected astrophysical parameters.\footnote{In calculating both the \LCDM\ mock data and the BDM predictions, the standard cosmological parameters are fixed to the reference \LCDM\ values given in Eq.~\eqref{eq:CosmoParameters}.}
 
For computational tractability, we vary two BDM parameters in Eq.~\eqref{eq:BDM_primary_parameters} and three
astrophysical parameters in Eq.~\eqref{eq:astro_params_varied}, while keeping the remaining parameters fixed to their reference values. The parameter vector is
\begin{equation}
\theta
=
\left\{
r_1,\,
\sigma_{\rm self1}/m_1,\,
\log_{10}\left(\frac{M_{\rm turn}}{M_{\odot}}\right),\,
\log_{10}\left(\frac{L_X^{\rm II}}{{\rm SFR}}\right),\,
\log_{10}N_{\gamma}
\right\}.
\label{eq:Scan_param}
\end{equation}
Here $r_1$ is the fractional abundance of the
light dark-matter component, and $\sigma_{\rm self1}/m_1$ is the
self-interaction cross section of $\chi_1$ per unit mass. Among the
astrophysical parameters, $M_{\rm turn}$ sets the turnover mass below which
star formation in halos is suppressed, $L_X^{\rm II}/{\rm SFR}$ controls the X-ray luminosity
efficiency, and $N_{\gamma}$ determines the number of ionizing photons
produced per baryon. We impose uniform priors on the variables listed in
Table~\ref{tab:priors}.

\begin{table}[t]
\centering
\begin{tabular}{ccc}
\hline
Parameter & Prior range & Description \\
\hline
$r_1$
& $(0.05,0.95)$
& Light-component fraction \\

$\sigma_{\rm self1}/m_1$
& $(10^{-1},10^1)\,{\rm cm^2\,g^{-1}}$
& Self-interaction cross section per unit mass \\

$\log_{10}(M_{\rm turn}/M_\odot)$
& $(8,9)$
& Turnover mass for star-forming halos \\

$\log_{10}(L_X^{\rm II}/{\rm SFR})$
& $(40,41)$
& X-ray luminosity efficiency \\

$\log_{10}(N_{\gamma})$
& $(3,4)$
& Ionizing photons per baryon \\
\hline
\end{tabular}
\caption{
Prior ranges adopted in the Bayesian analysis. The scan varies two BDM
parameters and marginalizes over three astrophysical nuisance parameters that
control star formation, X-ray heating, and ionizing photon production.
}
\label{tab:priors}
\end{table}

Since direct calculations of the 21\,cm power spectrum at every
sampled parameter point are computationally expensive, we employ a
neural-network emulator~\cite{Lopez-Honorez:2020lno} to predict
$\Delta^2_{21}(k,z)$ for each parameter set in
Eq.~\eqref{eq:Scan_param}. Details of the network architecture,
training set, and validation procedure are provided in
Appendix~\ref{app:machine_learning}.
To account for the finite accuracy of the emulator, we assign a
prediction uncertainty of $15\%$, normalized to the fiducial
\LCDM\ power spectrum,
\begin{equation}
\sigma_{\rm emu}(k,z)
=
0.15\,\Delta^2_{21,\Lambda{\rm CDM}}(k,z).
\end{equation}
We also include a theoretical modeling uncertainty of $10\%$,
\begin{equation}
\sigma_{\rm theo}(k,z)
=
0.1\,\Delta^2_{21,\Lambda{\rm CDM}}(k,z).
\end{equation}
Adding the emulator and theoretical uncertainties in quadrature with the SKA instrumental uncertainty $\sigma_{\rm SKA}(k,z)$, we define the total uncertainty as
\begin{equation}
\sigma_{\rm tot}^2(k,z)
=
\sigma_{\rm SKA}^2(k,z)
+
\sigma_{\rm theo}^2(k,z)
+
\sigma_{\rm emu}^2(k,z).
\label{eq_noise_tot}
\end{equation}

The log-likelihood is then
\begin{equation}
\ln \mathcal{L}(\theta)
=
-\frac{1}{2}
\sum_{k,z}
\left[
\frac{
\left[
\Delta^2_{21,{\rm emu}}(k,z;\theta)
-
\Delta^2_{21,\Lambda{\rm CDM}}(k,z)
\right]^2
}
{\sigma_{\rm tot}^2(k,z)}
+
\ln\!\left(2\pi\sigma_{\rm tot}^2(k,z)\right)
\right].
\label{eq_likelihood}
\end{equation}
Here $\Delta^2_{21,{\rm emu}}(k,z;\theta)$ is the emulator prediction for the
BDM model, while $\Delta^2_{21,\Lambda{\rm CDM}}(k,z)$ is the \LCDM\ prediction adopted as the mock observation.
We use the 8 $k$-bins over $0.2\leq k\leq1.0\,{\rm Mpc}^{-1}$ and 9 redshift bins over $9\leq z\leq17$.
We sample the posterior using the \texttt{MultiNest} nested-sampling algorithm~\cite{Feroz:2008xx}.

\begin{figure}[t]
    \centering
    \includegraphics[width=1.0\linewidth]{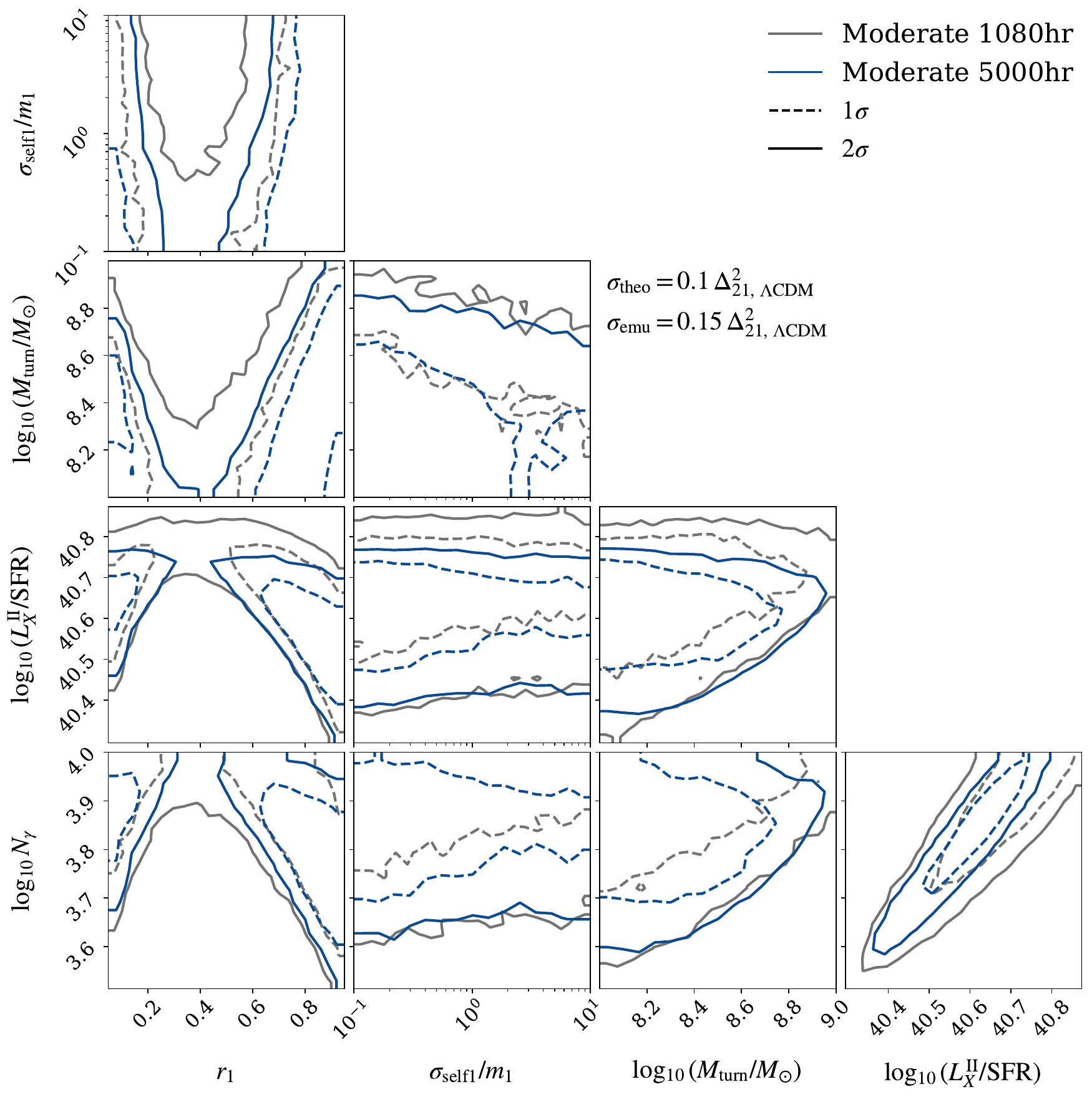}
    \caption{
    Two-dimensional marginalized posterior distributions of the BDM
    parameters $r_1$ and $\sigma_{\rm self1}/m_1$ and the
    astrophysical parameters $\log_{10}(M_{\rm turn}/M_\odot)$,
    $\log_{10}(L_X^{\rm II}/{\rm SFR})$, and $\log_{10}N_\gamma$ for
    the moderate forecast scenario. The fiducial \LCDM\ 21\,cm power
    spectrum is treated as the mock data. Gray and blue distributions
    correspond to observing times of $1080\,{\rm hr}$ and
    $5000\,{\rm hr}$, respectively. Dashed and solid contours enclose
    the $68\%$ and $95\%$ credible regions, respectively, after
    marginalizing over the remaining parameters. The total uncertainty
    entering the likelihood is defined in Eq.~\eqref{eq_noise_tot} and
    includes the SKA instrumental, theoretical modeling, and emulator
    uncertainties.
    }
    \label{fig:corner_plot_moderate}
\end{figure}

\begin{table}[t]
\centering
\small
\renewcommand{\arraystretch}{1.35}
\setlength{\tabcolsep}{5pt}

\begin{tabular}{lccccc}
\toprule
\multicolumn{1}{c}{Parameter}
& \multicolumn{1}{c}{\LCDM}
& \multicolumn{2}{c}{Moderate}
& \multicolumn{2}{c}{Optimistic} \\
\cmidrule(lr){2-2}
\cmidrule(lr){3-4}
\cmidrule(lr){5-6}
&
Fiducial
&
$1080\,{\rm hr}$
&
$5000\,{\rm hr}$
&
$1080\,{\rm hr}$
&
$5000\,{\rm hr}$ \\
\midrule

$\log_{10}(M_{\rm turn}/M_\odot)$
& $8.700$
& $8.335^{+0.335}_{-0.234}$
& $8.319^{+0.295}_{-0.224}$
& $8.339^{+0.272}_{-0.232}$
& $8.365^{+0.212}_{-0.237}$ \\

$\log_{10}(L_X^{\rm II}/{\rm SFR})$
& $40.500$
& $40.663^{+0.098}_{-0.141}$
& $40.608^{+0.081}_{-0.110}$
& $40.568^{+0.077}_{-0.099}$
& $40.497^{+0.061}_{-0.055}$ \\

$\log_{10}(N_\gamma)$
& $3.699$
& $3.874^{+0.088}_{-0.143}$
& $3.839^{+0.087}_{-0.116}$
& $3.805^{+0.084}_{-0.106}$
& $3.740^{+0.071}_{-0.062}$ \\

\bottomrule
\end{tabular}

\renewcommand{\arraystretch}{1.0}
\caption{
Fiducial \LCDM\ astrophysical parameter values used to generate the
mock data and marginalized posterior medians with their corresponding
$68\%$ credible intervals under the four forecast configurations. One-dimensional marginalized posterior summaries for the BDM
parameters are omitted because the $r_1$ posterior is bimodal and the
$\sigma_{\rm self1}/m_1$ posterior is truncated at the lower prior
boundary of $0.1\,{\rm cm^2\,g^{-1}}$.
}
\label{tab:posterior_summary}
\end{table}

Figure~\ref{fig:corner_plot_moderate} shows the two-dimensional
marginalized posterior distributions for the moderate forecast
scenario, assuming observing times of $1080\,{\rm hr}$ (gray) and
$5000\,{\rm hr}$ (blue).\footnote{Because the marginalized posterior
of $r_1$ is bimodal and the posterior of
$\sigma_{\rm self1}/m_1$ is truncated by the lower prior boundary, $\sigma_{\rm self1}/m_1=0.1\,{\rm cm^2\,g^{-1}}$, we
do not quote one-dimensional marginalized constraints on the
individual BDM parameters.} The posterior density increases toward the interior of the contours. Regions of high posterior density correspond to combinations of BDM and astrophysical parameters whose predicted 21\,cm power spectra remain statistically consistent with the fiducial \LCDM\ mock data.

The posterior in the $(r_1,\sigma_{\rm self1}/m_1)$ plane extends toward both ends of the $r_1$, which reflects the reduction of the BDM effect when either DM component becomes effectively subdominant. 
The larger posterior support on the high-$r_1$ branch can be
understood from the matter power spectrum in Figure~\ref{fig:bdm_pk}. Around
$k\sim\mathcal O(10)\,h\,{\rm Mpc}^{-1}$, which contributes strongly
to the variance on source-hosting halo mass scales, the matter power
suppression for $r_1=0.9$ is milder than that for $r_1=0.1$. The
high-$r_1$ models consequently produce halo abundances and 21\,cm
power spectra that more closely resemble those in \LCDM. Intermediate
$r_1$ produces a larger delay in the buildup of radiative sources and
a more distinguishable departure from the \LCDM\ mock signal.

The dependence on $\sigma_{\rm self1}/m_1$ is strongly correlated with
$r_1$. Models with large self-interaction strengths are most strongly
disfavored at intermediate $r_1$, where the BDM-induced suppression is
intrinsically largest. On the other hand, a broader range of self-interaction strengths remains compatible with the mock data near the two ends of the $r_1$
prior. 

The astrophysical parameters shift in the directions expected to
partially compensate for the delayed formation of radiative sources
in BDM.\footnote{Their displacement from the fiducial values in
Eq.~\eqref{eq:astro_params_schematic} reflects the extent to which
astrophysical variations compensate for the BDM-induced changes.}
Lower values of $M_{\rm turn}$ permit star formation in lower-mass
halos and thereby increase the abundance of active sources at earlier
times. Larger values of $L_X^{\rm II}/{\rm SFR}$ increase the X-ray
luminosity per unit star-formation rate and promote earlier heating of
the intergalactic medium. Larger values of $N_\gamma$ increase the
number of ionizing photons produced per stellar baryon and accelerate
the growth of ionized regions. Within the restricted set of
astrophysical parameters varied here, the posterior shifts in
Table~\ref{tab:posterior_summary} indicate that the dominant
compensation occurs through a reduction in $M_{\rm turn}$. The
posteriors of $L_X^{\rm II}/{\rm SFR}$ and $N_\gamma$ also generally
shift toward values above their fiducial values, but their
displacements are smaller, suggesting that the BDM-induced delay is
more readily compensated by adjusting $M_{\rm turn}$ within the
adopted source model.

\begin{figure}[t]
    \centering
    \includegraphics[width=1.0\linewidth]{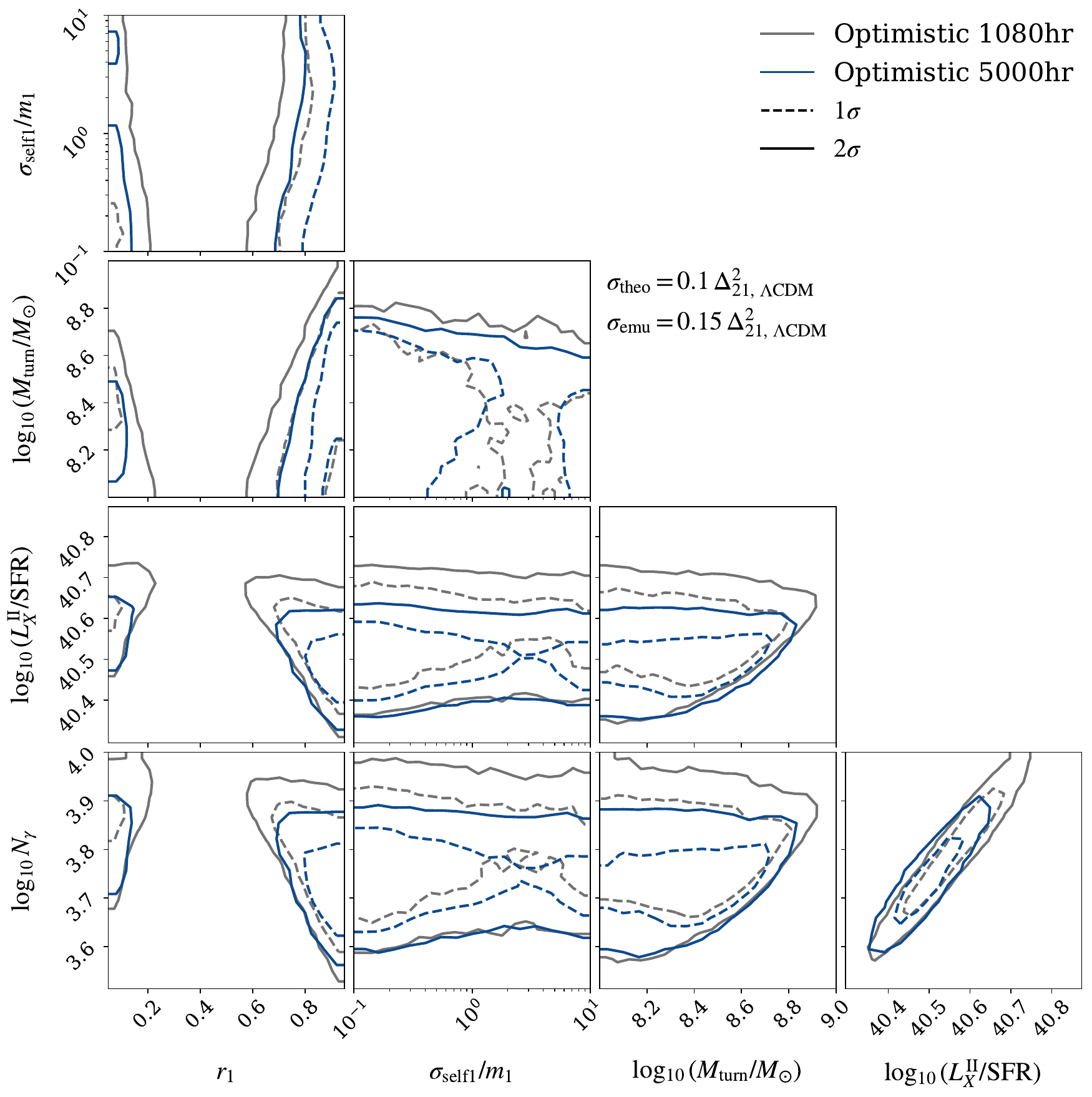}
    \caption{
    Two-dimensional marginalized posterior distributions for the
    optimistic forecast scenario. All plotting conventions are the
    same as in Figure~\ref{fig:corner_plot_moderate}.
    }
    \label{fig:corner_plot_optimistic}
\end{figure}

Figure~\ref{fig:corner_plot_optimistic} shows the corresponding
two-dimensional posterior distributions for the optimistic forecast
scenario. Relative to the moderate scenario, the improved
instrumental sensitivity reduces the allowed parameter regions. This
improvement is most clearly visible in the
$(r_1,\sigma_{\rm self1}/m_1)$ plane. For an observing time of
$5000\,{\rm hr}$, the posterior becomes strongly concentrated toward
the high-$r_1$ branch, while the intermediate-$r_1$ region is more
strongly disfavored.
For $\sigma_{\rm self1}/m_1\simeq0.1\,{\rm cm^2\,g^{-1}}$, models in the range
\begin{equation}
0.2\lesssim r_1\lesssim0.7
\label{eq:r1_excluded_optimistic}
\end{equation}
lie outside the $95\%$ credible region of the two-dimensional
posterior. These models are therefore disfavored by the fiducial
\LCDM\ mock data at the $95\%$ credible level under the optimistic
forecast and within the adopted prior ranges.

The astrophysical posteriors exhibit the same compensating behavior
identified in the moderate forecast. Increasing the observing time
from $1080\,{\rm hr}$ to $5000\,{\rm hr}$ narrows the allowed regions,
particularly for $L_X^{\rm II}/{\rm SFR}$ and $N_\gamma$. Their
posterior values move closer to the fiducial values as the BDM
posterior becomes concentrated toward the high-$r_1$ branch, where the
modification of the 21\,cm power spectrum is weaker.

\begin{figure}
    \centering
    \includegraphics[width=1.0\linewidth]{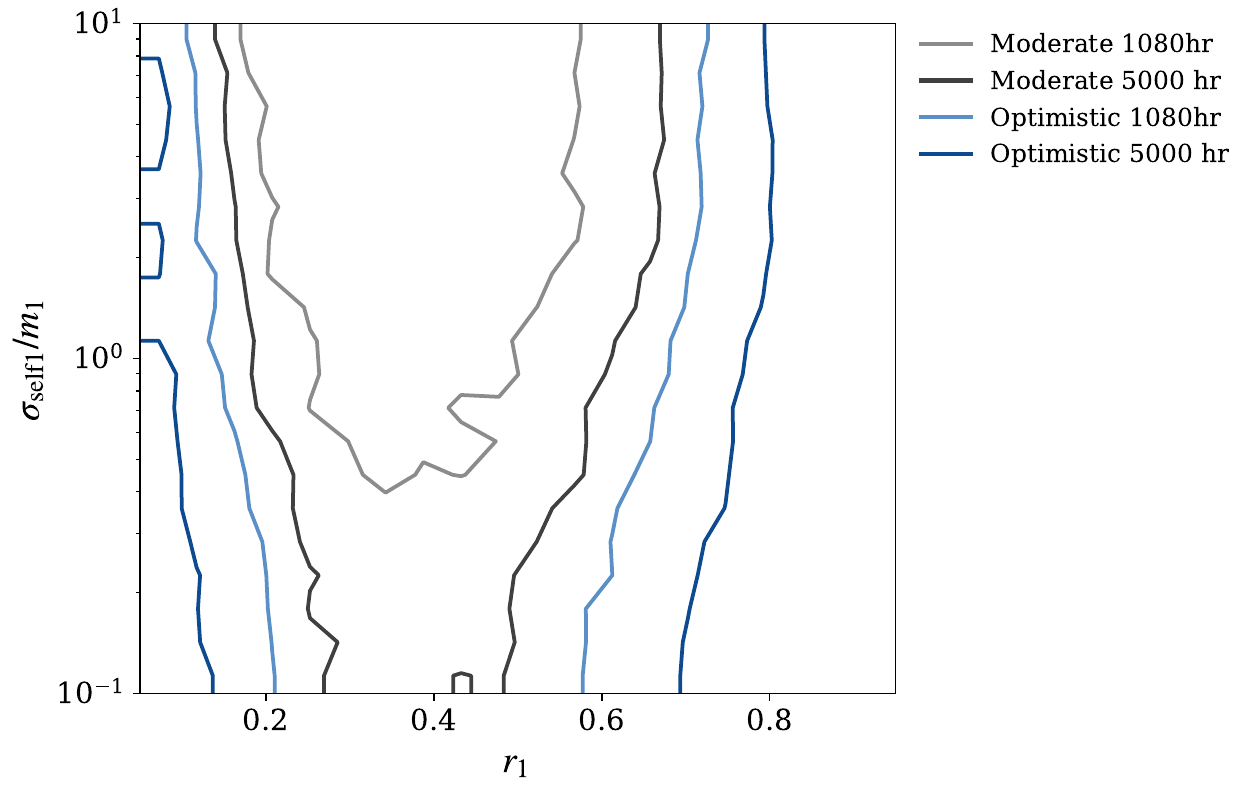}
    \caption{
    Projected $95\%$ exclusion boundaries in the
    $(r_1,\sigma_{\rm self1}/m_1)$ plane, obtained from the marginalized
    posterior distributions in
    Figures~\ref{fig:corner_plot_moderate} and
    \ref{fig:corner_plot_optimistic}. The BDM masses are fixed to
    $m_1=5\,{\rm MeV}$ and $m_2=30\,{\rm MeV}$, and the dark-photon mass
    is fixed to $m_{\gamma'}=30\,{\rm MeV}$. Gray and black curves
    correspond to the moderate forecast scenario with observing times of
    $1080\,{\rm hr}$ and $5000\,{\rm hr}$, respectively. Light and dark
    blue curves show the corresponding results for the optimistic
    forecast scenario. For each configuration, the intermediate-$r_1$
    region between the corresponding left and right boundaries is
    projected to be distinguishable from the fiducial \LCDM\ mock signal
    at the $95\%$ level.
    }
    \label{fig:r1_sigma_constraints}
\end{figure}

Figure~\ref{fig:r1_sigma_constraints} collects the two-dimensional marginalized posterior contours in the $(r_1,\sigma_{\rm self1}/m_1)$ plane from
Figures~\ref{fig:corner_plot_moderate} and
\ref{fig:corner_plot_optimistic} and presents the four forecast configurations together.
We compare the projected 21\,cm reach with the sub-GeV direct-detection constraints from SENSEI~\cite{SENSEI:2020dpa,SENSEI:2023zdf}. 
The reference cross section $\overline{\sigma}_e$ for the elastic scattering process
$\chi_1e^-\to\chi_1e^-$ scales as
\begin{equation}
\overline{\sigma}_e \propto \frac{g'^{\,2}\epsilon^2}{m_{\gamma'}^4}.
\label{eq:electron_scattering_scaling}
\end{equation}
For the fixed $\Omega_{\rm DM}$ and particle masses adopted here, the relic-abundance calculation determines $g'^{\,2}\epsilon^2$ as a function of $r_1$, as discussed in connection with Eq.~\eqref{eq:annihilation_cross_section_scalings}. 
Hence, $\overline{\sigma}_e$ is determined by $r_1$ independently of $\sigma_{\rm self1}$.
We assume that the local fraction of $\chi_1$ follows its cosmological abundance, $\rho_{\chi_1}^{\rm local}/\rho_{\rm DM}^{\rm local}=r_1$.\footnote{Previous $N$-body simulations have shown that self-interactions of $\chi_1$ can reduce its central halo density in the Milky Way, potentially causing its local fraction to differ from the cosmological abundance $r_1$~\cite{Kim:2024ltz}. 
However, we neglect this effect in this analysis.}
Under this assumption, the expected SENSEI event rate scales approximately as $R_{\rm SENSEI}\propto r_1\overline{\sigma}_e(r_1)$.

For the benchmark parameter space analyzed in this work, with $m_{1}$ in the 5--10 MeV range and $m_{\gamma'} = 30~\text{MeV}$, the coupling combination $g'^{2}\epsilon^{2}$ required to reproduce the observed relic abundance yields a $\chi_{1}$--electron elastic scattering cross section of order $\bar{\sigma}_{e} \sim 10^{-41}~\text{cm}^{2}$. 
This predicted value lies several orders of magnitude below the current sensitivity of SENSEI, which probes cross sections down to $\bar{\sigma}_{e} \sim 10^{-37}\text{--}10^{-36}~\text{cm}^{2}$ for dark matter masses of 1--10 MeV in the heavy-mediator regime~\cite{SENSEI:2020dpa,SENSEI:2023zdf}. 
Consequently, directly detecting the non-relativistic halo $\chi_{1}$ population in our benchmark scenario remains challenging with current low-threshold direct-detection experiments.

In addition to direct searches for non-relativistic halo DM, our scenario can be tested through the relativistic $\chi_{1}$ flux produced today by the annihilation of the heavier relic component, $\chi_{2}\bar{\chi}_{2} \to \chi_{1}\bar{\chi}_{1}$, in the Milky Way halo. 
The resulting all-sky flux of boosted $\chi_{1}$ at the Earth is approximately given by~\cite{Agashe:2014yua, Kim:2018veo}
\begin{equation}
\mathcal{F} = 1.6 \times 10^{-4}\,\text{cm}^{-2}\text{s}^{-1} \times r_{2}^{2} \times \left(\frac{\langle\sigma v\rangle_{\chi_{2}\bar{\chi}_{2}\to\chi_{1}\bar{\chi}_{1}}}{5\times 10^{-26}\,\text{cm}^{3}\text{s}^{-1}}\right) \times \left(\frac{\text{GeV}}{m_{2}}\right)^{2}, \label{eq:flux}
\end{equation}
where $r_{2} = 1 - r_{1}$. 
For a representative benchmark with $r_{1} = 0.1$, $m_{2} = 30~\text{MeV}$, and $m_{1} = 5~\text{MeV}$, integrating over the total $\chi_{1}e^{-} \to \chi_{1}e^{-}$ cross section yields at most $\sim 50$ electron-recoil events at SK over a 15-year exposure (and fewer for larger $r_{1}$). 
Because the boosted $\chi_{1}$ particles carry energies of $E_{1} \simeq m_{2} = 30~\text{MeV}$, the resulting recoil electrons are soft ($E_{e} \lesssim 16\text{ MeV}$), falling well below the 100 MeV threshold typically required in SK BDM searches to suppress atmospheric neutrino backgrounds~\cite{Agashe:2014yua, Super-Kamiokande:2017dch}. 
Lowering the analysis threshold into the $\mathcal{O}(1\text{--}10)\text{ MeV}$ regime exposes the search to severe backgrounds from solar neutrinos, cosmic-ray muon spallation, and low-energy atmospheric neutrino interactions (such as Michel electrons from invisible muon decays), rendering a signal of $\lesssim 50$ events over 15 years extremely difficult to resolve at SK. 

Similar challenges persist at next-generation large-volume neutrino detectors, including HK, JUNO, and DUNE. 
While HK~\cite{Hyper-Kamiokande:2018ofw, Hyper-Kamiokande:2016srs} benefits from a roughly eight-times larger fiducial volume than SK, it shares the same water-Cherenkov detection principle; thus, the irreducible low-energy solar, spallation, and atmospheric backgrounds scale up commensurately with the detector mass, limiting the sensitivity to such a small signal rate in the sub-20 MeV window. 
Meanwhile, JUNO~\cite{JUNO:2021vlw} and DUNE~\cite{DUNE:2020ypp, DUNE:2020fgq} offer lower energy thresholds, superior energy resolution, or enhanced event-topology reconstruction using liquid-scintillator and liquid-argon technologies, respectively. 
However, their fiducial masses (20 kton for JUNO and 40 kton for DUNE) are comparable to that of SK (22.5 kton), restricting the raw signal yield to only a few tens of events over a decade of operation. 
Combined with solar, reactor, and cosmogenic backgrounds in the MeV regime—as well as limited directional resolution in liquid scintillator for suppressing isotropic backgrounds at JUNO—probing this benchmark parameter space via boosted $\chi_{1}$ searches remains challenging across these facilities. 
This highlights the unique complementarity of high-redshift 21 cm cosmology, which can effectively test MeV-scale BDM scenarios that lie beyond the reach of both low-threshold direct-detection and large-volume neutrino experiments.

%%%%%%%%%%%%%%%%%%%%%%%%%%%%%%%%%%%%%%%%%%%%%%%%%%%
\section{Summary and Outlook}
\label{sec:summary}
%%%%%%%%%%%%%%%%%%%%%%%%%%%%%%%%%%%%%%%%%%%%%%%%%%%

The 21\,cm signal probes the formation of the first low-mass halos at high redshift, before nonlinear evolution and baryonic processes become dominant. 
In this work, we used this high-redshift observable to study BDM. 
In BDM cosmology, the annihilation of the heavier component converts part of the mass difference into the kinetic energy of the lighter component. 
Self-interactions redistribute this energy and maintain pressure support within the light component, suppressing small-scale density perturbations even when the DM masses are in the MeV range.

We propagated the BDM-modified matter and baryon perturbations through the thermal and ionization evolution of the intergalactic medium. 
The reduced abundance of low-mass halos postpones the formation of the sources that shape the 21\,cm signal. 
Consequently, the global absorption feature occurs at a lower redshift and is slightly shallower than in \LCDM. 
The 21\,cm power spectrum exhibits a corresponding shift in its redshift evolution. 
At high redshift, its amplitude is suppressed because spatial fluctuations sourced by the earliest galaxies have not yet developed as strongly as in \LCDM. 
At later times, the delayed source population can produce a more inhomogeneous radiation field and enhance the broad peak of the power spectrum.

We assessed the projected sensitivity of SKA1-Low using the 21\,cm power spectrum and allowing selected astrophysical parameters to vary. 
Models near either end of the $r_1$ range can remain close to \LCDM, while intermediate values produce the most distinguishable 21\,cm predictions. 
For the optimistic $5000\,{\rm hr}$ forecast and $\sigma_{\rm self1}/m_1\simeq0.1\,{\rm cm^2\,g^{-1}}$, models with approximately $0.2\lesssim r_1\lesssim0.7$ can be distinguished from the fiducial \LCDM\ prediction at the $95\%$ confidence level. 
Astrophysical variations can partially compensate for the delayed formation of low-mass halos. 
Within the parameter set considered here, this compensation occurs primarily through a lower $M_{\rm turn}$, which allows star formation in less massive halos.
The remaining structure of the posterior shows that these astrophysical variations do not fully remove the BDM dependence of the 21\,cm power spectrum.

We also compared the projected cosmological sensitivity with the reach of terrestrial searches for both non-relativistic halo DM and relativistic BDM. 
For our benchmark parameter space ($m_{1} \sim 5\text{--}10~\text{MeV}$ and $m_{\gamma'} = 30~\text{MeV}$), the relic-abundance requirement fixes the $\chi_{1}$--electron elastic scattering cross section to $\bar{\sigma}_{e} \sim 10^{-41}~\text{cm}^{2}$, which lies several orders of magnitude below the current sensitivity of low-threshold direct-detection experiments such as SENSEI. 
Furthermore, although present-day $\chi_{2}\bar{\chi}_{2} \to \chi_{1}\bar{\chi}_{1}$ annihilation in the Milky Way halo produces a flux of boosted $\chi_{1}$, the resulting electron recoils for $m_{2} = 30~\text{MeV}$ are soft ($E_{e} \lesssim 16~\text{MeV}$) and yield only a few tens of events over a decade of exposure at ten-kiloton-scale detectors. 
Consequently, resolving this signal is severely hindered by low-energy solar, reactor, cosmogenic spallation, and atmospheric backgrounds at current and next-generation neutrino observatories, including Super-Kamiokande, Hyper-Kamiokande, JUNO, and DUNE. 
High-redshift 21 cm observations with SKA1-Low therefore offer a unique, complementary window onto MeV-scale BDM scenarios that remain beyond the reach of terrestrial detectors.

For future work, dedicated BDM $N$-body and hydrodynamical simulations will be required to calibrate the conditional halo mass function and the unresolved gas-density distribution. 
The prescriptions adopted here were calibrated using \LCDM\ simulations.
We also plan to explore a broader astrophysical parameter space to determine whether astrophysical variations can mimic the BDM signal.
We will further investigate whether the 21\,cm signatures of BDM can be distinguished from those of other dark-matter scenarios, particularly WDM.

%%%%%%%%%%%%%%%%%%%%%%%%%%%%%%%%%%%%%%%%%%%%%%%%%%%
\acknowledgments
%%%%%%%%%%%%%%%%%%%%%%%%%%%%%%%%%%%%%%%%%%%%%%%%%%%
JHK thanks Hyunbae Park and Kyungjin Ahn for helpful discussions. 
We thank APCTP, Pohang, Korea, for its hospitality during the TRP program APCTP-2026-T02, from which this work benefited.
JHK and ML are supported in part by the National Research Foundation of Korea (NRF) Grant No. RS-2026-25484206, the BK-21 FOUR program through NRF, and the Institute of Information \& Communications Technology Planning \& Evaluation (IITP)-Information Technology Research Center (ITRC) Grant No. IITP-2025-RS-2024-00437284. 
JCP is supported by the NRF grant funded by the Ministry of Science and ICT (RS-2024-00356960) and by the Global - Learning \& Academic research institution for Master's/PhD students and Postdocs (G-LAMP) Program of the NRF grant funded by the Ministry of Education (RS-2025-25442707). 
KK is supported in part by the National Science Foundation through Grant PHY-2609759. 
%%%%%%%%%%%%%%%%%%%%%%%%%%%%%%%%%%%%%%%%%%%%%%%%%%%
\appendix
\section{Appendix}
\label{sec:appendix}
%%%%%%%%%%%%%%%%%%%%%%%%%%%%%%%%%%%%%%%%%%%%%%%%%%%

%%%%%%%%%%%%%%%%%%%%%%%%%%%%%%%%%%%%%%%%%%%%%%%%%%%
\subsection{Kinetic temperature evolution}
\label{sec_21cm_Tk}
%%%%%%%%%%%%%%%%%%%%%%%%%%%%%%%%%%%%%%%%%%%%%%%%%%%

This appendix describes the evolution of the gas kinetic temperature $T_k$, which appears in the spin-temperature relation given in Eq.~\eqref{eq_Ts_spin_bdm}. Applying the first law of thermodynamics, we express the evolution of $T_k$ with redshift as
\begin{align}
\frac{dT_k(\mathbf r,z)}{dz}
=
&
\frac{2T_k(\mathbf r,z)}{1+z}
+
\frac{2T_k(\mathbf r,z)}
{3\left[1+\delta_b^{\rm NL}(\mathbf r,z)\right]}
\frac{d\delta_b^{\rm NL}(\mathbf r,z)}{dz}
-
\frac{T_k(\mathbf r,z)}
{1+f_{\rm He}+x_e^{\rm X}(\mathbf r,z)}
\frac{d x_e^{\rm X}(\mathbf r,z)}{dz}
\nonumber\\
&
+
\frac{2}{3k_B\left[1+f_{\rm He}+x_e^{\rm X}(\mathbf r,z)\right]}
\frac{dt}{dz}
\left[
\epsilon_{\rm comp}(\mathbf r,t)
+
\epsilon_X(\mathbf r,t)
\right],
\label{eq_Tk_evolution}
\end{align}
where $dt/dz=-[(1+z)H(z)]^{-1}$,
$f_{\rm He}\equiv \bar n_{\rm He}/\bar n_{\rm H} = Y_p/[4(1-Y_p)]$, 
and $x_e^{\rm X}(\mathbf r,z)$ is the local free-electron fraction per hydrogen nucleus, which contains both residual and X-ray-induced ionization. Appendix~\ref{subsec:xe_evolution} describes how it evolves.

In Eq.~\eqref{eq_Tk_evolution}, the Hubble expansion produces adiabatic cooling through the first term, while local compression of the baryons raises the gas kinetic temperature through the second term. The third term accounts for changes in the gas heat capacity as the local free-electron fraction evolves. When ionization increases $x_e^{\rm X}(\mathbf r,z)$, the same thermal energy is distributed among more particles, causing this term to lower $T_k(\mathbf r,z)$.
The last line accounts for Compton heating by CMB photons and X-ray heating from astrophysical sources.\footnote{In the default implementation used in this work, the radiative recoil and Ly$\alpha$ heating corrections are not included. The radiative recoil heating term is usually a very small correction to the gas thermal history. The gas heating rate from Ly$\alpha$ scattering is also typically subdominant compared with X-ray heating in standard source models. This should be distinguished from Ly$\alpha$ coupling through the Wouthuysen--Field effect, which can make $x_\alpha$ large and drive $ T_\alpha\simeq T_k$.}
For a given channel $\lambda\in\{\mathrm{comp},\,X\}$, the net energy transferred to the gas per hydrogen nucleus per unit time is represented by $\epsilon_\lambda(\mathbf r,t)$, in units of $\rm eV\,s^{-1}$. We define the corresponding quantity per unit redshift as 
\begin{equation}
    \tilde\epsilon_\lambda(\mathbf r,z)\,dz
    \equiv
    -
    \epsilon_\lambda(\mathbf r,t)\,dt .
\label{eq_eps_redshift_def}
\end{equation}
The units of $\tilde\epsilon_\lambda$ are thus ${\rm eV}$ per hydrogen nucleus.

The heating rate due to Compton scattering between free electrons and CMB photons is given by
\begin{equation}
\epsilon_{\rm comp}(\mathbf r,t)
=
\frac{3}{2}k_B x_e^{\rm X}(\mathbf r,t)
\frac{
\bar T_\gamma(t)-T_k(\mathbf r,t)
}{
t_\gamma(t)
},
\qquad
t_\gamma(t)
=
\frac{3m_e c}
{8\sigma_T a_R \bar T_\gamma^4(t)} .
\label{eq_eps_comp_time}
\end{equation}
Here $t_\gamma(t)$ is the characteristic timescale, in seconds, for Compton energy exchange.
The Thomson cross section has units of ${\rm cm}^2$ and is written as $\sigma_T=(8\pi/3)r_e^2$, where $r_e=\alpha\hbar/(m_e c)$ is the classical electron radius and $\alpha$ is the fine-structure constant. The radiation constant $a_R$ is defined by the blackbody energy density $u_\gamma(t)=a_R\bar T_\gamma^4(t)$ and is given by $a_R=\pi^2 k_B^4/(15\hbar^3c^3)$. With the unit convention adopted here, $u_\gamma$ has units of ${\rm eV}\,{\rm cm}^{-3}$ and $a_R$ has units of ${\rm eV}\,{\rm cm}^{-3}\,{\rm K}^{-4}$. 
In the BDM cosmology, $\epsilon_{\rm comp}(\mathbf r,t)$ can be modified indirectly through $x_e^{\rm X}(\mathbf r,t)$ and $T_k(\mathbf r,t)$.

The relative importance of the heating terms depends strongly on the redshift.
At high redshift, roughly $z\gtrsim100$, Compton scattering is the dominant thermal coupling between the gas and the CMB. Residual free electrons transfer energy between CMB photons and the gas, keeping $T_k$ close to $\bar T_\gamma$. As the Universe expands, the residual free-electron fraction and the CMB energy density decrease, and Compton heating rapidly becomes inefficient. 
In the intermediate range, roughly $30\lesssim z\lesssim100$, the diffuse IGM therefore cools mainly through adiabatic expansion.
Once the astrophysical sources have formed at lower redshifts, the gas temperature is raised mainly by X-ray heating $\epsilon_X(\mathbf r,t)$, and this heating drives the thermal transition of the 21\,cm signal.
In BDM cosmology, the main impact on the heating history arises from the delayed and spatially redistributed formation of early sources and is reflected most directly in the X-ray heating term $\epsilon_X(\mathbf r,t)$, which we describe in Appendix~\ref{subsec_xray}.

%%%%%%%%%%%%%%%%%%%%%%%%%%%%%%%%%%%%%%%%%%%%%%%%%%%
\subsection{X-ray heating}
\label{subsec_xray}
%%%%%%%%%%%%%%%%%%%%%%%%%%%%%%%%%%%%%%%%%%%%%%%%%%%

In Eq.~\eqref{eq_Tk_evolution}, the X-ray contribution to $T_k$ is represented by $\epsilon_X(\mathbf r,t)$, the rate of thermal energy deposition per hydrogen nucleus.
Since the semi-numerical calculation is performed on redshift slices, we use its redshift-space counterpart defined in Eq.~\eqref{eq_eps_redshift_def} where $dt/dz=-[(1+z)H(z)]^{-1}$. Thus $\tilde\epsilon_X$ has units of energy per hydrogen nucleus. 

X-ray photons with long mean free paths can affect the IGM at redshift $z$ after being emitted at an earlier redshift $z'\geq z$. The X-ray background at $(\mathbf r,z)$ accumulates contributions from sources in shells at different emission redshifts. Integrating over the frequencies of these photons gives the resulting heating term,
\begin{align}
\tilde \epsilon_X(\mathbf r,z)
=&
\int_{\nu_0}^{\infty} d\nu
\sum_{i}
\left(h\nu-E_i^{\rm th}\right)
f_{\rm heat}
\left(
h\nu-E_i^{\rm th},
x_e^{\rm X}(\mathbf r,z)
\right)
x_i^X(\mathbf r,z)
\sigma_i(\nu)
\widetilde{\mathcal J}_X(\mathbf r,\nu,z).
\label{eq_eps_X_shell}
\end{align}
Here $i$ runs over the absorbing species ${\rm HI}$, ${\rm HeI}$, and ${\rm HeII}$. The quantity $E_i^{\rm th}$ is the ionization threshold energy of species $i$, and $h\nu-E_i^{\rm th}$ is the kinetic energy of the primary photoelectron produced by photoionization. The dimensionless function $f_{\rm heat}$ gives the fraction of this kinetic energy deposited as heat in the IGM, depending on the local ionized fraction $x^{\rm X}_e(\mathbf r,z)$. The quantity $\sigma_i(\nu)$ is the photoionization cross section of species $i$ for an X-ray photon with frequency $\nu$, with units of ${\rm cm}^2$. For $E_i^{\rm th}$, $f_{\rm heat}$, and $\sigma_i(\nu)$, we follow the standard implementation adopted in \FirstClass. The dimensionless quantity $x_i^X(\mathbf r,z)$ denotes the local number abundance of the absorbing species $i$ per hydrogen nucleus,
\begin{eqnarray}
x_{\rm HI}^X(\mathbf r,z) &=& 1-x_e^{\rm X}(\mathbf r,z), \\
x_{\rm HeI}^X(\mathbf r,z) &=& f_{\rm He}\left[1-x_e^{\rm X}(\mathbf r,z)\right],\\
x_{\rm HeII}^X(\mathbf r,z) &=& f_{\rm He}x_e^{\rm X}(\mathbf r,z).
\label{eq_xabs_approx}
\end{eqnarray}

The X-ray photon field $\widetilde{\mathcal J}_X$ appearing in Eq.~\eqref{eq_eps_X_shell} has units of $\rm cm^{-2}\,Hz^{-1}$ and represents the arriving X-ray photons at the target cell $(\mathbf r,z)$ from all source shells,
\begin{equation}
\widetilde{\mathcal J}_X(\mathbf r,\nu,z)
\equiv
\int_z^{\infty} dz'\;
\frac{1}{4\pi r_p^2(z,z')}
\frac{d\widetilde\phi_X(\mathbf r,\nu,z,z')}{dz'} .
\label{eq_JX_def}
\end{equation}
For the absorption redshift $z$ and the emission redshift $z'$, the proper radial separation is
\begin{equation}
r_p(z,z')
=
\frac{c}{1+z}
\int_z^{z'}
\frac{d\bar z}{H(\bar z)} .
\label{eq_rp_def}
\end{equation}

In Eq.~\eqref{eq_JX_def}, we model the redshift-space number of X-ray photons reaching $(\mathbf r,z)$ from sources within the shell $[z',z'+dz']$ as
\begin{equation}
\frac{d\widetilde\phi_X(\mathbf r,\nu,z,z')}{dz'}
=
\left[-\frac{dt}{dz}\right]_{z}
\frac{d\dot N_X(\mathbf r,z,z')}{dz'}
\alpha_X\nu_0^{-1}
\left(
\frac{\nu}{\nu_0}
\right)^{-\alpha_X-1}
\left(
\frac{1+z'}{1+z}
\right)^{-\alpha_X-1}
e^{-\tau_X(\nu,z,z')} ,
\label{eq_dphidxdz_shell}
\end{equation}
where the redshift-time Jacobian is evaluated at the absorption redshift $z$,
\begin{equation}
\left[-\frac{dt}{dz}\right]_{z}
=
\frac{1}{(1+z)H(z)} .
\end{equation}
Here $d\dot N_X/dz'$ has units of ${\rm s}^{-1}$, while $\left[-dt/dz\right]_z$ converts the arrival rate at the absorption redshift into a redshift-space photon number. Therefore $d\widetilde\phi_X/dz'$ has units of ${\rm Hz}^{-1}$.
The normalized photon number spectrum of the X-ray sources is given by the factor $\alpha_X\nu_0^{-1}(\nu/\nu_0)^{-\alpha_X-1}$, whose units are ${\rm Hz}^{-1}$.
It distributes the total X-ray photon production rate over frequency for a power law spectrum above the cutoff $\nu_0$. In our fiducial setup, we take $\alpha_X=1.0$ and $h\nu_0=500~{\rm eV}$.
Cosmological redshifting between emission at $z'$ and absorption at $z$ is taken into account by the factor that contains $(1+z')/(1+z)$. The dimensionless factor $e^{-\tau_X}$ accounts for the attenuation of X-ray photons due to absorption during their propagation through the IGM. The X-ray optical depth accumulated by a photon observed with frequency $\nu$ at redshift $z$ after emission at redshift $z'$ is
\begin{equation}
\tau_X(\nu,z,z')
=
\int_z^{z'}
d\tilde z\,
\frac{c}{(1+\tilde z)H(\tilde z)}
\bar n_{\rm H}(\tilde z)
\sigma_{\rm eff}(\tilde z,\tilde\nu),
\label{eq_tauX_meanIGM}
\end{equation}
where the photon frequency at the intermediate redshift $\tilde z$ is given by $\tilde\nu=\nu(1+\tilde z)/(1+z)$. For the effective photoionization cross section per hydrogen nucleus, we use
\begin{equation}
\sigma_{\rm eff}(\tilde z,\tilde\nu)
=
\sum_{i\in{{\rm HI},{\rm HeI},{\rm HeII}}}
\left\langle x_i^X(\mathbf r,\tilde z)\right\rangle
\sigma_i(\tilde\nu).
\label{eq_sigma_eff}
\end{equation}
The averaged fraction $\langle x_i^X(\mathbf r,z)\rangle$ inherits the BDM dependence through the local ionized fraction.

In Eq.~\eqref{eq_dphidxdz_shell}, we express the total rate of X-ray photon emission from sources within the shell as
\begin{equation}
\frac{d\dot N_X(\mathbf r,z,z')}{dz'}
=
f_*^{\rm II} \zeta_X 
\Bigg(\Omega_b\rho_{\rm crit,0}
\left[
1+\delta_{b,R_{\rm sh}}^{\rm NL}(\mathbf r,z')
\right]
\frac{dV(z,z')}{dz'} \Bigg)
\frac{df^{\rm eff}_{\rm coll,\star}(\mathbf r,z',R_{\rm sh})}{dt}.
\label{eq_dNxdz_source}
\end{equation}
Here, $f_*^{\rm II}$ is the Pop II star-formation efficiency, $\Omega_b$ is the present-day baryon density parameter, and $\rho_{\rm crit,0}=3H_0^2/(8\pi G)$ is the present-day critical density.
The effective collapsed fraction $f_{\rm coll,\star}^{\rm eff}$ is defined through Eqs.~\eqref{eq_fcoll_eff_definition} and \eqref{eq_fcoll_star}.
Its time derivative describes the rate at which matter collapses into star-forming halos and thereby determines the radiation emission rate. In the numerical implementation, the time derivative is approximated as $df_{\rm coll,\star}^{\rm eff}(\mathbf r,z',R_{\rm sh})/dt \simeq f_{\rm coll,\star}^{\rm eff}(\mathbf r,z',R_{\rm sh})/[t_\ast H^{-1}(z')]$ where $t_\ast$ is the star-formation timescale as a fraction of the Hubble time defined in Eq.~\eqref{eq:astro_params_schematic}.
Its suppression in BDM cosmology is illustrated in Figure~\ref{fig:eff_f_coll_ratio}. 
This is one of
the main channels through which BDM cosmology strongly affects the
radiation backgrounds.
The factor $\zeta_X$ is the number of X-ray photons emitted per unit stellar mass, with units of $M_{\odot}^{-1}$, defined as
\begin{equation}
\zeta_X
\equiv
\frac{L_X^{\rm II}/{\rm SFR}}{h\nu_0},
\label{eq:zeta_X}
\end{equation}
where $h\nu_0=500~{\rm eV}$ is the characteristic energy of the emitted X-ray photons.
We write $\delta_{b,R_{\rm sh}}^{\rm NL}$ for the smoothed nonlinear baryon overdensity using Eq.~\eqref{eq_delta_smoothed_top_hat} on the shell scale
\begin{equation}
R_{\rm sh}(z,z')
=
\int_z^{z'}
c\frac{d\bar z}{H(\bar z)} .
\label{eq_Rsh_def}
\end{equation}
The corresponding comoving shell volume is
\begin{equation}
\frac{dV(z,z')}{dz'}
=
4\pi R_{\rm sh}^2(z,z')
\frac{c}{H(z')} .
\label{eq_dVdz_shell}
\end{equation}
The baryonic mass in the source shell, in solar-mass units, is therefore given by the combination $\Omega_b\rho_{\rm crit,0}[1+\delta_{b,R_{\rm sh}}^{\rm NL}]\,dV/dz'$.

\begin{figure}[t]
\centering
\includegraphics[width=1\linewidth]{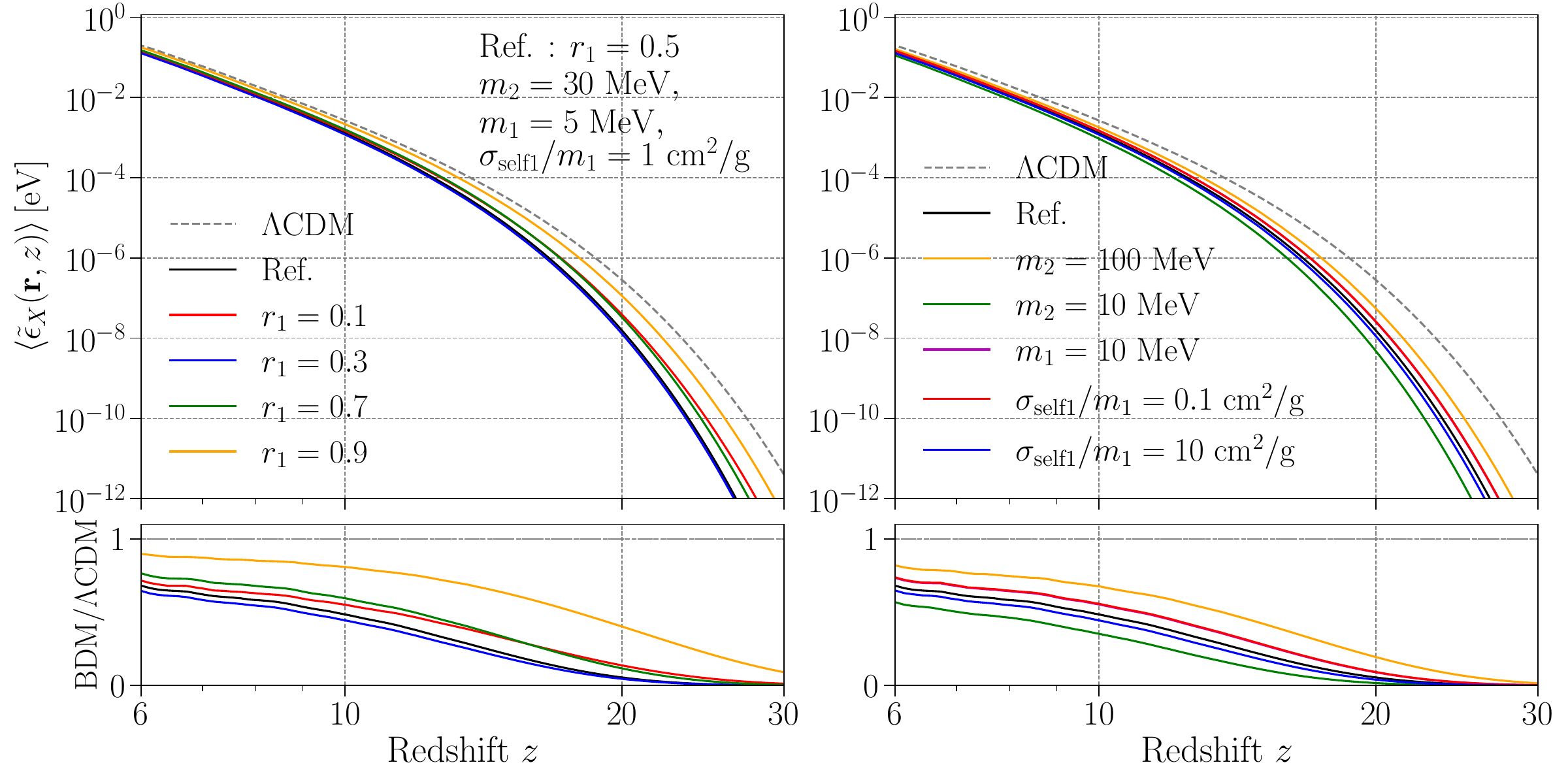}
\caption{Redshift evolution of the volume averaged X-ray heating rate $\langle \tilde\epsilon_X(\mathbf{r},z) \rangle$ (see Eq.~\eqref{eq_eps_X_shell}) for $\Lambda$CDM and representative BDM parameter choices. The left panel shows the dependence on the light component fraction $r_1$, while the right panel shows one-parameter variations of $m_2$, $m_1$, and $\sigma_{\rm self1}/m_1$ around the reference model. Compared with $\Lambda$CDM, BDM models yield a suppressed X-ray heating rate at fixed redshift, reflecting the delayed formation of low-mass source-hosting halos and the corresponding reduction of the X-ray heating background. This suppression becomes more pronounced at higher redshifts, where the X-ray emissivity is most sensitive to the abundance of the earliest star-forming halos.}
\label{fig:epsilon_X_bdm}
\end{figure}

Figure \ref{fig:epsilon_X_bdm} shows the redshift evolution of the volume-averaged X-ray heating rate $\langle\tilde \epsilon_X(\mathbf r,z)\rangle$ for representative BDM parameter choices.
In BDM cosmology, the X-ray heating field is modified through several ingredients in Eq.~\eqref{eq_eps_X_shell}. The local fractions $x_i^{\rm X}(\mathbf r,z)$ affect the absorption and energy deposition at the target cell, while the smoothed overdensity $\delta_{b,R_{\rm sh}}^{\rm NL}(\mathbf r,z')$ changes the baryonic mass assigned to each source shell. However, the dominant effect comes from the time variation of the collapsed fraction, $df_{\rm coll,\star}^{\rm eff}/dt$. Since BDM suppresses small-scale matter fluctuations, it reduces the abundance and growth rate of source hosting halos, lowering $df_{\rm coll,\star}^{\rm eff}/dt$ at fixed redshift. The resulting X-ray photon field builds up more slowly than in \LCDM, which weakens X-ray heating and keeps the gas kinetic temperature lower.

%%%%%%%%%%%%%%%%%%%%%%%%%%%%%%%%%%%%%%%%%%%%%%%%%%%
\subsection{Evolution of the X-ray Free-Electron Fraction}
\label{subsec:xe_evolution}
%%%%%%%%%%%%%%%%%%%%%%%%%%%%%%%%%%%%%%%%%%%%%%%%%%%

This appendix describes how $x_e^{\rm X}(\mathbf r,z)$, the X-ray contribution to the free-electron fraction, evolves.
It includes the residual ionization initialized at recombination and the subsequent ionization and recombination driven by X-rays. The UV ionized component is treated separately through $x_{\rm HII}^{\rm UV}$ in Eq.~\eqref{eq_xHII_total}.

We take a spatially homogeneous initial condition, which is given by the residual free-electron fraction from \CLASS\ introduced in Eq.~\eqref{eq_xe_initial},
\begin{equation}
x_e^{\rm X}(\mathbf r,z_{\rm rec})
=
\bar x_e(z_{\rm rec}) =0.1 \;.
\label{eq_xeX_initial}
\end{equation}
Starting from this value, the local X-ray free-electron fraction is evolved on redshift slices according to
\begin{equation}
\frac{d x_e^{\rm X}(\mathbf r,z)}{dz}
= 
-
\widetilde{\Lambda}_{\rm ion,X}(\mathbf r,z)
+
\widetilde{\Lambda}_{\rm rec,X}(\mathbf r,z) .
\label{eq_xeX_evolution_z}
\end{equation}
To make the physical dependence on recombination explicit, we write the
case-B recombination contribution as
\begin{equation}
\widetilde{\Lambda}_{\rm rec,X}(\mathbf r,z)
\propto
\frac{1}{(1+z)H(z)}
\alpha_{B}\!\left[T_k(\mathbf r,z)\right]\,
n_{\rm H}(\mathbf r,z)\,
\left[x_e^{\rm X}(\mathbf r,z)\right]^2 .
\label{eq_xe_local_rec}
\end{equation}
Here $\alpha_{B}$ is the case-B recombination coefficient, with units of
${\rm cm}^3\,{\rm s}^{-1}$, 
and $n_{\rm H}(\mathbf r,z)=\bar n_{\rm H}(z)[1+\delta_b^{\rm NL}(\mathbf r,z)]$ is the
local total hydrogen nuclei density, with units of ${\rm cm}^{-3}$.
The redshift-time Jacobian $-dt/dz=[(1+z)H(z)]^{-1}$ converts the time-domain recombination rate into a
redshift-space contribution, so $\widetilde{\Lambda}_{\rm rec,X}$ is
dimensionless.
Eq.~\eqref{eq_xe_local_rec} shows that, in the BDM scenario, the recombination rate can be modified through changes in $T_k(\mathbf r,z)$, $\delta_b^{\rm NL}(\mathbf r,z)$, and $x_e^{\rm X}(\mathbf r,z)$. In the numerical implementation, we evaluate the recombination rate using \HyRec\ instead of the case-B approximation to account for the recombination processes more accurately. Although the recombination rate is computed with \HyRec, its BDM dependence still enters indirectly through the same quantities.

The redshift-space X-ray ionization term is obtained from
\begin{align}
\widetilde{\Lambda}_{\rm ion,X}(\mathbf r,z)
=&
\int_{\nu_0}^{\infty} d\nu
\sum_{i\in\{{\rm HI},{\rm HeI},{\rm HeII}\}}
x_i^X(\mathbf r,z)\,
\sigma_i(\nu)\,
F_i(\nu,\mathbf r,z)\,
\widetilde{\mathcal J}_X(\mathbf r,\nu,z).
\label{eq_Lambda_ion_X_local}
\end{align}
Here $x_i^X(\mathbf r,z)$ are the local absorber abundances per hydrogen nucleus introduced in Appendix~\ref{subsec_xray}.
The total ionization yield per absorbed X-ray photon is given by the dimensionless function $F_i$, which counts the primary ionization of the absorber together with the secondary ionizations generated by the energetic photoelectron,
\begin{equation}
F_i(\nu,\mathbf r,z)
=
1+
\sum_{j\in\{{\rm HI},{\rm HeI},{\rm HeII}\}}
f_{{\rm ion},j}
\left(
h\nu-E_i^{\rm th},
x_e^{\rm X}(\mathbf r,z)
\right)
\frac{h\nu-E_i^{\rm th}}{E_j^{\rm th}} .
\label{eq_Fi_definition}
\end{equation}
The first term, equal to unity, corresponds to the primary ionization of species $i$. The second term adds the secondary ionizations of species $j$ that are caused by the photoelectron with kinetic energy $h\nu-E_i^{\rm th}$. The dimensionless function $f_{{\rm ion},j}(E,x_e)$ gives the fraction of the primary photoelectron energy that is deposited into secondary ionizations of species $j$. It is evaluated at the local free-electron fraction. Both $E_i^{\rm th}$ and $f_{{\rm ion},j}$ are taken from the standard implementation of \FirstClass.

Eq.~\eqref{eq_Lambda_ion_X_local} is the ionization analogue of the X-ray heating term in Eq.~\eqref{eq_eps_X_shell}. Both expressions contain the same redshift-space X-ray photon field $\widetilde{\mathcal J}_X(\mathbf r,\nu,z)$, and only the deposition kernel is different. The weight on the photon field is the deposited thermal energy $(h\nu-E_i^{\rm th})f_{\rm heat}$ in the heating equation and the ionization yield $F_i$ in the ionization equation. The neutral IGM is thus heated and ionized at the same time by the same X-ray background.

In BDM cosmology, the evolution of $x_e^{\rm X}$ is modified primarily
through the X-ray source field $\widetilde{\mathcal J}_X(\mathbf r,\nu,z)$. The suppression of small-scale matter
fluctuations reduces the collapsed fraction and its time variation, thereby
lowering the X-ray emissivity. This weakens
$\widetilde{\Lambda}_{\rm ion,X}$ at fixed redshift. Additional BDM
dependence enters through the local absorber fractions $x_i^X(\mathbf r,z)$,
the free-electron fraction that controls $f_{{\rm ion},j}$, and the local
hydrogen density entering the recombination term.

%%%%%%%%%%%%%%%%%%%%%%%%%%%%%%%%%%%%%%%%%%%%%%%%%%%
\subsection{Ly$\alpha$ background and Wouthuysen--Field coupling}
\label{subsec_xalpha}
%%%%%%%%%%%%%%%%%%%%%%%%%%%%%%%%%%%%%%%%%%%%%%%%%%%

We now describe the Ly$\alpha$ background that controls the Wouthuysen--Field coupling $x_\alpha(\mathbf r,z)$ in Eq.~\eqref{eq_Ts_spin_bdm}.  
We write $I_\nu(\mathbf r,\hat{\mathbf n},z)$ for the specific intensity of the UV background field along the direction $\hat{\mathbf n}$, at position $\mathbf r$ and redshift $z$.
Thus $I_\nu$ has units of ${\rm eV}\,{\rm s}^{-1}\,{\rm cm}^{-2}\,{\rm Hz}^{-1}\,{\rm sr}^{-1}$. We define the angle averaged specific intensity by
\begin{equation}
J_\nu(\mathbf r,z)
\equiv
\frac{1}{4\pi}
\int d\Omega\,
I_\nu(\mathbf r,\hat{\mathbf n},z).
\label{eq_Jnu_def}
\end{equation}
The Ly$\alpha$ background intensity field is evaluated at the Ly$\alpha$ resonance,
\begin{equation}
J_\alpha(\mathbf r,z)
\equiv
J_\nu(\mathbf r,z)\big|_{\nu=\nu_\alpha}.
\label{eq_Jalpha_def}
\end{equation}
It is useful to work with the corresponding photon number intensity field,
\begin{equation}
\widehat J_\alpha(\mathbf r,z)
\equiv
\frac{J_\alpha(\mathbf r,z)}{h\nu_\alpha},
\label{eq_Jalpha_hat_def}
\end{equation}
which has units of ${\rm cm}^{-2}\,{\rm s}^{-1}\,{\rm Hz}^{-1}\,{\rm sr}^{-1}$.
Here $\nu_\alpha$ is the Ly$\alpha$ resonance frequency of the
$1s\leftrightarrow 2p$ transition.

We split the total Ly$\alpha$ background field into two parts, a stellar UV contribution and an X-ray-induced contribution,
\begin{equation}
\widehat J_\alpha(\mathbf r,z)
=
\widehat J_{\alpha,\star}(\mathbf r,z)
+
\widehat J_{\alpha,X}(\mathbf r,z).
\label{eq_Jalpha_split}
\end{equation}
Here $\widehat{J}_{\alpha,\star}$ is the Ly$\alpha$ background that comes from stellar UV photons emitted between Ly$\alpha$ and the Lyman limit, where the Lyman limit is the ionization threshold of ground-state hydrogen. Two processes give rise to this contribution. In the first process, photons are emitted at $z'$ with frequencies between Ly$\alpha$ ($2p\leftrightarrow 1s$) and Ly$\beta$ ($3p\leftrightarrow 1s$). These photons meet no other Lyman-series resonance and redshift directly into the local Ly$\alpha$ resonance at $z$. In the second process, photons are emitted at $z'$ with frequencies between Ly$\beta$ and the Lyman limit. In most cases, they are absorbed in a higher Lyman-series transition, Ly$n$ ($1s\leftrightarrow np$) with $n\ge 3$, and this is followed by a cascade through lower levels at $z$. For this reason, we write the UV contribution as a sum over Lyman-series resonances,
\begin{equation}
\widehat J_{\alpha,\star}(\mathbf r,z)
=
\sum_{n=2}^{n_{\rm max}}
\widehat J_\alpha^{(n)}(\mathbf r,z),
\label{eq_Jalpha_star_sum}
\end{equation}
with
\begin{equation}
\widehat J_\alpha^{(n)}(\mathbf r,z)
=
f_{\rm recycle}(n)
\int_z^{z_n^{\rm max}} dz'\,
\frac{1}{4\pi}
\frac{1}{4\pi r_p^2(z,z')}
\frac{d\phi_{*,e}(\nu'_n,\mathbf r,z,z')}{dz'} .
\label{eq_Jalpha_star_n}
\end{equation}
Here, $n_{\rm max}$ is the maximum principal quantum number included in the stellar UV contribution. The factor $f_{\rm recycle}(n)$ gives the probability that a Ly$\alpha$ photon is finally produced through a radiative cascade after a photon is absorbed in the Ly$n$ resonance. For the direct Ly$\alpha$ channel, this probability is $f_{\rm recycle}(2)=1$. For higher Lyman resonances, the recycling probabilities are smaller, and they are fixed by atomic branching ratios.
We follow the standard implementation of $f_{\rm recycle}(n)$ and $n_{\rm max}=23$ adopted in \FirstClass.

A photon emitted above a certain source redshift would encounter the next higher Lyman resonance before it redshifts into the Ly$n$ resonance at redshift $z$. The upper limit $z_n^{\rm max}$ is this maximum source redshift.
In the standard hydrogenic case,
\begin{equation}
1+z_n^{\rm max}(z)
=
(1+z)\,
\frac{1-(n+1)^{-2}}{1-n^{-2}} .
\label{eq_zmax_Lyn}
\end{equation}

The lines of the hydrogenic Lyman series are the transitions $1s\leftrightarrow np$ between the ground state and the excited $np$ levels, and their frequencies follow the Rydberg relation
\begin{equation}
\nu_n
=
\nu_\infty\left(1-\frac{1}{n^2}\right),
\label{eq:nu_n_rydberg}
\end{equation}
where $\nu_\infty = E_{\rm HI}^{\rm th} /h$ denotes the ionization threshold frequency of ground-state hydrogen.
For $n=2$, Eq.~(\ref{eq:nu_n_rydberg}) gives the Ly$\alpha$ frequency,
\begin{equation}
\nu_{n=2}
\equiv
\nu_\alpha
=
\frac{3}{4}\,\nu_\infty .
\end{equation}
The emission frequency $\nu_n'$ at the source redshift $z'$ is fixed by cosmological redshifting,
\begin{equation}
\nu'_n
=
\nu_n
\left(\frac{1+z'}{1+z}\right).
\label{eq_nuprime_Lyn}
\end{equation}
The quantity $d\phi_{*,e}(\nu'_n,\mathbf r,z,z')/dz'$ in Eq.~\eqref{eq_Jalpha_star_n} is the stellar UV photon emission rate per unit redshift, in units of ${\rm s}^{-1}\,{\rm Hz}^{-1}$. It collects the photons from all sources in the shell $[z',z'+dz']$ that are emitted at frequency $\nu'_n$ and afterwards redshift into the local Ly$n$ resonance at the point $(\mathbf r,z)$,
\begin{equation}
\frac{d\phi_{*,e}(\nu'_n,\mathbf r,z,z')}{dz'}
=
f_*^{\rm II}\,
\epsilon_\star(\nu'_n)
\Bigg(
\frac{\Omega_b\rho_{\rm crit,0}}{m_p}
\left[
1+\delta_{b,R_{\rm sh}}^{\rm NL}(\mathbf r,z')
\right]
\frac{dV(z,z')}{dz'}
\Bigg)
\frac{df_{\rm coll,\star}^{\rm eff}(\mathbf r,z',R_{\rm sh})}{dt}.
\label{eq_dphistar_dz}
\end{equation}
The function $\epsilon_\star(\nu'_n)$, in units of ${\rm Hz}^{-1}$, gives the number of stellar UV photons emitted per unit frequency per stellar baryon.
As described in Ref.~\cite{Barkana:2004vb}, it is approximated by a separate power law for each frequency interval between successive Lyman-series resonances. With the normalization adopted in \FirstClass, $\epsilon_\star(\nu'_n)$ is written as
\begin{equation}
\epsilon_\star(\nu'_n)
=
\frac{q_m N_\gamma}{\nu_\alpha}
\frac{\beta_m+1}
{\widetilde{\nu}_{m+1}^{\beta_m+1}
-\widetilde{\nu}_{m}^{\beta_m+1}}
\left(\frac{\nu'_n}{\nu_\alpha}\right)^{\beta_m},
\qquad
\nu_m\leq\nu'_n<\nu_{m+1},
\label{eq:stellar_uv_spectrum}
\end{equation}
where $\widetilde{\nu}_m\equiv\nu_m/\nu_\alpha$. The relative photon yield $q_m$ and the spectral index $\beta_m$ for each interval are taken from the tabulated Pop II spectrum used in \FirstClass.

The baryon number contained in the source shell per unit source redshift is given by the combination $\left[\Omega_b\rho_{\rm crit,0}/m_p\right] \left[1+\delta_{b,R_{\rm sh}}^{\rm NL}(\mathbf r,z')\right]dV/dz'$.
The shell scale $R_{\rm sh}$, the volume element $dV/dz'$, and the collapsed fraction $f_{\rm coll,\star}^{\rm eff}$ are defined in Eqs.~\eqref{eq_Rsh_def}, \eqref{eq_dVdz_shell}, and \eqref{eq_fcoll_star}.

The X-ray-induced Ly$\alpha$ background field $\widehat J_{\alpha,X}(\mathbf r,z)$ is generated by secondary excitations of neutral hydrogen. 
As described in Appendices~\ref{subsec_xray} and \ref{subsec:xe_evolution}, the X-ray photon field $\widetilde{\mathcal J}_X(\mathbf r,\nu,z)$ defined in Eq.~\eqref{eq_JX_def} heats and ionizes the IGM. Through photoionization of HI, HeI, and HeII, the same field also produces energetic primary electrons. These primary electrons then deposit a fraction of their kinetic energy into atomic excitations, and some of these excitations lead in the end to HI Ly$\alpha$ photons.
We write
\begin{align}
\widehat J_{\alpha,X}(\mathbf r,z)
=
&
\frac{c\,n_{\rm H}(\mathbf r,z)(1+z)}
{4\pi \nu_\alpha}
\int_{\nu_0}^{\infty} d\nu
\sum_{i\in\{{\rm HI},{\rm HeI},{\rm HeII}\}}
x_i^X(\mathbf r,z)\,
\sigma_i(\nu)\,
\widetilde{\mathcal J}_X(\mathbf r,\nu,z)
\nonumber\\
&
\times
\left(h\nu-E_i^{\rm th}\right)
\frac{
f_{\rm Ly\alpha}
\left(
h\nu-E_i^{\rm th},
x_e^{\rm X}(\mathbf r,z)
\right)
}
{h\nu_\alpha}.
\label{eq_Jalpha_X_hat}
\end{align}
The fraction of the primary photoelectron energy that is deposited into Ly$\alpha$ photons is denoted by the dimensionless function $f_{\rm Ly\alpha}(E,x_e)$, which we evaluate at the local free-electron fraction. We use the standard tabulated prescription of \FirstClass\ for $f_{\rm Ly\alpha}$.

The expressions above specify the two components of the total Ly$\alpha$
background field. In BDM cosmology, these quantities are modified through
the local density $n_{\rm H}(\mathbf r,z)$ and ionization fields
$x_i(\mathbf r,z)$. More importantly, the suppression of small-scale matter
fluctuations changes the growth rate of the collapsed fraction,
$df_{\rm coll,\star}^{\rm eff}/dt$, which enters both the stellar UV photon production rate
in Eq.~\eqref{eq_dphistar_dz} and the X-ray photon field
$\widetilde{\mathcal J}_X(\mathbf r,\nu,z)$.

\begin{figure}[t!]
    \centering
    \includegraphics[width=1\linewidth]{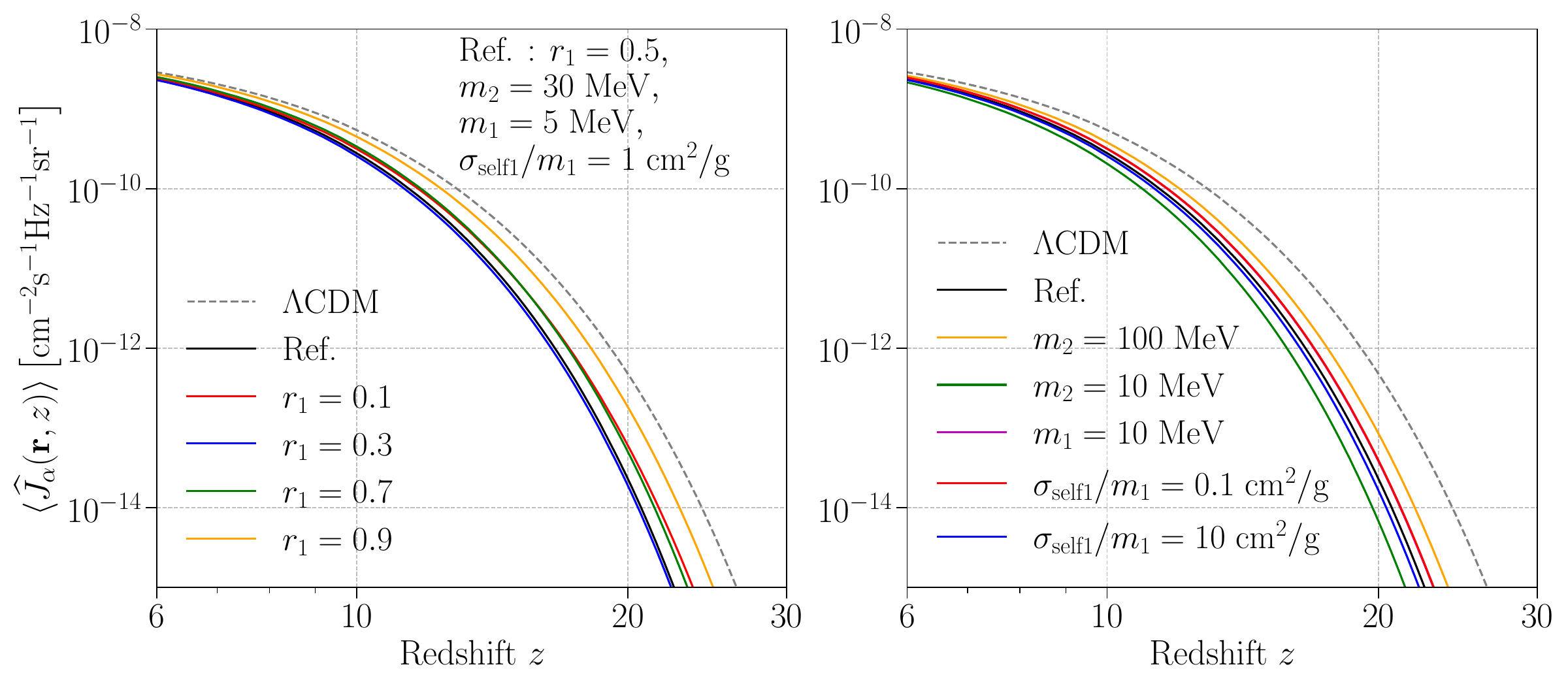}
    \caption{
    Redshift evolution of the volume averaged Ly$\alpha$ photon number
    intensity $\langle \widehat J_\alpha(\mathbf r,z)\rangle$ (see Eq.~\eqref{eq_Jalpha_split}) for \LCDM\
    and representative BDM parameter choices. The left panel shows the
    dependence on the light component fraction $r_1$, while the right panel
    illustrates the effects of varying $m_2$, $m_1$, and
    $\sigma_{\rm self1}/m_1$ around the reference model. In BDM cosmology,
    the suppression of small-scale matter fluctuations delays the formation
    of radiative sources. As a result, the Ly$\alpha$ background builds up
    more slowly than in \LCDM, leading to a smaller
    $\langle \widehat J_\alpha\rangle$ at fixed redshift.
    }
    \label{fig:Jalpha_bdm}
\end{figure}

We show in Figure~\ref{fig:Jalpha_bdm} the volume averaged total Ly$\alpha$
photon number intensity $\langle \widehat J_\alpha(\mathbf r,z)\rangle$.
Relative to \LCDM, BDM models produce a weaker Ly$\alpha$ background at fixed
redshift. This follows directly from the delayed formation of source hosting
halos. Since fewer halos contribute stellar UV photons and X-ray photons at
a given redshift, both $\widehat J_{\alpha,\star}$ and
$\widehat J_{\alpha,X}$ build up more slowly. The parameter dependence
follows the same trend as the small-scale matter power spectrum, with
stronger small-scale suppression leading to a smaller
$\langle \widehat J_\alpha\rangle$.

We next convert the Ly$\alpha$ background field into the scattering rate
that enters the Wouthuysen--Field coupling. For details of this conversion,
see Appendix~\ref{app_lya_line_profile}. In the approximation used here, the
Ly$\alpha$ scattering rate per neutral hydrogen atom, with units of ${\rm s}^{-1}$, can be written as
\begin{equation}
P_\alpha(\mathbf{r},z)
\simeq
4\pi\,
\sigma_{\alpha,0}\,
\widehat J_\alpha(\mathbf r,z)\,
S_\alpha(\mathbf{r},z).
\label{eq:Palpha_narrowline}
\end{equation}
The overall normalization $\sigma_{\alpha,0}$ has units of
${\rm cm}^2\,{\rm Hz}$ and is determined by the oscillator strength of the
Ly$\alpha$ transition,
\begin{equation}
\sigma_{\alpha,0}
=
\frac{\pi e^2}{m_e c}\,
f_\alpha ,
\label{eq_sigma_alpha0}
\end{equation}
where $f_\alpha\simeq0.4162$. The dimensionless factor
$S_\alpha(\mathbf r,z)$ is an order-unity correction that accounts for the
detailed spectral shape of the radiation field near the Ly$\alpha$ resonance.

Substituting Eq.~\eqref{eq:Palpha_narrowline} into the definition of the Wouthuysen-Field coupling coefficient, Eq.~\eqref{eq_xalpha_bdm}, we obtain
\begin{equation}
x_\alpha(\mathbf r,z)
\simeq
\frac{16\pi}{27}\,
\frac{T_\star}
{\bar T_\gamma(z)A_{10}}\,
\sigma_{\alpha,0}\,
S_\alpha(\mathbf r,z)\,
\widehat J_\alpha(\mathbf r,z).
\label{eq_xalpha_expanded}
\end{equation}
Eq.~\eqref{eq_xalpha_expanded} shows that the BDM dependence of the
Wouthuysen--Field coupling enters primarily through the Ly$\alpha$ photon
number intensity $\widehat J_\alpha(\mathbf r,z)$. As shown
in Figure~\ref{fig:Jalpha_bdm}, BDM models predict a smaller volume averaged
$\langle\widehat J_\alpha(\mathbf r,z)\rangle$ than \LCDM\ at fixed redshift,
because the suppression of small-scale matter fluctuations delays the
formation of radiative sources. Consequently, the Wouthuysen--Field coupling
is weakened and its onset is shifted to lower redshift.

%%%%%%%%%%%%%%%%%%%%%%%%%%%%%%%%%%%%%%%%%%%%%%%%%%%
\subsection{UV Ionization and the Effective Collapsed Fraction} 
\label{app:UV_fraction}
%%%%%%%%%%%%%%%%%%%%%%%%%%%%%%%%%%%%%%%%%%%%%%%%%%%

In this appendix, we describe how the UV photon contribution $x_{\rm HII}^{\rm UV}(\mathbf r,z)$ to the local ionized hydrogen fraction in Eq.~\eqref{eq_xHII_total} is calculated. Following the excursion-set prescription~\cite{Furlanetto_2004,Zahn:2011} implemented in \FirstClass, the UV ionization module evaluates the ionization state at each position $\mathbf r$ by comparing the number of ionizing photons with the number of hydrogen atoms within a spherical region of radius $R$.
To determine the matter overdensity within this region, we first smooth $\delta_m$ using the top-hat window function defined in Eq.~\eqref{eq_tophat_window},
\begin{equation}
\delta_{m,R}(\mathbf r,z)
=
\int d^3\mathbf y\,
W_R(|\mathbf r-\mathbf y|)
\delta_m(\mathbf y,z).
\label{eq_delta_smoothed_top_hat}
\end{equation}
The number of ionizing photons per hydrogen nucleus produced within a spherical region of radius $R$ is~\cite{Park:2018ljd}
\begin{equation}
n_{\rm ion}(\mathbf r,z,R)=
\int_{M_{\min}}^{M_R} dM_h
\frac{M_h}{\bar\rho_{m}(0)}
\frac{dn_{\rm cond}}{dM_h}
\left(\mathbf r,z,M_h,R\right)
f_*(M_h)\;
f_{\rm esc}(M_h)\;
f_{\rm duty}(M_h)\;
N_{\gamma}.
\label{eq_ionizing_photon_per_baryon}
\end{equation}
Here, $M_R=(4\pi/3)R^3\bar\rho_m(0)$ is the total matter mass enclosed within the region, where $\bar\rho_m(0)$ is the mean matter density at $z=0$. 
For the lower integration limit we use $M_{\min}=M_{\rm turn}/50$, where $M_{\rm turn}$ is the parameter defined in Eq.~\eqref{eq:astro_params_schematic}.
The abundance of halos in the mass range $[M_h,M_h+dM_h]$ within the region $R$ is described by the conditional halo mass function $dn_{\rm cond}/dM_h$. See Appendix~\ref{app:conditional_halo_mass_function} for details.
The quantity $N_\gamma$ is the number of hydrogen-ionizing photons produced per stellar baryon. 
The functions $f_*(M_h)$ and $f_{\rm esc}(M_h)$ are the fraction of halo baryons converted into stars and the escape fraction of ionizing photons into the IGM, respectively. 
The duty-cycle factor $f_{\rm duty}(M_h)$ gives the fraction of the halos mass $M_h$ that host star-forming galaxies. It suppresses the contribution from low-mass halos, where inefficient cooling and feedback hinder gas accumulation and star formation. These mass-dependent source functions are parameterized in terms of the astrophysical parameters defined in Eq.~\eqref{eq:astro_params_schematic} as
\begin{equation}
f_*(M_h)=f_*^{\rm II} \left(\frac{M_h}{10^{10}M_\odot}\right)^{\alpha_*^{\rm II}},
f_{\rm esc}(M_h)=f_{\rm esc}^{\rm II}\left(\frac{M_h}{10^{10}M_\odot}\right)^{\alpha_{\rm esc}^{\rm II}},
f_{\rm duty}(M_h)=\exp\left(-\frac{M_{\rm turn}}{M_h}\right).
\label{eq_mass_dependent_source_functions}
\end{equation}
In Eq.~\eqref{eq_ionizing_photon_per_baryon}, we factor out the mass-independent ionizing efficiency
\begin{equation}
\zeta_{\rm ion}
\equiv
f_*^{\rm II}f_{\rm esc}^{\rm II}N_{\gamma},
\end{equation}
and collect the remaining mass-dependent factors into the dimensionless effective collapsed fraction $f_{\rm coll}^{\rm eff}$ defined by
\begin{equation}
\begin{aligned}
f_{\rm coll}^{\rm eff}(\mathbf r,z,R)
&\equiv
\int_{M_{\min}}^{M_R} dM_h\,
\frac{M_h}{\bar\rho_m(0)}
\frac{dn_{\rm cond}}{dM_h}
\left(\mathbf r,z,M_h,R\right)
\left(
\frac{M_h}{10^{10}M_\odot}
\right)^{\alpha_*^{\rm II}+\alpha_{\rm esc}^{\rm II}}
\exp\left(-\frac{M_{\rm turn}}{M_h}\right).
\end{aligned}
\label{eq_fcoll_eff_definition}
\end{equation}
This gives 
\begin{equation}
n_{\rm ion}(\mathbf r,z,R)
=
\zeta_{\rm ion}f_{\rm coll}^{\rm eff}(\mathbf r,z,R).
\label{eq_fcoll_eff}
\end{equation}
In addition, the effective collapsed fraction with no escape-fraction weighting is defined as
\begin{equation}
f_{\rm coll,\star}^{\rm eff}(\mathbf r,z,R)
\equiv
\left.
f_{\rm coll}^{\rm eff}(\mathbf r,z,R)
\right|_{\alpha_{\rm esc}^{\rm II}=0}.
\label{eq_fcoll_star}
\end{equation}

\begin{figure}
    \centering
    \includegraphics[width=0.8\linewidth]{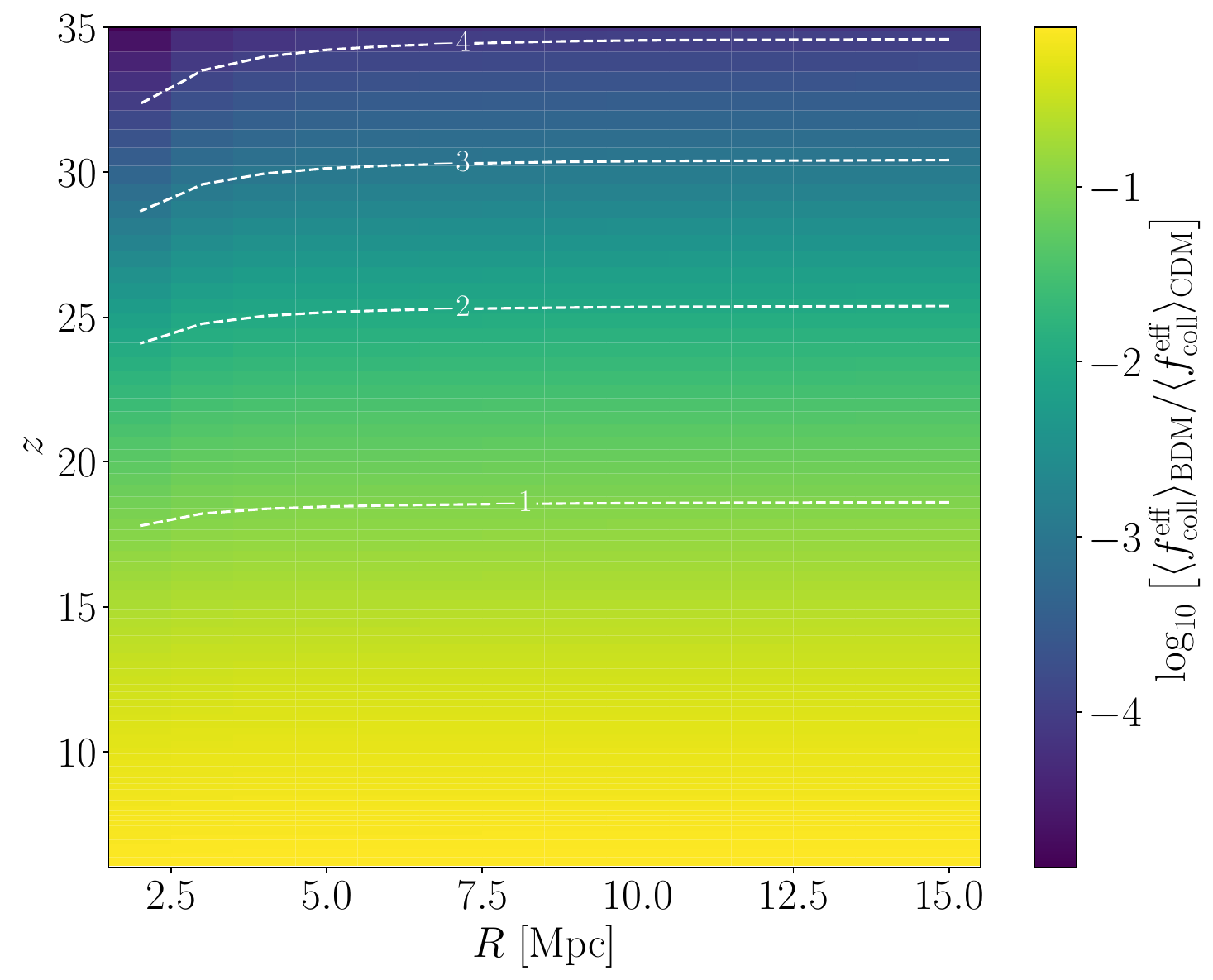}
    \caption{
    Ratio of the volume-averaged effective collapsed fraction (see Eq.~\eqref{eq_fcoll_eff_definition}) in the reference BDM model to that in $\Lambda$CDM as a function of smoothing scale $R$ and redshift $z$. The reference BDM parameters are $m_2=30~{\rm MeV}$, $m_1=5~{\rm MeV}$, $r_1=0.5$, and $\sigma_{{\rm self}1}/m_1=1.0~{\rm cm^2\,g^{-1}}$. The suppression becomes stronger toward higher redshift, where source-hosting halos are rare and their abundance is highly sensitive to the reduced matter variance. It is also mildly enhanced at smaller $R$, where the smoothed density field receives greater contributions from the large-$k$ modes affected by BDM.}
    \label{fig:eff_f_coll_ratio}
\end{figure}

Figure~\ref{fig:eff_f_coll_ratio} shows the ratio
$\langle f_{\rm coll}^{\rm eff}\rangle_{\rm BDM}/
\langle f_{\rm coll}^{\rm eff}\rangle_{\Lambda{\rm CDM}}$
for the reference BDM model, where $\langle\cdots\rangle$ denotes the volume average over the simulation box. For this comparison, the
astrophysical parameters are fixed to the values given in
Eq.~\eqref{eq:astro_params_schematic}. The ratio remains below unity throughout the plotted range, showing that BDM reduces the effective collapsed fraction of source-hosting halos. 
These halos correspond to rare density peaks, and the abundance of such peaks depends exponentially on the matter variance. When the small-scale variance is reduced in BDM, the halo abundance therefore decreases by a much larger fraction. This is why the suppression grows toward higher redshift.
The suppression is also slightly stronger at smaller $R$, where the smoothed density field receives greater contributions from the large-$k$ modes affected by BDM. At fixed ionizing efficiency $\zeta_{\rm ion}$, the resulting decrease in $f_{\rm coll}^{\rm eff}$ directly reduces the UV photon production through Eq.~\eqref{eq_fcoll_eff}.

The number of ionizing photons per hydrogen nucleus computed in Eq.~\eqref{eq_fcoll_eff} is used to determine the UV contribution $x_{\rm HII}^{\rm UV}(\mathbf r,z)$ to the local ionized hydrogen fraction. Before doing so, several additional effects need to be taken into account. 
When a galaxy emits an ionizing UV photon, the photon may travel to an ionization front and ionize neutral hydrogen outside the existing HII region. In this case, the HII region grows. The photon may instead be absorbed inside the HII region, by a hydrogen atom that was ionized earlier and has since recombined. Then the gas returns to its ionized state, but the ionization front does not advance. Because of recombinations, a part of the source photon budget is thus spent on keeping already ionized gas ionized, and not on the growth of HII regions. This extra photon consumption matters most in dense structures that recombine rapidly. We model it with the inhomogeneous-recombination prescription described in Appendix~\ref{app:inhomogeneous_recombination}.

Let $N_{H,R}$ denote the total number of hydrogen nuclei within this
region. Since X-rays have already ionized a fraction
$x_{e,R}^X(\mathbf r,z)$, the number of hydrogen nuclei remaining
to be ionized by UV photons is
\begin{equation}
\left[1-x_{e,R}^X(\mathbf r,z)\right]N_{H,R}.
\end{equation}
Here, $x_{e,R}^X(\mathbf r,z)$ is obtained by smoothing the X-ray contribution to the free-electron fraction over a spherical region of radius $R$,
\begin{equation}
x_{e,R}^X(\mathbf r,z)
=
\int d^3\mathbf y\,
W_R(|\mathbf r-\mathbf y|)
x_e^{\rm X}(\mathbf y,z).
\label{eq_xeX_smoothed}
\end{equation}
If the mean number of recombinations per hydrogen nucleus is $n_{{\rm rec},R}(\mathbf r,z)$ (see Appendix~\ref{app:inhomogeneous_recombination} for details), 
the total number of UV photons required to ionize the remaining hydrogen nuclei is therefore
\begin{equation}
N_{\gamma,{\rm required}}^{\rm UV}
=
\left(1-x_{e,R}^X(\mathbf r,z)\right)
\left(1+n_{{\rm rec},R}(\mathbf r,z)\right)
N_{H,R}.
\label{eq_uv_required_photons}
\end{equation}
Dividing Eq.~\eqref{eq_uv_required_photons} by
$N_{H,R}$ gives the required number of UV photons per hydrogen
nucleus,
\begin{equation}
\frac{N_{\gamma,{\rm required}}^{\rm UV}}{N_{H,R}}
=
\left(1-x_{e,R}^X(\mathbf r,z)\right)
\left(1+n_{{\rm rec},R}(\mathbf r,z)\right).
\label{eq_uv_required_per_hydrogen}
\end{equation}

The ionization condition is evaluated on progressively smaller smoothing
scales. We start from the maximum smoothing scale $R_{\max}=50\,{\rm Mpc}$, which is the maximum mean free path of ionizing UV photons adopted here, and go down to the cell scale $R_{\rm cell}\simeq\Delta x_{\rm lo}=2\,{\rm Mpc}$. For each cell centered at $\mathbf r$, the code checks if the following condition holds for some smoothing scale $R\in[R_{\rm cell},R_{\max}]$,
\begin{equation}
n_{\rm ion}(\mathbf r,z,R)
\geq
\left(1-x_{e,R}^X(\mathbf r,z)\right)
\left(1+n_{{\rm rec},R}(\mathbf r,z)\right).
\label{eq_ionization_condition}
\end{equation}
The central cell is marked as fully ionized when Eq.~\eqref{eq_ionization_condition} holds on at least one smoothing scale. Since the X-ray contribution has already ionized a fraction
$x_e^X(\mathbf r,z)$ of the hydrogen in the cell, the additional contribution assigned to UV photons is $x_{\rm HII}^{\rm UV}(\mathbf r,z)=1-x_e^X(\mathbf r,z)$.

When no smoothing scale satisfies Eq.~\eqref{eq_ionization_condition}, we compute the partial UV ionized fraction on the cell scale,
\begin{equation}
x_{\rm HII}^{\rm UV}(\mathbf r,z)
=
n_{\rm ion}(\mathbf r,z,R_{\rm cell}).
\label{eq_xHII_UV_partial}
\end{equation}
In this cell-scale prescription, no additional correction for
inhomogeneous recombinations is applied to the partial UV ionized
fraction.
In summary, the UV contribution is therefore
\begin{equation}
x_{\rm HII}^{\rm UV}(\mathbf r,z)
=
\begin{cases}
1-x_e^{\rm X}(\mathbf r,z),
&
\text{if Eq.~\eqref{eq_ionization_condition} is satisfied for some }R,
\\[6pt]
\displaystyle
n_{\rm ion}(\mathbf r,z,R_{\rm cell})
&
\text{otherwise}.
\end{cases}
\label{eq_xHII_UV}
\end{equation}

%%%%%%%%%%%%%%%%%%%%%%%%%%%%%%%%%%%%%%%%%%%%%%%%%%%
\subsection{Conditional halo mass function}
\label{app:conditional_halo_mass_function}
%%%%%%%%%%%%%%%%%%%%%%%%%%%%%%%%%%%%%%%%%%%%%%%%%%%

A summary of the conditional halo mass function $dn_{\rm cond}/dM_h(\mathbf r,z,M_h,R)$ that enters Eq.~\eqref{eq_ionizing_photon_per_baryon} is given in this section. This function is the comoving number density of halos with mass in the interval $[M_h,M_h+dM_h]$ inside a spherical region of radius $R$ centered at $\mathbf r$.

Our starting point is the extended Press--Schechter conditional halo mass function~\cite{Lacey:1994su,Somerville:1997df}
\begin{equation}
\frac{dn_{\rm cond}^{\rm EPS}}{dM_h}
\left(\mathbf r,z,M_h,R\right)
=
\frac{\bar\rho_m(0)}{M_h}
f\left(\mathbf r,z,M_h,R\right)
\left|
\frac{d\sigma^2(M_h,z)}{dM_h}
\right|,
\label{eq:eps_conditional_hmf}
\end{equation}
where $f(\mathbf r,z,M_h,R)$ denotes the conditional first-crossing distribution per unit variance, given by
\begin{equation}
f\left(\mathbf r,z,M_h,R\right)
=
\frac{1}{\sqrt{2\pi}}
\frac{\delta_c-\delta_{m,R}(\mathbf r,z)}
{\left[\sigma^2(M_h,z)-\sigma^2(M_R,z)\right]^{3/2}}
\exp\left[
-\frac{
\left[\delta_c-\delta_{m,R}(\mathbf r,z)\right]^2
}{
2\left[\sigma^2(M_h,z)-\sigma^2(M_R,z)\right]
}
\right].
\label{eq:conditional_function}
\end{equation}
Here, the total matter mass of the smoothing region is $M_R = (4\pi/3)R^3\bar\rho_m(0)$. In the same way, the comoving radius $R_h$ of a halo of mass $M_h$ follows from $M_h=(4\pi/3)R_h^3\bar\rho_m(0)$. The halo mass is limited to $M_h\leq M_R$, which is the same as $R_h\leq R$, so that $\sigma^2(M_h,z)-\sigma^2(M_R,z)\geq0$ always holds. The smoothed linear matter overdensity $\delta_{m,R}(\mathbf r,z)$ and the variance $\sigma^2(M,z)$ are those defined in Eqs.~\eqref{eq_delta_smoothed_top_hat} and \eqref{eq_sigma_R}, respectively. For the critical linear overdensity for spherical collapse, we use the fixed value $\delta_c=1.686$.

Numerical \LCDM\ simulations are fitted better by the Sheth--Tormen mass function. We therefore rescale the extended Press--Schechter conditional mass function as~\cite{Park:2018ljd,Mesinger:2010ne}
\begin{equation}
\frac{dn_{\rm cond}}{dM_h}
\left(\mathbf r,z,M_h,R\right)
=
\frac{
\displaystyle \frac{dn^{\rm ST}}{dM_h}(M_h,z)
}{
\displaystyle \frac{dn^{\rm PS}}{dM_h}(M_h,z)
}
\frac{dn_{\rm cond}^{\rm EPS}}{dM_h}
\left(\mathbf r,z,M_h,R\right).
\label{eq:renormalized_conditional_hmf}
\end{equation}
For the Press--Schechter and Sheth--Tormen halo mass functions in this expression, we use~\cite{Jenkins:2000bv}
\begin{equation}
\frac{dn^{\rm PS}}{dM_h}(M_h,z)
=
\frac{\bar\rho_m(0)}{M_h}
\frac{1}{\sqrt{2\pi}}
\frac{\delta_c}{\sigma^3(M_h,z)}
\exp\left[
-\frac{\delta_c^2}{2\sigma^2(M_h,z)}
\right]
\left|
\frac{d\sigma^2(M_h,z)}{dM_h}
\right|,
\label{eq:PS_halo_mass_function}
\end{equation}
and
\begin{equation}
\begin{aligned}
\frac{dn^{\rm ST}}{dM_h}(M_h,z)
={}&
A_{\rm ST}
\frac{\bar\rho_m(0)}{M_h}
\sqrt{\frac{a_{\rm ST}}{2\pi}}
\frac{\delta_c}{\sigma^3(M_h,z)}
\left[
1+
\left(
\frac{a_{\rm ST}\delta_c^2}{\sigma^2(M_h,z)}
\right)^{-p_{\rm ST}}
\right]
\\
&\times
\exp\left[
-\frac{a_{\rm ST}\delta_c^2}{2\sigma^2(M_h,z)}
\right]
\left|
\frac{d\sigma^2(M_h,z)}{dM_h}
\right|,
\end{aligned}
\label{eq:ST_halo_mass_function}
\end{equation}
respectively. The Sheth--Tormen fitting parameters are set to $A_{\rm ST}=0.3222$, $a_{\rm ST}=0.707$, and $p_{\rm ST}=0.3$.

\FirstClass\ adopts the halo mass function prescription in Eq.~\eqref{eq:renormalized_conditional_hmf}. This prescription is calibrated against $\Lambda$CDM simulations, and it has not been validated for BDM cosmologies.
We nevertheless use this prescription as an approximation in the present analysis. In future work, we plan to perform dedicated BDM $N$-body simulations, building on Refs.~\cite{Kim:2023onk,Kim:2024ltz}, to calibrate the halo mass function directly in BDM cosmology.

BDM cosmology modifies both the smoothed overdensity and the matter
variances entering Eq.~\eqref{eq:conditional_function}. Because the
conditional halo mass function depends exponentially on these
quantities, even modest changes can substantially alter the abundance
of source-hosting halos, particularly at high redshift. The resulting
suppression of $f_{\rm coll}^{\rm eff}$ is shown in
Figure~\ref{fig:eff_f_coll_ratio}. Since this quantity determines the
formation of the radiation sources used throughout the 21\,cm
calculation, its modification provides a primary source of the
difference between the $\Lambda$CDM and BDM predictions.

%%%%%%%%%%%%%%%%%%%%%%%%%%%%%%%%%%%%%%%%%%%%%%%%%%%
\subsection{Sub-Grid Prescription for Inhomogeneous Recombinations}
\label{app:inhomogeneous_recombination}
%%%%%%%%%%%%%%%%%%%%%%%%%%%%%%%%%%%%%%%%%%%%%%%%%%%

The cumulative recombination correction $n_{{\rm rec},R}(\mathbf r,z)$ in Eq.~\eqref{eq_ionization_condition} is computed with the sub-grid prescription implemented in \FirstClass. This appendix gives a short summary of that prescription.

First, the dimensionless sub-grid hydrogen overdensity is defined by
\begin{equation}
\Delta
\equiv
\frac{n_{\rm H}^{\rm sg}}
{n_{\rm H}(\mathbf r,z)},
\label{eq:subgrid_hydrogen_density}
\end{equation}
where the hydrogen-nucleus number density is written as $n_{\rm H}^{\rm sg}$ for an unresolved gas element and as $n_{\rm H}(\mathbf r,z)$ for the resolved cell in Eq.~\eqref{eq_local_nH}.

The sub-grid overdensity $\Delta$ should be distinguished from the resolved cell-averaged overdensity, cf.\ Eq.~\eqref{eq_local_nH},
\begin{equation}
\Delta_{\rm cell}(\mathbf r,z)
\equiv
\frac{n_{\rm H}(\mathbf r,z)}
{\bar n_{\rm H}(z)}
=
1+\delta_b^{\rm NL}(\mathbf r,z).
\label{eq:resolved_cell_overdensity}
\end{equation}
While $\Delta_{\rm cell}$ specifies the mean density of a resolved cell, $\Delta$ describes the range of unresolved gas densities within that cell.

To model dense gas structures below the simulation resolution, we use the dimensionless, volume-weighted Miralda-Escudé--Haehnelt--Rees (MHR) gas-density distribution~\cite{Miralda-Escude:1998adl,Sobacchi:2014rua,Park:2018ljd},
\begin{equation}
P_V(\mathbf r,z,\Delta)
=
A(\mathbf r,z)
\exp\left[
-\frac{
\left[\Delta^{-2/3}-C_0(\mathbf r,z)\right]^2
}{
2\left[2\delta_0(\mathbf r,z)/3\right]^2
}
\right]
\Delta^{-\beta(\mathbf r,z)}.
\label{eq:MHR_pdf}
\end{equation}
Unresolved gas with an overdensity between $\Delta$ and $\Delta+d\Delta$ fills the fraction $P_V(\mathbf r,z,\Delta)d\Delta$ of the cell volume.
The parameter $\delta_0$ controls the width of the low-density part of the distribution and is evaluated as
\begin{equation}
\delta_0(\mathbf r,z)
=
\frac{7.61}{(1+z_{\rm eff}(\mathbf r,z))}
=
\frac{7.61}
{(1+z)\Delta_{\rm cell}^{1/3}(\mathbf r,z)}.
\label{eq:MHR_delta0}
\end{equation}
The variable $z_{\rm eff}$ is the redshift at which the cosmic background mean hydrogen density equals the resolved cell hydrogen density, $\bar{n}_{\rm H}(z_{\rm eff})=n_{\rm H}(\mathbf r,z)$.
The parameters $\beta(\mathbf r,z)$ and $C_0(\mathbf r,z)$ are obtained by interpolating their tabulated values at the effective redshift $z_{\rm eff}(\mathbf r,z)$, and the coefficient $A(\mathbf r,z)$ is determined by normalization conditions,
\begin{equation}
\int_0^\infty d\Delta\, P_V(\mathbf r,z,\Delta) = 1.
\label{eq:MHR_normalization}
\end{equation}

The recombination rate at a given sub-grid overdensity depends on both
the gas density and its ionization state. The latter is determined by
the ionizing UV background incident on the cell. We first estimate the
mean photoionization rate inside a candidate UV-ionized region of
radius $R$ as
\begin{equation}
\Gamma_{\rm HII}(\mathbf r,z,R)
\simeq
(1+z)^2
\lambda_{\rm mfp}
\sigma_{\rm HI,0}
\frac{\alpha_{\rm UV}}
{\alpha_{\rm UV}+\beta_{\rm HI}}
\bar n_{\rm H}(0)
\frac{dn_{\rm ion}}{dt}(\mathbf r,z,R).
\label{eq:Gamma_HII}
\end{equation}
The quantity $\Gamma_{\rm HII}$ has units of ${\rm s}^{-1}$ and gives
the mean photoionization probability per unit time for a neutral
hydrogen atom exposed to the UV background. The scale $R$ is the
excursion-set smoothing scale used in
Eq.~\eqref{eq_ionization_condition}, with
$R_{\rm cell}\leq R\leq R_{\max}$. It characterizes the size of the
candidate HII region. In this prescription, the
comoving mean free path of ionizing photons is approximated by the
size of this region $\lambda_{\rm mfp}=R$.
The quantity $n_{\rm ion}(\mathbf r,z,R)$ defined in
Eq.~\eqref{eq_ionizing_photon_per_baryon} is the cumulative number of
ionizing photons produced per hydrogen nucleus within the region.
Its time derivative $dn_{\rm ion}/dt$ is therefore the ionizing photon
production rate per hydrogen nucleus. In the numerical implementation, the time derivative is approximated as $dn_{\rm ion}(\mathbf r,z,R)/dt \simeq n_{\rm ion}(\mathbf r,z,R)/[t_\ast H^{-1}(z)]$.

The ionizing UV spectrum and the hydrogen photoionization cross
section are approximated by
\begin{equation}
J_\nu
=
J_{\nu,0}
\left(
\frac{\nu}{\nu_{\rm HI}}
\right)^{-\alpha_{\rm UV}},
\qquad
\sigma_{\rm HI}(\nu)
=
\sigma_{\rm HI,0}
\left(
\frac{\nu}{\nu_{\rm HI}}
\right)^{-\beta_{\rm HI}},
\qquad
\nu\geq\nu_{\rm HI},
\label{eq:UV_spectrum_cross_section}
\end{equation}
where $\nu_{\rm HI}$ is the hydrogen ionization-threshold frequency. The normalization $J_{\nu,0}$ is the specific intensity at this threshold, and $\sigma_{\rm HI,0}=6.3\times10^{-18}\,{\rm cm^2}$ is the photoionization cross section at the same frequency. We adopt $\alpha_{\rm UV}=5.0$ and $\beta_{\rm HI}=2.75$. 
Averaging the photoionization cross section over the ionizing photon spectrum gives the factor $\sigma_{\rm HI,0}\alpha_{\rm UV}/(\alpha_{\rm UV}+\beta_{\rm HI})$ in Eq.~\eqref{eq:Gamma_HII}.

Eq.~\eqref{eq:Gamma_HII} gives the photoionization rate before
attenuation by unresolved dense gas. Such gas can become
self-shielded from the external UV background. Physically, the
neutral hydrogen near the surface of a dense gas structure absorbs
a substantial fraction of the incident ionizing photons. The UV
intensity reaching its interior is consequently reduced, allowing
the inner gas to remain more neutral. This attenuation becomes
important when the neutral hydrogen column density is large enough
for the optical depth to ionizing photons to approach or exceed
unity.
The attenuated photoionization rate with
overdensity $\Delta$ is denoted by
$\Gamma_{\rm local}(\mathbf r,z,\Delta,R)$. We model it using
\begin{equation}
\begin{aligned}
\frac{
\Gamma_{\rm local}(\mathbf r,z,\Delta,R)
}{
\Gamma_{\rm HII}(\mathbf r,z,R)
}
={}&
0.98
\left[
1+
\left(
\frac{\Delta}
{\Delta_{\rm ss}(\mathbf r,z,R)}
\right)^{1.64}
\right]^{-2.28}
\\
&+
0.02
\left[
1+
\frac{\Delta}
{\Delta_{\rm ss}(\mathbf r,z,R)}
\right]^{-0.84},
\end{aligned}
\label{eq:Gamma_local}
\end{equation}
where $\Delta_{\rm ss}$ is the characteristic overdensity at which
self-shielding becomes important,
\begin{equation}
\Delta_{\rm ss}(\mathbf r,z,R)
=
26.7
\left(
\frac{T_{\rm ion}}{10^4\,{\rm K}}
\right)^{0.17}
\left(
\frac{1+z_{\rm eff}(\mathbf r,z)}{10}
\right)^{-3}
\left(
\frac{\Gamma_{\rm HII}(\mathbf r,z,R)}
{10^{-12}\,{\rm s}^{-1}}
\right)^{2/3}.
\label{eq:Delta_ss}
\end{equation}
Here, $T_{\rm ion}$ is the characteristic temperature of gas inside UV-ionized regions, for which we adopt $T_{\rm ion}=10^4\,{\rm K}$. 
When $\Delta\ll\Delta_{\rm ss}$, the gas is nearly optically thin and $\Gamma_{\rm local}\simeq\Gamma_{\rm HII}$. When $\Delta\gtrsim\Delta_{\rm ss}$, $\Gamma_{\rm local}$ falls more and more below the mean value inside the HII region because of self-shielding.

The neutral fraction of a sub-grid gas element with overdensity
$\Delta$ is defined as
\begin{equation}
x_{\rm HI}^{\rm sg}(\mathbf r,z,\Delta,R)
\equiv
\frac{
n_{\rm HI}^{\rm sg}(\mathbf r,z,\Delta,R)
}{
\Delta\;{n_{\rm H}(\mathbf r,z)}
},
\label{eq:subgrid_neutral_fraction}
\end{equation}
where $n_{\rm HI}^{\rm sg}$ is the neutral hydrogen number density of
that gas element. The neutral fraction is determined by balancing
photoionizations and recombinations,
\begin{equation}
x_{\rm HI}^{\rm sg}(\mathbf r,z,\Delta,R)
\Gamma_{\rm local}(\mathbf r,z,\Delta,R)
=
\chi_{\rm HeII}\,
\Delta {n_{\rm H}(\mathbf r,z)}\,
\alpha_{\rm B}(T_{\rm ion})
\left[
1-x_{\rm HI}^{\rm sg}(\mathbf r,z,\Delta,R)
\right]^2.
\label{eq:subgrid_photoionization_equilibrium}
\end{equation}
The left-hand side is the photoionization rate per hydrogen nucleus. The right-hand side is the recombination rate per hydrogen nucleus. The factor $1-x_{\rm HI}^{\rm sg}(\mathbf r,z,\Delta,R)$ is the ionized hydrogen fraction, and its square arises because the recombination rate is proportional to the product of the free-electron and ionized hydrogen densities. 
Singly ionized helium supplies additional free electrons, and these are included through the factor $\chi_{\rm HeII}\simeq1.08$.

The quantity $\alpha_{\rm B}(T_{\rm ion})$ is the case-B hydrogen
recombination coefficient. At $T_{\rm ion}=10^4\,{\rm K}$, we use
$\alpha_{\rm B}\simeq2.9\times10^{-13}\,{\rm cm^3\,s^{-1}}$. The
case-B coefficient excludes direct recombinations to the hydrogen
ground state because the ionizing photons produced by these
recombinations are assumed to be absorbed locally. Solving
Eq.~\eqref{eq:subgrid_photoionization_equilibrium} determines
$x_{\rm HI}^{\rm sg}$ and hence
$n_{\rm HI}^{\rm sg}$ through
Eq.~\eqref{eq:subgrid_neutral_fraction}.

The sub-grid recombination rate per hydrogen nucleus is then obtained
by averaging over the MHR density distribution,
\begin{equation}
\frac{dn_{\rm rec}}{dt}(\mathbf r,z,R)
=
\int_0^\infty d\Delta\,
P_V(\mathbf r,z,\Delta)\,
\Delta^2\,{n_{\rm H}(\mathbf r,z)}\,
\alpha_{\rm B}(T_{\rm ion})
\left[
1-x_{\rm HI}^{\rm sg}(\mathbf r,z,\Delta,R)
\right]^2.
\label{eq:subgrid_recombination_rate}
\end{equation}
The integral adds the recombination contributions from the different unresolved density environments within the cell, weighted by their volume fractions $P_V(\mathbf r,z,\Delta)d\Delta$. 
A larger $\Delta$ means a larger hydrogen density, and so the recombination rate goes up. However, self-shielding also makes $x_{\rm HI}^{\rm sg}$ larger, which leaves a smaller ionized fraction available for recombination.
Eq.~\eqref{eq:subgrid_recombination_rate} incorporates both effects.

If a cell is first ionized at $z_{\rm ion}(\mathbf r)$, its cumulative number of recombinations per hydrogen nucleus is given by
\begin{equation}
n_{\rm rec}(\mathbf r,z)
=
\int_z^{z_{\rm ion}(\mathbf r)}
dz'\,
\frac{
dn_{\rm rec}/dt
\left(
\mathbf r,z',R_{\rm HII}(\mathbf r,z')
\right)
}{
(1+z')H(z')
}.
\label{eq:cumulative_recombinations}
\end{equation}
Here, $R_{\rm HII}(\mathbf r,z')$ is the largest smoothing scale on which the UV ionization condition in Eq.~\eqref{eq_ionization_condition} holds for the cell at redshift $z'$.
It therefore lies within the range $R_{\rm cell}\leq R_{\rm HII}(\mathbf r,z')\leq R_{\max}$.
The redshift interval is changed into a time interval by the factor $[(1+z')H(z')]^{-1}$.
Eq.~\eqref{eq:cumulative_recombinations} therefore counts all recombinations that occur between the first ionization of the cell and the redshift of interest. For a cell that has not yet been UV ionized, $n_{\rm rec}(\mathbf r,z)$ is set to zero.

Since the UV ionization condition is evaluated over a region of
radius $R$, the cumulative recombination field is smoothed over the
same region,
\begin{equation}
n_{{\rm rec},R}(\mathbf r,z)
=
\frac{
\displaystyle
\int d^3\mathbf y\,
W_R(\lvert\mathbf r-\mathbf y\rvert)
n_{\rm rec}(\mathbf y,z)}
{\displaystyle
\int d^3\mathbf y\,
W_R(\lvert\mathbf r-\mathbf y\rvert)\Delta_{\rm cell}(\mathbf y,z)
}.
\label{eq:smoothed_recombinations}
\end{equation}
The resulting quantity $n_{{\rm rec},R}(\mathbf r,z)$ is the
cumulative number of recombinations per hydrogen nucleus, smoothed
over the spherical region of radius $R$.

\begin{figure}
    \centering
    \includegraphics[width=1\linewidth]{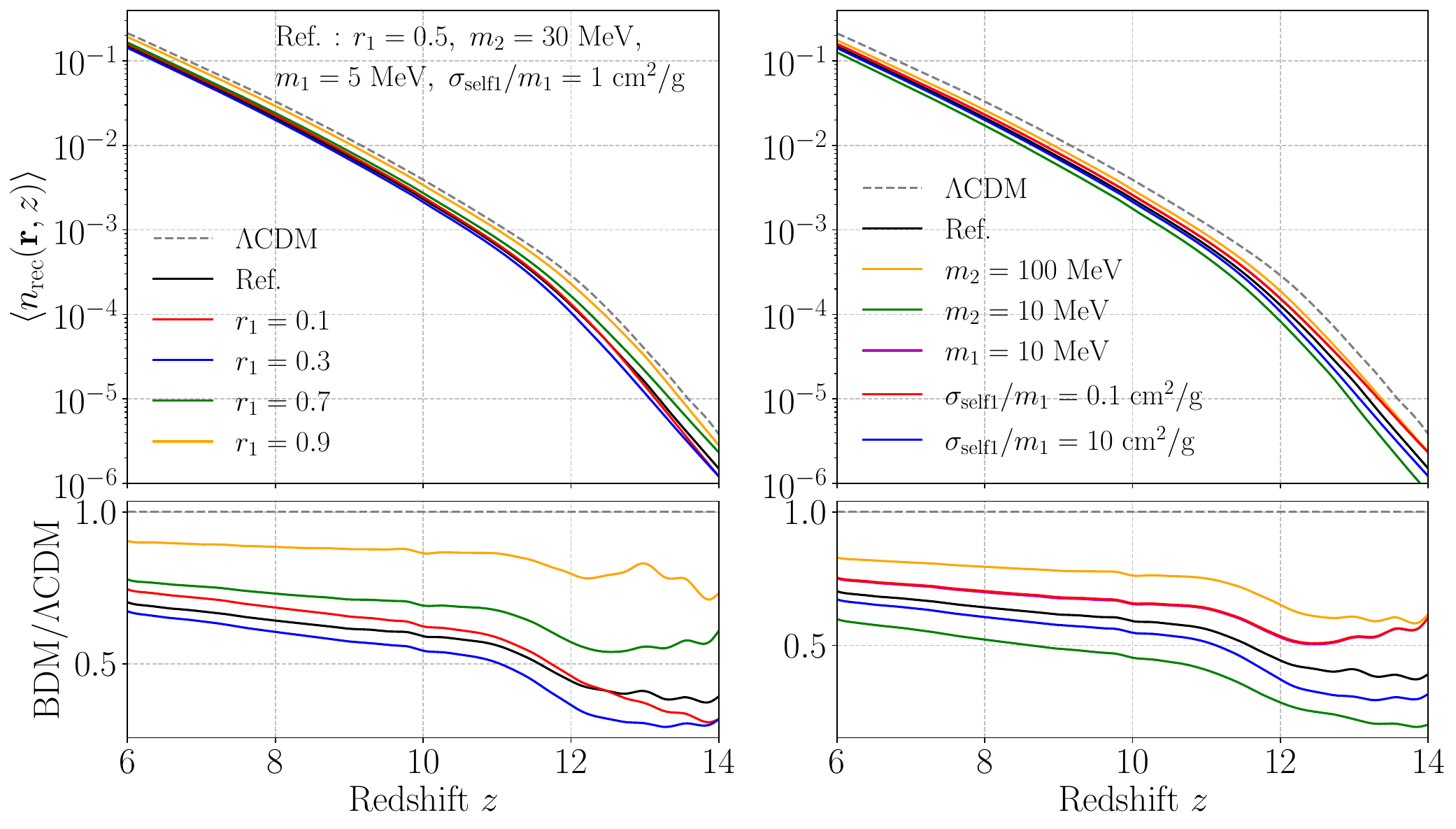}
    \caption{Redshift evolution of the volume-averaged cumulative number of recombinations per hydrogen nucleus (see Eq.~\eqref{eq:cumulative_recombinations}) for $\Lambda$CDM and representative BDM parameter choices. The upper panels show $\langle n_{\rm rec}(\mathbf r,z)\rangle$, while the lower panels show the ratio of the BDM prediction to the corresponding $\Lambda$CDM result. The left panels illustrate the dependence on the light-component fraction $r_1$, whereas the right panels show the effects of varying $m_2$, $m_1$, and $\sigma_{\rm self1}/m_1$ around the reference model. The cumulative recombination count is lower in BDM because suppressed baryon clustering reduces the amount of dense ionized gas and delayed ionization shortens the period over which recombinations accumulate.
    \label{fig:global_nrec} }
\end{figure}

We use the same MHR density distribution and self-shielding
prescription for BDM and \LCDM. The BDM dependence included in our
calculation enters through several resolved quantities. The modified
baryon density field $\delta_b^{\rm NL}(\mathbf r,z)$ changes
$\Delta_{\rm cell}(\mathbf r,z)$ and thereby modifies the local
parameters of $P_V(\mathbf r,z,\Delta)$. The suppression of
$f_{\rm coll}^{\rm eff}$ changes the ionizing photon production rate
$dn_{\rm ion}/dt$, which propagates to $\Gamma_{\rm HII}$,
$\Delta_{\rm ss}$, and $x_{\rm HI}^{\rm sg}$. BDM also modifies
$z_{\rm ion}(\mathbf r)$ and $R_{\rm HII}(\mathbf r,z)$, changing
both the instantaneous recombination rate and the period over which
recombinations accumulate. 

As shown in Figure~\ref{fig:global_nrec}, the representative BDM models
predict fewer cumulative recombinations than \LCDM. The suppression of
baryon clustering reduces the amount of dense ionized gas, where
recombinations occur most efficiently, while the reduced abundance of
star-forming halos delays ionization and shortens the period over which
recombinations accumulate.

The adopted MHR distribution was calibrated using standard CDM
simulations and has not been validated for BDM cosmology. The
suppression of small-scale density fluctuations in BDM may alter the
abundance of unresolved dense gas structures. This change can affect the recombination
correction  and is not captured by
the present prescription. A dedicated calibration using
high-resolution hydrodynamical simulations of BDM cosmology is left
for future work.

%%%%%%%%%%%%%%%%%%%%%%%%%%%%%%%%%%%%%%%%%%%%%%%%%%%
\subsection{Ly$\alpha$ line profile}
\label{app_lya_line_profile}
%%%%%%%%%%%%%%%%%%%%%%%%%%%%%%%%%%%%%%%%%%%%%%%%%%%

In this appendix, we summarize the line-profile ingredients that lead to the
narrow-line expression used in Eq.~\eqref{eq:Palpha_narrowline}. We begin
from the exact Ly$\alpha$ scattering rate per neutral hydrogen atom,
\begin{equation}
P_\alpha(\mathbf r,z)
=
4\pi
\int d\nu\,
\frac{J_\nu(\mathbf r,z)}{h\nu}\,
\sigma_\alpha(\nu,\mathbf r,z).
\label{eq_Palpha_def_app}
\end{equation}
Here $\sigma_\alpha(\nu,\mathbf r,z)$ is the Ly$\alpha$ scattering cross
section, with units of ${\rm cm}^2$. The factor $J_\nu/(h\nu)$ has units of
${\rm s}^{-1}\,{\rm cm}^{-2}\,{\rm Hz}^{-1}$, so
Eq.~\eqref{eq_Palpha_def_app} gives a transition rate per atom after integrating
over frequency.

It is customary to factor the frequency dependence of the Ly$\alpha$ cross
section into an overall strength and a normalized line profile,
\begin{equation}
\sigma_\alpha(\nu,\mathbf r,z)
=
\sigma_{\alpha,0}\,
\phi_\alpha(\nu-\nu_\alpha,\mathbf r,z),
\qquad
\int d\nu\,
\phi_\alpha(\nu-\nu_\alpha,\mathbf r,z)
=
1 .
\label{eq_sigma_alpha_profile_app}
\end{equation}
The line profile $\phi_\alpha$ has units of ${\rm Hz}^{-1}$ and describes
how the absorption probability is distributed around the Ly$\alpha$
line-center frequency $\nu_\alpha$ of neutral hydrogen.
The normalization in
Eq.~\eqref{eq_sigma_alpha_profile_app} is imposed at fixed $\mathbf r$ and $z$,
so the line profile integrates to unity.
The overall normalization $\sigma_{\alpha,0}$ has units of
${\rm cm}^2\,{\rm Hz}$ and is determined by the oscillator strength of the
Ly$\alpha$ transition,
\begin{equation}
\sigma_{\alpha,0}
=
\frac{\pi e^2}{m_e c}\,
f_\alpha ,
\label{eq_sigma_alpha0_app}
\end{equation}
where $f_\alpha\simeq0.4162$.

We specify the line profile by including Doppler broadening from the thermal
motion of hydrogen atoms and natural broadening from the finite lifetime of
the excited state. In the local rest frame of the gas, hydrogen atoms have
thermal velocities set by the kinetic temperature $T_k(\mathbf r,z)$. A
photon with frequency $\nu$ in this frame is therefore seen by individual
atoms at slightly Doppler-shifted frequencies. As a result, atoms with
different line-of-sight thermal velocities can resonantly scatter photons
whose frequencies differ slightly from the line-center frequency
$\nu_\alpha$. This thermal motion broadens the line into a Gaussian profile
with a width set by the Doppler width $\Delta\nu_D(\mathbf r,z)$. In
addition, the finite lifetime of the excited $2p$ state produces natural
broadening, which gives a Lorentzian contribution controlled by the decay
rate $\Gamma_\alpha$. The combined effect of thermal Doppler broadening and
natural broadening is described by the Voigt profile~\cite{Furlanetto:2006jb}
\begin{equation}
\phi_\alpha(\nu-\nu_\alpha,\mathbf r,z)
=
\frac{1}{\sqrt{\pi}\,\Delta\nu_D(\mathbf r,z)}
H\!\left[
a_\alpha(\mathbf r,z),
u_\alpha(\nu,\mathbf r,z)
\right],
\label{eq_lya_voigt_profile_app}
\end{equation}
where
\begin{equation}
u_\alpha(\nu,\mathbf r,z)
\equiv
\frac{\nu-\nu_\alpha}{\Delta\nu_D(\mathbf r,z)},
\qquad
a_\alpha(\mathbf r,z)
\equiv
\frac{\Gamma_\alpha}{4\pi\Delta\nu_D(\mathbf r,z)} .
\label{eq_lya_voigt_variables_app}
\end{equation}
The Doppler width is
\begin{equation}
\Delta\nu_D(\mathbf r,z)
=
\frac{\nu_\alpha}{c}
\sqrt{
\frac{2k_B T_k(\mathbf r,z)}{m_{\rm H}}
},
\label{eq_lya_doppler_width_app}
\end{equation}
and the dimensionless Voigt function is
\begin{equation}
H(a,u)
=
\frac{a}{\pi}
\int_{-\infty}^{\infty}
dy\,
\frac{e^{-y^2}}{(u-y)^2+a^2}.
\label{eq_voigt_function_app}
\end{equation}
With this convention, $\int du\,H(a,u)=\sqrt{\pi}$ and hence
$\int d\nu\,\phi_\alpha(\nu-\nu_\alpha,\mathbf r,z)=1$ at fixed
$\mathbf r$ and $z$.

Substituting Eq.~(\ref{eq_sigma_alpha_profile_app}) into Eq.~\eqref{eq_Palpha_def_app}, we obtain
\begin{equation}
P_\alpha(\mathbf{r},z)
=
4\pi\,\sigma_{\alpha,0}
\int d\nu\,
\frac{J_\nu(\mathbf{r},z)}{h\nu}\,
\phi_\alpha\!\bigl(\nu-\nu_\alpha,\mathbf{r},z\bigr).
\label{eq:Palpha_profile_form}
\end{equation}
The Ly$\alpha$ line profile has a sharp peak at the resonance frequency $\nu_\alpha$, so most of the integral comes from a narrow interval near line center.
In BDM cosmology, the lower gas kinetic temperature further reduces the Doppler width, $\Delta\nu_D(\mathbf r,z)\propto T_k^{1/2}(\mathbf r,z)$, making the line profile more localized around $\nu_\alpha$. Therefore the approximation that $J_\nu(\mathbf r,z)/(h\nu)$ is nearly constant across the line profile is at least as well justified as in \LCDM. 
The normalization of the line profile then gives the narrow-line estimate
\begin{equation}
P_\alpha(\mathbf{r},z)
\simeq
4\pi\,
\sigma_{\alpha,0}\,
\widehat J_\alpha(\mathbf r,z)\,
S_\alpha(\mathbf{r},z),
\label{eq:Palpha_narrowline_app}
\end{equation}
where $S_\alpha(\mathbf{r},z)$ is a dimensionless factor of order unity that accounts for the fact that, after many Ly$\alpha$ scatterings, the radiation spectrum near the line center is no longer perfectly smooth. Recoil slightly shifts photons to lower frequencies, while thermal motions of atoms cause photons to diffuse in frequency space. Together, these effects modify the local spectral shape around $\nu_\alpha$ and $S_\alpha(\mathbf{r},z)$ summarizes their impact.

%%%%%%%%%%%%%%%%%%%%%%%%%%%%%%%%%%%%%%%%%%%%%%%%%%%
\subsection{SKA1-Low sensitivity calculation}
\label{app_ska_sensitivity}
%%%%%%%%%%%%%%%%%%%%%%%%%%%%%%%%%%%%%%%%%%%%%%%%%%%

This appendix explains how the instrumental uncertainty $\sigma_{\rm SKA}(k,z)$ is obtained with \Sense. The calculation begins with the array layout of the interferometer. Consider two antenna stations with position vectors $\mathbf r_i$ and $\mathbf r_j$ expressed in a common coordinate system. The baseline of this station pair is given by the position of station $i$ relative to station $j$,
\begin{equation}
\mathbf b_{ij}
=
\mathbf r_i-\mathbf r_j .
\label{eq_baseline_vector}
\end{equation}
We introduce $\hat{\mathbf e}_u$ and $\hat{\mathbf e}_v$, two orthonormal basis vectors spanning the plane perpendicular to the line-of-sight direction $\hat{\mathbf n}$. The components of the baseline along these two vectors, measured in units of the observing wavelength, are the dimensionless interferometric coordinates
\begin{equation}
u_{ij}
=
\frac{\mathbf b_{ij}\cdot \hat{\mathbf e}_u}{\lambda_{\rm obs}},
\qquad
v_{ij}
=
\frac{\mathbf b_{ij}\cdot \hat{\mathbf e}_v}{\lambda_{\rm obs}},
\label{eq_uv_coordinates}
\end{equation}
where $\lambda_{\rm obs}=\lambda_{21}(1+z)$.
For notational brevity, we omit the station-pair indices $ij$ and denote the corresponding baseline coordinate by
\begin{equation}
\mathbf u=(u,v).
\end{equation}
The plane spanned by these coordinates is called the $uv$ plane.

For radiation arriving from a direction separated by an angle $\theta$, the path length to the two stations differs slightly,
producing a relative phase difference
$\Delta\phi\simeq 2\pi |\mathbf u|\theta$.
The baseline becomes sensitive to angular variations on the scale for which this phase changes by order unity, giving the characteristic angular scale
$\theta\sim |\mathbf u|^{-1}$. Thus larger $|\mathbf u|$ probes smaller
angular structure on the sky.

At redshift $z$, an angular scale $\theta$ on the sky corresponds to a
transverse comoving scale
\begin{equation}
\lambda_\perp^{\rm com}
=
X(z)\theta
\simeq
\frac{X(z)}{|\mathbf u|},
\end{equation}
where $X(z)$ is the transverse comoving distance to redshift $z$. The
transverse wavenumber associated with the baseline is therefore
\begin{equation}
k_\perp
=
\frac{2\pi}{\lambda_\perp^{\rm com}}
\simeq
\frac{2\pi |\mathbf u|}{X(z)} .
\label{eq_kperp_u}
\end{equation}
In a spatially flat cosmology,
\begin{equation}
X(z)=\chi(z).
\label{eq_X_def}
\end{equation}

The line-of-sight direction is encoded in the observing frequency because
the observed 21\,cm frequency determines the redshift,
$\nu_{\rm obs}=\nu_{21}/(1+z)$. Therefore
$z=\nu_{21}/\nu_{\rm obs}-1$, and
\begin{equation}
\frac{dz}{d\nu_{\rm obs}}
=
-\frac{\nu_{21}}{\nu_{\rm obs}^2}
=
-\frac{(1+z)^2}{\nu_{21}} .
\end{equation}
The minus sign only indicates that a higher observing frequency corresponds
to a lower redshift. Since the sensitivity calculation uses interval
lengths, we define the positive frequency-to-distance Jacobian factor
\begin{equation}
Y(z)
\equiv
\left|
\frac{d\chi}{d\nu_{\rm obs}}
\right|
=
\left|
\frac{d\chi}{dz}
\frac{dz}{d\nu_{\rm obs}}
\right|
=
\frac{c(1+z)^2}{\nu_{21}H(z)} ,
\label{eq_Y_conversion}
\end{equation}
where $d\chi/dz=c/H(z)$. Thus a bandwidth $\Delta\nu_{\rm obs}$ corresponds
to a comoving line-of-sight depth
\begin{equation}
\Delta r_\parallel
=
Y(z)\,\Delta\nu_{\rm obs}.
\end{equation}

The Fourier variable conjugate to the observing frequency $\nu_{\rm obs}$ is written as $\eta$. Its units are ${\rm Hz}^{-1}$, and it gives the inverse frequency scale over which the signal varies.
For a Fourier mode with frequency-space scale $\Delta\nu_{\rm obs}\sim \eta^{-1}$, the corresponding comoving line-of-sight wavelength is
\begin{equation}
\lambda_\parallel^{\rm com}
=
Y(z)\Delta\nu_{\rm obs}
\simeq
\frac{Y(z)}{\eta}.
\end{equation}
Since the cosmological wavenumber is $2\pi$ divided by the corresponding
comoving wavelength, the line-of-sight wavenumber is
\begin{equation}
k_\parallel
=
\frac{2\pi}{\lambda_\parallel^{\rm com}}
=
\frac{2\pi\eta}{Y(z)} .
\label{eq_kparallel_eta}
\end{equation}
This gives the mapping from the instrumental coordinates $(u,v,\eta)$ to the transverse and line-of-sight cosmological wavenumbers $(k_\perp,k_\parallel)$.
The spherically averaged power spectrum is then expressed in terms of
\begin{equation}
k
=
\sqrt{k_\perp^2+k_\parallel^2}.
\end{equation}

For a single $uv\eta$ cell, the thermal-noise uncertainty on the dimensional 21\,cm power spectrum is~\cite{Morales:2004ca,McQuinn:2005hk,Pober:2013jna}
\begin{equation}
\epsilon_N(\mathbf u,\eta,z)
=
X^2(z)Y(z)\,
\frac{\Omega_p^2}{\Omega_{pp}}\,
\frac{T_{\rm sys}^2(\nu_{\rm obs})}
{2\,t_{\mathbf u}^{\rm tot}} .
\label{eq_21cmsense_noise_cell}
\end{equation}
Here $\epsilon_N$ has the same units of ${\rm mK}^2\,{\rm Mpc}^3$ as the dimensional 21\,cm power spectrum
$P_{21}$.
The beam factors are
\begin{equation}
\Omega_p
=
\int d\Omega\,
A_p(\hat{\mathbf n}),
\qquad
\Omega_{pp}
=
\int d\Omega\,
A_p^2(\hat{\mathbf n}) .
\label{eq_beam_factors}
\end{equation}
Here $A_p(\hat{\mathbf n})$ is the dimensionless primary-beam response of a
station in the direction $\hat{\mathbf n}$. It describes how strongly the
station receives radiation from each direction on the sky. The quantity
$\Omega_p$ is the effective beam solid angle, while $\Omega_{pp}$ is the
corresponding squared-beam solid angle that enters the variance of the
power-spectrum estimate. Thus the combination $\Omega_p^2/\Omega_{pp}$ acts
as the effective angular area of the observation after primary-beam
weighting. The factor $X^2(z)$ converts this effective angular area into a
transverse comoving area.
The quantity $t_{\mathbf u}^{\rm tot}$ is the total integration time
accumulated in the corresponding $uv$ cell after summing over all baselines
and observing days. The factor $1/t_{\mathbf u}^{\rm tot}$ provides the frequency dimension, while $Y(z)$
converts frequency intervals into comoving line-of-sight distances.

In Eq.~\eqref{eq_21cmsense_noise_cell}, since the thermal-noise contribution
scales as $1/t_{\mathbf u}^{\rm tot}$, cells with larger accumulated
integration time have smaller thermal noise. The value of
$t_{\mathbf u}^{\rm tot}$ is computed by \Sense\ from the
SKA1-Low AA4 baseline distribution and the observing strategy. 
In this work, we consider two total observing times,
\begin{equation}
t_{\rm obs}
=
180\times 6\,{\rm hr}
=
1080\,{\rm hr},
\label{eq_tobs_SKA_1080}
\end{equation}
and
\begin{equation}
t_{\rm obs}
=
1000\times 5\,{\rm hr}
=
5000\,{\rm hr}.
\label{eq_tobs_SKA_5000}
\end{equation}
These values are used independently of the moderate and optimistic
sky-temperature assumptions discussed below.
The detailed gridding of the baseline tracks and the resulting calculation of $t_{\mathbf u}^{\rm tot}$ follow the \Sense\ sensitivity algorithm, and we refer to Refs.~\cite{Pober:2013jna,Murray:2024} for details.

The same equation shows that the thermal-noise contribution grows as
$T_{\rm sys}^2$. Following the SKA1-Low design prescription for receiver temperature,\footnote{See the SKA1 imaging science performance document: \url{https://indico.skatelescope.org/event/270/attachments/1919/2406/SKA1_Science_PerformanceRevA_draft2.pdf}.

}
\begin{equation}
T_{\rm sys}(\nu_{\rm obs})
=
T_{\rm sky}(\nu_{\rm obs})
+
T_{\rm rcv}(\nu_{\rm obs}),
\qquad
T_{\rm rcv}(\nu_{\rm obs})
=
40\,{\rm K}
+
0.1T_{\rm sky}(\nu_{\rm obs}),
\label{eq_Tsys_SKA_components}
\end{equation}
or equivalently,
\begin{equation}
T_{\rm sys}(\nu_{\rm obs})
=
40\,{\rm K}
+
1.1T_{\rm sky}(\nu_{\rm obs}).
\label{eq_Tsys_SKA}
\end{equation}

Here $T_{\rm sky}$ denotes the effective sky temperature entering
through the station beam. In addition to the observing time, the
sensitivity estimate depends on the adopted sky-temperature model
and foreground treatment. We therefore consider two forecast
scenarios, moderate and optimistic, which differ in both assumptions,
to compute the effective uncertainty bands in
Figure~\ref{fig:21power_fix_k}.

For the moderate scenario, we adopt the sky-temperature
model~\cite{2017isra.book,Mesinger:2016ddl,Greig:2015qca}
\begin{equation}
T_{\rm sky}^{\rm mod}(\nu_{\rm obs})
=
60\,{\rm K}
\left(
\frac{\nu_{\rm obs}}{300\,{\rm MHz}}
\right)^{-2.55},
\label{eq_Tsky_moderate}
\end{equation}
while for the optimistic scenario, we
adopt~\cite{DeBoer:2016tnn,Mason:2022obt}
\begin{equation}
T_{\rm sky}^{\rm opt}(\nu_{\rm obs})
=
120\,{\rm K}
\left(
\frac{\nu_{\rm obs}}{150\,{\rm MHz}}
\right)^{-2.55}.
\label{eq_Tsky_optimistic}
\end{equation}
At the low observing frequencies relevant to this work, the sky-noise
contribution is dominated by diffuse Galactic synchrotron emission.
Since $\nu_{\rm obs}=\nu_{21}/(1+z)$, higher redshift corresponds to
lower observing frequency, where $T_{\rm sky}$ is larger.
The quadratic dependence on $T_{\rm sys}$ in
Eq.~\eqref{eq_21cmsense_noise_cell} therefore increases the thermal-noise
contribution at higher redshift, reducing the instrumental sensitivity.

In addition to the sensitivity limitations imposed by thermal noise, foreground contamination restricts the range of Fourier modes available for 21\,cm power spectrum analysis. Radiation arriving from directions away from the pointing center produces a geometric time delay that depends on the baseline and the direction of incidence. The resulting frequency-dependent phase difference causes even spectrally smooth foreground emission to contaminate a wedge-shaped region in $(k_{\perp},k_{\parallel})$ space~\cite{Parsons_2012,Pober:2013jna}. The geometric boundary of this foreground wedge is approximately given by 
\begin{equation}
    k_{\parallel,\rm wedge}
    =
    \frac{X(z)H(z)}{c(1+z)}
    \sin\theta_{\rm max}\,k_\perp,
\end{equation}
where $\theta_{\rm max}$ is the maximum angular separation from the
pointing center considered in the foreground treatment.
Fourier modes satisfying
$|k_{\parallel}|<k_{\parallel,\rm wedge}$
are excluded from the sensitivity calculation.

For the moderate scenario, we adopt the horizon limit corresponding to $\theta_{\rm max}=90^\circ$. To account for foreground contamination extending beyond this geometric boundary, we additionally apply a horizon buffer of $\Delta k_{\parallel,\rm buf}=0.1\,h\,{\rm Mpc}^{-1}$. The exclusion condition for this scenario becomes $|k_\parallel|<k_{\parallel,\rm wedge}+\Delta k_{\parallel,\rm buf}$.
For the optimistic scenario, we adopt the primary-beam limit implemented in \Sense, with no additional foreground buffer.
The sky-temperature models and foreground treatments adopted for the moderate and optimistic forecast scenarios are summarized in Table~\ref{tab:observational_setting}.

\begin{table}[t]
\centering
\renewcommand{\arraystretch}{1.5}
\begin{tabular}{@{}lcc@{}}
\toprule
Setting & Moderate & Optimistic \\
\midrule
Sky temperature, $T_{\rm sky}$
&
$60\,{\rm K}
\left(\frac{\nu_{\rm obs}}{300\,{\rm MHz}}\right)^{-2.55}$
&
$120\,{\rm K}
\left(\frac{\nu_{\rm obs}}{150\,{\rm MHz}}\right)^{-2.55}$
\\
Foreground limit
&
Horizon limit
&
Primary-beam limit
\\
Foreground buffer, $\Delta k_{\parallel,\rm buf}$
&
$0.1\,h\,{\rm Mpc}^{-1}$
&
$0$
\\
\bottomrule
\end{tabular}
\caption{
Sky-temperature and foreground assumptions used to compute the effective uncertainty bands in Figure~\ref{fig:21power_fix_k}.
Modes satisfying $|k_\parallel|<k_{\parallel,\rm wedge}+\Delta k_{\parallel,\rm buf}$ are excluded. The foreground wedge boundary is determined by the horizon limit for the moderate scenario and by the primary-beam limit for the optimistic scenario.
}
\label{tab:observational_setting}
\end{table}

The cell-level thermal noise in Eq.~\eqref{eq_21cmsense_noise_cell} is first
defined in the instrumental coordinates $(\mathbf u,\eta)$. We label a cell by $\alpha$, with coordinates
$(\mathbf u_\alpha,\eta_\alpha)$. The corresponding wavenumbers are
\begin{equation}
k_{\perp,\alpha}
=
\frac{2\pi |\mathbf u_\alpha|}{X(z)},
\qquad
k_{\parallel,\alpha}
=
\frac{2\pi \eta_\alpha}{Y(z)},
\qquad
k_\alpha
=
\sqrt{
k_{\perp,\alpha}^2+k_{\parallel,\alpha}^2
}.
\label{eq_cell_k_mapping}
\end{equation}
For this cell labeled by $\alpha$, the thermal-noise contribution to the dimensionless 21\,cm power spectrum is given by
\begin{equation}
\Delta_N^2(k_\alpha,z)
=
\frac{k_\alpha^3}{2\pi^2}\,
\epsilon_N(\mathbf u_\alpha,\eta_\alpha,z).
\label{eq_DeltaN_cell_def}
\end{equation}
For a spherical bin centered at $k$, \Sense\ combines all available Fourier cells whose $k_\alpha$ values fall inside that bin. We denote this set of cells by ${\cal B}(k,z)$.
In addition to thermal noise, the sensitivity calculation includes sample variance arising from the finite number of independent cosmological modes sampled by the observation.
The resulting SKA1-Low uncertainty can be approximately expressed as
\begin{equation}
\sigma_{\rm SKA}(k,z)
=
\left[
\sum_{\alpha\in{\cal B}(k,z)}
\frac{1}
{\left[\Delta_N^2(k_\alpha,z)+\Delta^2_{21,\Lambda\rm CDM}(k_\alpha,z)\right]^2}
\right]^{-1/2}.
\label{eq_sigma_SKA_bin}
\end{equation}
Here $\Delta^2_{21,\Lambda\rm CDM}(k_\alpha,z)$ denotes the fiducial \LCDM\ 21\,cm power spectrum used to estimate the sample-variance contribution. The numerical calculation follows the mode-averaging prescription implemented in \Sense.

%%%%%%%%%%%%%%%%%%%%%%%%%%%%%%%%%%%%%%%%%%%%%%%%%%%
\subsection{Machine Learning}
\label{app:machine_learning}
%%%%%%%%%%%%%%%%%%%%%%%%%%%%%%%%%%%%%%%%%%%%%%%%%%%

\begin{figure}[t]
\centering
\includegraphics[width=\textwidth]{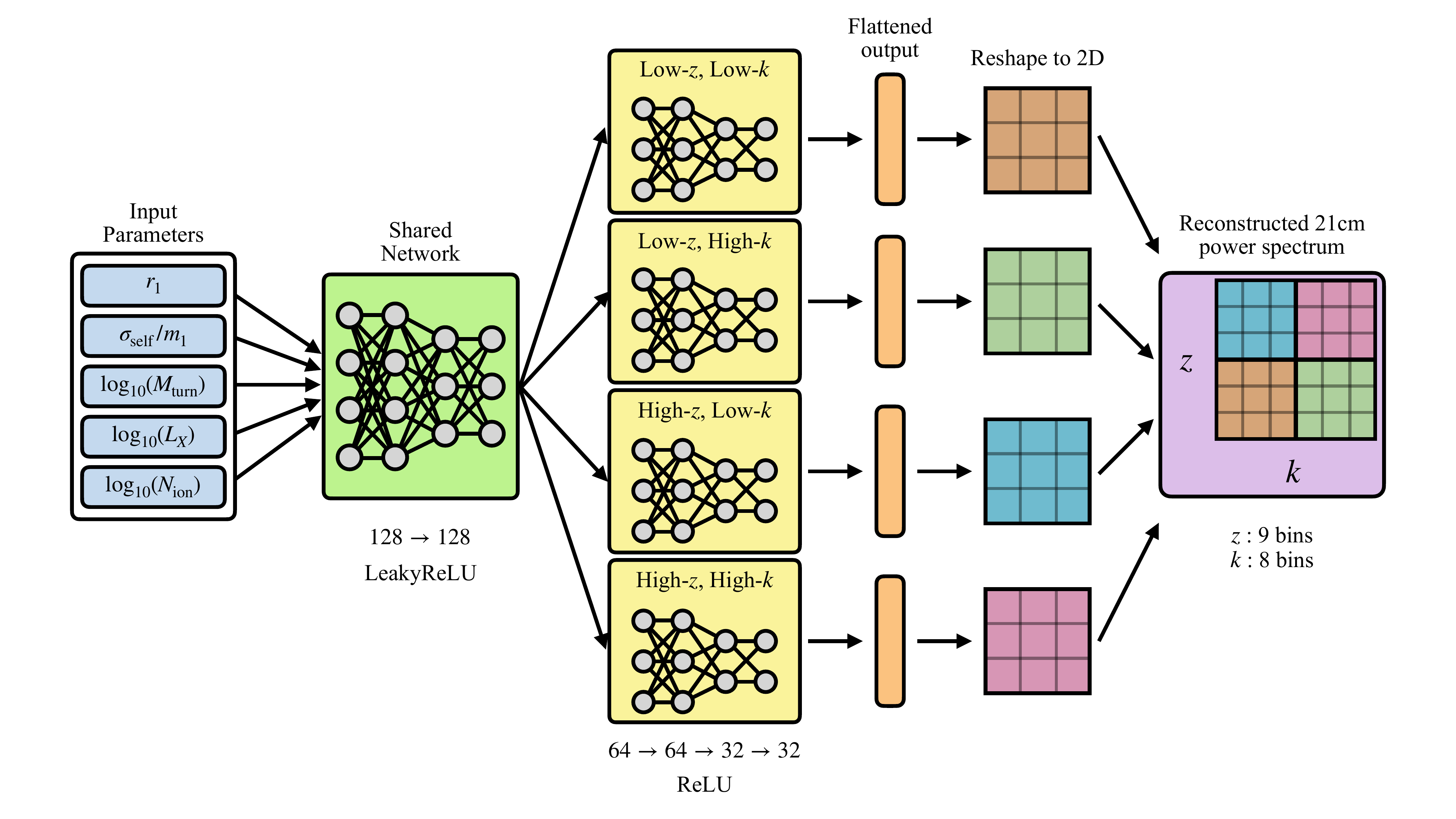}
\vspace*{-0.8cm}
\caption{Schematic diagram of the emulator network architecture used in this study. A five-dimensional input parameter vector is first processed by a shared MLP $(128 \rightarrow 128)$ with LeakyReLU activation, and the resulting latent features are passed to four quadrant-specific branch networks $(64 \rightarrow 64 \rightarrow 32 \rightarrow 32)$ with ReLU activation. Each branch predicts a flattened output for one quadrant of the $(k,z)$ plane, which is reshaped into a two-dimensional patch. The four patches are then combined to reconstruct the full 21 cm power spectrum over 9 redshift bins and 8 wavenumber bins.}
\label{fig:ml_net}
\end{figure}

In this section, we describe the machine learning (ML) architecture used in Sec.~\ref{sec:bayesian}. Figure \ref{fig:ml_net} illustrates the network architecture of the emulator. The inputs to the network are the parameters used in the Bayesian analysis, \{$r_1$, $\sigma_{\rm self1}/m_1$, $\log_{10}(M_{\rm turn})$, $\log_{10}(L_X^{\rm II}/{\rm SFR})$, $\log_{10}(N_{\gamma})$\}. This vector is first processed by a shared Multi-Layer Perceptron (MLP), with two hidden layers $(128,128)$ and LeakyReLU activations with negative slope 0.1, to extract common latent features. The target 21\,cm power spectrum on the $(k,z)$ plane is divided into four quadrants, each of which is modeled by an independent branch network. Each branch has hidden layers $(64,64,32,32)$ with ReLU activations and outputs a flattened vector for its assigned quadrant. These vectors are reshaped into two-dimensional power-spectrum patches and then assembled into the full $\Delta^2_{21}(k,z)$ map. This quadrant-based architecture allows the emulator to capture distinct signal behavior in different redshift and wavenumber regions while sharing a common latent representation of the physical input parameters.

\begin{figure}[t]
\centering
\includegraphics[width=0.7\textwidth]{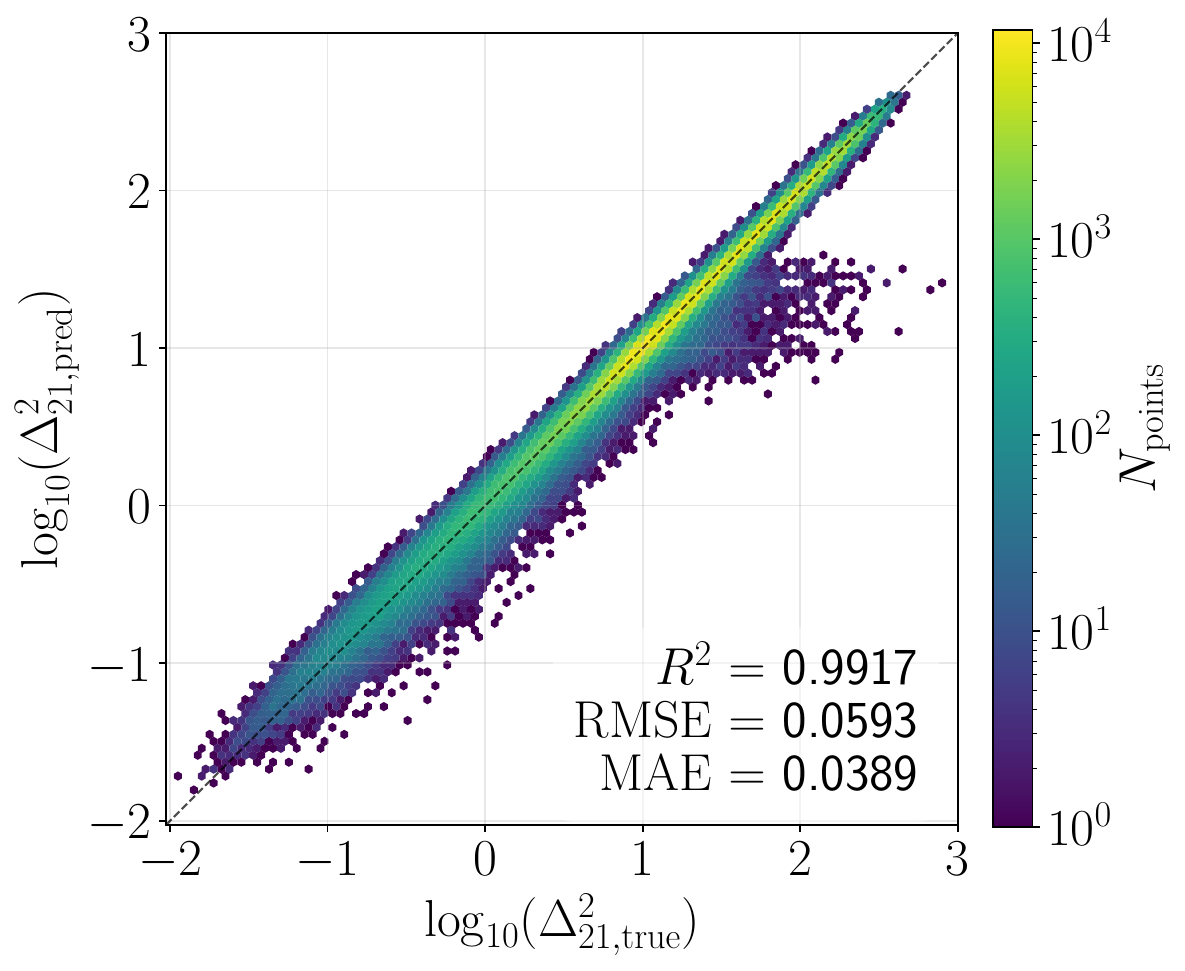}
\vspace*{-0.5cm}
\caption{Comparison between the true and predicted 21\,cm power spectra on the test samples. Each hexagonal bin contains the number of grid points falling in the corresponding region of the $(\log_{10}\Delta^2_{21,\rm true},\log_{10}\Delta^2_{21,\rm pred})$ plane, as indicated by the color scale. The dashed line denotes the ideal prediction $y=x$. The strong clustering of points around this line indicates that the emulator accurately reproduces the overall amplitude and variation of the 21\,cm power spectrum across the test samples. The resulting performance metrics are $R^2=0.9917,{\rm RMSE}=0.0593,{\rm MAE}=0.0389$ in log-space.}
\label{fig:ml_net2}
\end{figure}
To train the emulator, we use the typical MSE loss function, 
\begin{eqnarray}    
L  &=& \frac{1}{N_{\rm batch}N_{\rm out}}\sum_{n=1}^{N_{\rm batch}}\sum_{a=1}^{N_{\rm out}}\left[\hat{Y}^{\rm std}_{n,a}-Y^{\rm std}_{n,a}\right]^2 \, , 
\end{eqnarray}
where $N_{\rm batch}$ denotes the number of training samples in a mini-batch and $N_{\rm out}$ is the number of output components in the flattened 21\,cm power-spectrum vector. 
For an output grid with $N_z$ redshift bins and $N_k$ wavenumber bins, one has $N_{\rm out}=N_zN_k$. 
$Y^{\rm std}_{n,a}$ and $\hat{Y}^{\rm std}_{n,a}$ denote the standardized values of $y=\log_{10}(\Delta^2_{21,\rm true})$ and $\hat{y}=\log_{10}(\Delta^2_{21,\rm pred})$ for the $n$-th sample at the $a$-th flattened grid point $(k,z)$, respectively.

The data set contains $\mathcal{O}(5\times 10^4)$ samples, of which 70\%, 10\%, and 20\% are allocated to training, validation and test sets, respectively. The output power spectrum is represented on a discrete grid spanning $9 \le z \le 17$ with $N_z=$9 bins and $0.2\,{\rm Mpc}^{-1} \le k \le 1.0\,{\rm Mpc}^{-1}$ with $N_k=$8 bins. We employ an Adam optimizer \cite{Kingma:2014vow} with a batch size of $N_{\rm batch}=$32, a learning rate of $10^{-4}$ and a weight decay of $10^{-6}$.  The training is performed for up to 400 epochs, and the checkpoint at epoch 241, which corresponds to the minimum validation loss, is selected as the final emulator. The accuracy of the trained emulator is evaluated on the independent test set. We compare the predicted and true values of $\log_{10}(\Delta^2_{21})$ for all $(k,z)$ grid points, as shown in Figure~\ref{fig:ml_net2}. The predictions are tightly clustered around the ideal relation, shown by the dashed line, demonstrating that the emulator reproduces the simulation outputs with high accuracy over the full range of 21\,cm power spectrum amplitudes. We obtain $R^2=0.9917$, ${\rm RMSE}=0.0593$, and ${\rm MAE}=0.0389$ in log-space for the correlation coefficient ($R^2$), Root Mean Squared Error (${\rm RMSE}$) and Mean Absolute Error (${\rm MAE}$), respectively. The corresponding typical multiplicative error is approximately $15\%$ based on the $\text{RMSE}$ in $\log_{10}\Delta^2_{21}$.
Figure~\ref{fig:ml_net2} shows a small deviation from the ideal one-to-one correlation near ($\log_{10}(\Delta^2_{21, \rm true})$, $\log_{10}(\Delta^2_{21, \rm pred})$) = (2,1). These points account for only about 0.1\% of the full dataset and are therefore not of practical concern. Our investigation tells us that the majority of these outliers are associated with samples lying near the boundaries of the chosen $(k,z)$ parameter space. Expanding the scanning range is expected to reduce or eliminate this boundary effect.

\bibliographystyle{JHEP}
\bibliography{draft}

\end{document}